\documentclass[twocolumn]{aastex63}
\usepackage{amsmath}
\usepackage{multirow}
\hypersetup{linkcolor=red,citecolor=magenta,filecolor=cyan,urlcolor=magenta}

\defcitealias{2020AJ....160..279K}{K20}
\defcitealias{2023A&A...673A.114H}{H23}
\defcitealias{2024A&A...686A..42H}{H24}

\submitjournal{APJs}

\shorttitle{Planets in clusters}
\shortauthors{Yuan-Zhe Dai et al.}
\graphicspath{{./}{figures/}}

\begin{document}

\title{Understanding the Planetary Formation and Evolution in Star Clusters(UPiC)-II: Catalog of planets/candidates in Open Clusters and Moving Groups} 

\correspondingauthor{Hui-Gen Liu}
\email{huigen@nju.edu.cn}

\author{Yuan-Zhe Dai}
\affiliation{School of Astronomy and Space Science, Nanjing University, 163 Xianlin Avenue, Nanjing, 210023, People's Republic of China}
\affiliation{Key Laboratory of Modern Astronomy and Astrophysics, Ministry of Education, Nanjing, 210023, People's Republic of China}
\author{Hui-Gen Liu}
\affiliation{School of Astronomy and Space Science, Nanjing University, 163 Xianlin Avenue, Nanjing, 210023, People's Republic of China}
\affiliation{Key Laboratory of Modern Astronomy and Astrophysics, Ministry of Education, Nanjing, 210023, People's Republic of China}
\author{Xiaoying Pang}
\affiliation{Xi'an Jiaotong-Liverpool University, Suzhou, People's Republic of China}
 \affiliation{Shanghai Key Laboratory for Astrophysics, Shanghai Normal University, 100 Guilin Road, Shanghai 200234, P.R. China}
\author{Yueyue Jiang}
\affiliation{School of Astronomy and Space Science, University of Chinese Academy of Sciences, No. 19A, Yuquan Road, Beijing 100049, People's Republic of China}
\affiliation{Astrophysics Division, Shanghai Astronomical Observatory, Chinese Academy of Sciences, 80 Nandan Road, Shanghai 200030, People's Republic of China}
\author{Jerome~P.~de~Leon}
\affiliation{Komaba Institute for Science, The University of Tokyo, 3-8-1 Komaba, Meguro, Tokyo 153-8902, Japan}
\author{Jing Zhong}
\affiliation{Astrophysics Division, Shanghai Astronomical Observatory, Chinese Academy of Sciences, 80 Nandan Road, Shanghai 200030, People's Republic of China}
\author{Ji-Lin Zhou}
\affiliation{School of Astronomy and Space Science, Nanjing University, 163 Xianlin Avenue, Nanjing, 210023, People's Republic of China}
\affiliation{Key Laboratory of Modern Astronomy and Astrophysics, Ministry of Education, Nanjing, 210023, People's Republic of China}

\begin{abstract}
Detecting planets in open clusters offers a unique opportunity to test planet formation theories in clustered environments. The precisely determined ages of young open clusters make their planets particularly valuable for tracing the early evolution of planetary systems. As the second paper of the UPiC project, this study focuses on stars in stellar groups that host transiting planets or planetary candidates. We categorize these stellar groups into Open Clusters (OCs) and Moving Groups (MGs) based on the Jacobi radius to investigate potential differences in their planetary systems. By cross-matching the latest star cluster catalogs with catalogs of transiting planets and candidates, we have compiled the most extensive catalog to date, containing 106 confirmed planets and 168 candidates within OCs and MGs. We refitted the structural parameters of these stellar groups and identified substructures using the \texttt{HDBSCAN} and Gaussian Mixture Model (GMM) algorithms. Our analysis reveals the density evolution of both MGs and OCs during their first Gyr. We find that MGs consistently exhibit a significantly higher planet fraction than OCs, regardless of sample selection, particularly for Hot Jupiters. Furthermore, exoplanet radii show a clear dichotomy at early stages: most sub-Jupiters evolve into Neptune-sized planets within 100 Myr, while super-Jupiters undergo only minimal contraction. These results suggest that young sub-Jupiters (\textless 100 Myr) represent puffy, Neptune-mass planets undergoing vigorous photoevaporation, whereas Jupiter-mass planets can maintain their atmospheres. We also report evidence for the early emergence of the hot-Neptune desert at 100 Myr in both OCs and MGs.

\end{abstract}


\section{Introduction}
The stellar groups, such as open clusters (OCs) and moving groups (MGs), serve as bridges that connect galaxies, stars, and planets. Stellar groups provide a unique way to characterize the homogeneous parameters of stellar members, such as age and metallicity. Detecting and studying exoplanets in stellar groups can also reveal the influence of the stellar environments on planetary systems. Recent papers on the environmental influence on planet formation and evolution (e.g., \citealt{2020Natur.586..528W, 2020ApJ...905L..18K, 2021AJ....162...46D, 2021ApJ...911L..16L}) have renewed our perspective and placed planet formation and evolution within a broader context. 

In the early stages of planet formation, the intrinsic properties of protoplanetary disks and host stars, as well as external environments, constrain disk lifetimes and influence subsequent planet formation, particularly the formation of giant planets. Frequent stellar flybys during the planet-forming stage can truncate and accelerate disk dissipation \citep{2006ApJ...642.1140O, 2018MNRAS.478.2700W}. The majority protoplanetary disks are exposed to strong ultraviolet irradiation from nearby massive stars, which drives the wind, truncates the disk, and accelerates disk dissipation \citep{1998ApJ...499..758J, 2003ApJ...582..893M, 2018MNRAS.480.4080D, 2018MNRAS.478.2700W}. Strong external photoevaporation and disk truncation can suppress giant planet formation \citep{2022MNRAS.515.4287W,2024A&A...689A.338H}. A recent review paper on external photoevaporation \citep{2025arXiv250212255P} mentioned that studying young (< 100 Myr) exoplanetary systems can provide convincing observational evidence of a causal relationship between the environment and the properties of the planet. Therefore, all of these mechanisms play an important role during planet formation. 

Although mature planetary systems around field stars were originally formed in different stellar groups, they may differ from those in star clusters because systems formed in different environments are mixing together. E.g. frequent flybys in star clusters can significantly influence planetary architecture. Investigating planets in young OCs and MGs is a feasible way to learn the environmental influences on planetary evolution.
Especially those planetary systems younger than 100 Myr may preserve their original characteristics \citep{2024AJ....168..239D, 2024AJ....168..194T}. E.g. the young planetary system AU Mic \citep{2020Natur.582..497P} and V 1298 Tau \citep{2019ApJ...885L..12D}, which have recently exited the disk-hosting stage (< 10 Myr), can provide a relatively pristine orbital architecture, allowing us to study the system before longer-timescale effects—such as secular chaos or tidal realignment—have had sufficient time to operate. 

Additionally, planets in young OCs and MGs can offer important constraints on the evolution of the planetary radius by testing mechanisms such as planet cooling \citep{2007ApJ...659.1661F, 2021MNRAS.507.2094M}, atmospheric escape driven by photoevaporation \citep{2013ApJ...775..105O}, or core-powered mass loss \citep{2018MNRAS.476..759G}. Previous works \citep{2020MNRAS.495.4924N, 2023AJ....166..175F, 2024AJ....167..210V} have attempted to estimate the occurrence of young exoplanets and compare them with mature systems to illuminate planet formation and evolution processes. 

However, there are some challenges in detecting planets in young star clusters. Serious dilution effects in a dense star field will shallow the transit depths of the exoplanet, especially for Keper and TESS, both of which have large pixel scales. The activities of young stars generate larger photomitric jitters, and obscure the transiting singals. Until now, less than 30 transiting planets younger than 100 Myr have been detected. The small number limits the ability to conduct further statistical analysis to clarify the formation mechanisms of different planets. Thus, it is valuable to compile a relatively comprehensive catalog of young planets, including candidates in different stellar groups, particularly those in young OCs and MGs.

Thanks to the unprecedented accuracy of Gaia DR2 and DR3, more OCs have been discovered in the Milky Way \citep{2018A&A...618A..93C, 2019ApJS..245...32L, 2019AJ....158..122K, 2020AJ....160..279K,QIN2023}. With more stars in star clusters, we can expect to identify more planets in these clusters. Since most stars are born in star clusters \citep{2003ARA&A..41...57L} and 50\% of field stars have planets according to recent exoplanet census data \citep{2021ARA&A..59..291Z}, there should be many planets in star clusters. In our first series of papers, namely UPiC-I \citep{2023AJ....166..219D}, we compiled a catalog of transiting planets in star clusters, which includes 73 confirmed planets and 84 planet candidates. However, several efforts should be made to improve this catalog. First, we can expand our previous catalog in light of rapid progress in the field of open clusters, as recent efforts to search for new open clusters have significantly increased the available sample \citep{2023A&A...673A.114H}. Second, some of the stellar groups identified in \citep{2020AJ....160..279K} (the star cluster catalog used in UPiC-I \citep{2023AJ....166..219D}) are more appropriately classified as moving groups (MGs), which are usually younger (<200 Myr), with flattened radial density profiles \citep{2024A&A...686A..42H}. 


In this paper, UPiC-II, we aim to obtain a more comprehensive catalog of young planets in stellar groups than in UPiC-I and classify these stellar groups into open clusters (OCs) and moving groups (MGs), following the methods described in \cite{2024A&A...686A..42H}. Also we try to provide the parameters of the host clusters of these planets, e.g. the age, stellar density profile, for further studies. In Section \ref{method}, we introduce our methods for obtaining the catalog of confirmed planets and candidates in OCs and MGs. The method we used to obtain the parameters of the stellar groups is also described. In Section \ref{sec3}, we collect and describe the properties of the archived catalogs and compare the properties of OCs and MGs. In Section \ref{Planethosts}, we compare the planet host stars in OCs and MGs, and we indicate the evolution of OCs and MGs. In section \ref{results}, we compare the planets in OCs and MGs, indicate the evolution trend due to the radius evolution of planets with gas envelope, and discuss some possible explanation of the results. In Section \ref{discussion}, we clarify some limitations of the catalogs obtained. Finally, in Section \ref{summary}, we summarize the major conclusions. 

\section{Method} \label{method}
In this section, we will introduce the methodology for obtaining the catalog of young planets and candidates, as well as the way to obtain the updated cluster parameters, e.g. cluster mass $M_{cl}$, tidal radius $r_{t}$, core radius $r_{c}$, jacobi radius $r_{J}$(where the star cluster self-gravity balance the galactic potential), etc.

\subsection{Choice of cluster catalogues}
The cluster catalogs adopted in this paper are based on the two catalogs, i.e. \cite{2020AJ....160..279K}(Hereafter \citetalias{2020AJ....160..279K}) and \cite{2023A&A...673A.114H}(Hereafter \citetalias{2023A&A...673A.114H}). \citetalias{2020AJ....160..279K} focuses on the extended structures of the Milky Way, identifying 8,292 comoving groups within 3 kpc and galactic latitude $|b| < 30^\circ$ by applying the unsupervised machine learning algorithm Hierarchical Density-Based Spatial Clustering of Applications with Noise \texttt{HDBSCAN} \cite{McInnes2017hdbscanHD} to Gaia DR2's 5D data. Using updated astrometric data from Gaia DR3, \citetalias{2023A&A...673A.114H} performs an all-sky star cluster search, extending the Gaia magnitude cut to Gmag \textless 20 (compared to Gmag \textless 18 in \citetalias{2020AJ....160..279K}), and recovers 7,167 star clusters containing 729 million sources. Note, \citetalias{2020AJ....160..279K} find more stellar groups with extended structures, while \citetalias{2023A&A...673A.114H} focus on open clusters with more strict cuts. Thus, we adopted both catalogs in this paper to enlarge stars in stellar groups.

Since \citetalias{2020AJ....160..279K} identify stellar groups via GAIA DR2, we have systematically aligned all astrometric parameters with the Gaia DR3 reference frame, to ensure astrometric consistency with \citetalias{2023A&A...673A.114H}. We also revised the distance estimates for stars in both \citetalias{2020AJ....160..279K} and \citetalias{2023A&A...673A.114H} according to the distance catalog of \cite{2021AJ....161..147B}, thus improving the spatial accuracy of the data set. Then we used \texttt{HDBSCAN} to refit the star groups with planet host stars and identified 168 groups, including some sparse structures, which are considered substructures of one cluster in \citetalias{2020AJ....160..279K}. Note that the star population is exactly the same as \citetalias{2020AJ....160..279K}. The stellar groups identified in \citetalias{2023A&A...673A.114H} based on GAIA DR3 are completely adopted.

In UPiC-I, we extended the definition of open clusters to include not only core cluster members, but also coeval stars in tidal tails. This significantly increases the number of planets in star clusters and aids our investigation of the environmental influences on planet formation and evolution. However, there are no cluster morphology parameters in \citetalias{2020AJ....160..279K} and \citetalias{2023A&A...673A.114H}. Recent work \cite{2024A&A...686A..42H}(see below \citetalias{2024A&A...686A..42H}) provides a catalog with cluster morphology parameters including core radius and tidal radius based on \citetalias{2023A&A...673A.114H}, which includes a subset of $\sim$ 7,000 clusters. We will refit and calculate all morphology parameters in two catalogs as described in section \ref{sec-clparameter} and section \ref{OC_MG_classify}. 

\subsection{Cross match with planet hosts} \label{sec-cross}


After identifying stars in stellar groups, we cross-matched the stellar group catalogs from \citetalias{2020AJ....160..279K} and \citetalias{2023A&A...673A.114H} with multiple planet catalogs, including confirmed planets from the NASA Exoplanet Archive and candidates from KOI, K2, TOI, PATHOS \citep{2020MNRAS.495.4924N}, and CTOI. The results of this cross-matching are summarized in Table \ref{crossmatch}.

The initial cross-match yielded 79 confirmed planets from the \citetalias{2020AJ....160..279K} sample and 57 from the \citetalias{2023A&A...673A.114H} sample. We then applied the UPIC-I selection criterion, retaining only KOIs with a score greater than 0.9 and excluding false positives, which resulted in 43 and 12 KOIs from the respective catalogs. From the K2 candidates, we selected 4 from \citetalias{2020AJ....160..279K} and 35 from \citetalias{2023A&A...673A.114H}. For TOIs, we selected 75 and 37 from the respective catalogs after excluding those flagged as ``APC'', ``FP'', or ``FA''.

The sources of CTOIs are highly heterogeneous, as they are identified by different authors using various pipelines (e.g., Luke Bouma's CDIPS \citep{2019ApJS..245...13B} and PATHOS \citep{2019MNRAS.490.3806N}). Some have even been contributed by citizen scientists (e.g., TOI-6650). Therefore, caution is advised when selecting CTOIs for analysis. We assign a disposition to each CTOI in Table \ref{plt} to classify them into different groups (see Appendix \ref{pl} for definitions). The classification follows these rules:
\begin{itemize}
    \item CTOIs with existing disposition flags (e.g., ``PC'', ``FP'') in the catalog retain their original flags.
    \item For CTOIs without a disposition flag, we apply the criterion from \cite{2019MNRAS.490.3806N}, which suggests that most confirmed planets lie within 25 Earth radii, and candidates beyond this threshold are likely eclipsing binaries. Thus, we flag such CTOIs as ``FP''.
    \item CTOIs exhibiting only a single transit event or lacking an orbital period measurement are classified as ``Single Transit''.
\end{itemize}
It is beyond the scope of this paper to perform specific statistical validation for each CTOI. Our goal is to compile a comprehensive catalog of planets and candidates in OCs and MGs to avoid missing any potential targets. Nevertheless, we inspected the SPOC light curves of each CTOI and excluded obvious false positives, such as CTOI 198390707.01, CTOI 206544316.01, CTOI 206544316.02, CTOI 206544316.03, and CTOI 1438206544316.04 (see Appendix \ref{pl}). Additionally, the ``comments'' column in our catalog includes the original notes for each TOI and CTOI.

Cross-matching  yields 171 planet candidates from \citetalias{2023A&A...673A.114H} and 235 from \citetalias{2020AJ....160..279K}, corresponding to 112 planet-hosting stellar groups (58,128 stars total) and 156 stellar groups (96,523 stars total), respectively. After removing duplicate entries, we obtain a final sample of 346 planets and candidates residing in 280 unique stellar groups, as summarized in Table \ref{plclt}.

Since some stellar groups hosting planets identified in \citetalias{2020AJ....160..279K} contains only a small numbers of stars, and can not be distinguished from the filed star obviously, thus we also add a flag which indicates some stellar groups as high-quality groups (HQ), while others as low quality groups(LQ). For stellar groups from \citetalias{2024A&A...686A..42H}, we flag those stellar groups with a median CMD class greater than 0.5 and an astrometric signal-to-noise ratio (S/N) greater than 5$\sigma$ as high-quality groups (HQ), while those below this threshold are flagged as low-quality groups(following the criteria described in \citetalias{2024A&A...686A..42H}). For stellar groups from \citetalias{2020AJ....160..279K}, we flagged them based on their basic distribution of their astrometry data and color-magnitude diagram(CMD). Groups with astrometry data satisfying similar criteria (Eq. 5 and 6 in \cite{2021A&A...646A.104H}) and a clear main sequence in the CMD are classified as HQ; the rest are considered LQ.  



The catalog in Table \ref{plht} also includes membership probabilities for the planet-hosting stars. However, while \citetalias{2023A&A...673A.114H} provides these probabilities for individual stars, such data are unavailable in \citetalias{2020AJ....160..279K}. To address this, we introduce a flag to indicate high-probability (HP) or low-probability (LP) membership specifically for the planet hosts from \citetalias{2020AJ....160..279K}. We classify a host as HP if its star aligns with the main sequence in the CMD and with the cluster core in spatial/kinematic diagrams. Hosts are flagged as LP if they exhibit deviations in the CMD or significant offsets (e.g., $>$ 1\(\sigma\)) from the group center in spatial/kinematic dimensions (see column descriptions in Table \ref{combine}).

\begin{deluxetable}{ccc} \label{crossmatch}
\centering
\tabletypesize{\scriptsize}
\tablewidth{0pt} 
\tablenum{1}
\tablecaption{Comparison of H2023 and K2020 in cross-matching with planets/candidates}
\tablehead{
\colhead{Numbers}     &\colhead{H2023}      & \colhead{\citetalias{2020AJ....160..279K}} 
}
\colnumbers
\startdata 
Confirmed Planets(NASA) &57   & 79  \\
KOI & 12 & 43 \\
K2 & 35 & 4 \\
TOI & 37 & 75 \\
PATHOS & 40 & 43 \\
CTOI & 88 & 121 \\
Planet Host Stars & 171 & 235 \\
Planet Host Stellar Groups & 112 & 156 \\
Members of Stellar Groups with Planets/candidates  & 58128
& 96523
\enddata
\tablecomments{This is the preliminary results of planets/candidates in stellar groups. Details in Appendix \ref{catalogs}}
\end{deluxetable}

\subsection{Cluster parameter estimation} \label{sec-clparameter}
Although \citetalias{2024A&A...686A..42H} provides comprehensive cluster morphology parameters based on \citetalias{2023A&A...673A.114H}, we reprocess the structural parameters for all planet-hostingto ensure parameter homogeneity and minimize methodological bias, we employ a unified approach to fit all of the stellar groups from host planets in both \citetalias{2020AJ....160..279K} and \citetalias{2023A&A...673A.114H} using a unified methodology. This ensures homogeneity and minimizes methodological biases in our comparative analysis. The parameters are derived as follows: \\
 Physical Parameters: For groups from \citetalias{2024A&A...686A..42H} and \citetalias{2020AJ....160..279K}, we directly adopt their published values (e.g., distance D and age). However, we estimate stellar masses via isochrone fitting using the provided ages, stellar magnitudes, and an assumed solar metallicity.\\
Structural Parameters: For all groups, we consistently derive the core radius $r_c$, tidal radius $r_t$, central surface density $\rho_0$, and the surrounding stellar density for planet-hosting stars $\rho_p$ through a unified radial density profile analysis using both King and EFF models.

\subsubsection{Mass determination} \label{mass_cal}
We use isochrone fitting to estimate the mass of individual stars and sum them to derive the total cluster mass. However, this method is subject to several selection biases, such as detection completeness which depends on a cluster's distance and location. The binary fraction also significantly impacts the mass calculation \cite{Pang2023,Jiang2024}. 

We compared our total mass estimates with the bias-corrected values from \citetalias{2024A&A...686A..42H}. The mean deviation is approximately 20\%, which is consistent with the expected impact of the aforementioned biases. While this indicates that our mass values are likely underestimates (due to incompleteness and unresolved binaries), we proceed without correction to maintain internal consistency in our subsequent analysis. We will discuss the influence of this mass underestimation in Section \ref{sec6.1}, where we show that it does not significantly alter the stellar group classification. In Table \ref{plclt}, we list the total masses estimated in this work alongside the values from \citetalias{2024A&A...686A..42H} to demonstrate the differences for each group.

\subsubsection{Radial Density Profile} \label{RDPfitting}
The spatial distribution of stellar members is typically described by fitting the radial density profile (RDP). The empirical King model is commonly adopted, particularly for globular clusters and bounded open clusters. However, for young clusters, especially those in the Large Magellanic Cloud (LMC), the observed surface brightness profiles do not align well with the King model. Instead, these profiles are better described by the Elson, Fall, and Freeman (EFF) model \citep{1987ApJ...323...54E}, because some of these young open clusters are not tidally truncated, similar structure also discovered in nearby OCs \cite{2022ApJ...931..156P}. Given that our sample also includes young OCs, we employ both the King and EFF models to derive the structural parameters as an comparison. Additionally, using these parameters, we can calculate the surrounding stellar number density of planet-hosting stars.

According to \cite{King1962}, the empirical King model can be defined as:
\begin{equation}
\rho \left(r\right) = 
    \begin{cases}
        \rho_{0} \cdot \left(\frac{1}{\sqrt{1+(r/r_{c})^{2}}}-\frac{1}{\sqrt{1+(r_{t}/r_{c})^{2}}}\right)+c & \text{if } r < r_{t} \\
        c & \text{if } r \geq r_{t}
    \end{cases}
\end{equation}
where $\rho_{0}$ is a scaling constant related to the central surface density in unit of stars per squared parsec, $r_{c}$ is the core radius, $r_{t}$ is the tidal radius, and $\rho(r)$ is the radial surface density profile.

While for the EEF model, the RDP can be described as follows,
\begin{equation}
    \rho\left(r\right) = \rho_{0}\left[1+\left(\frac{r}{r_{\rm EEF}}\right)^{2}\right]^{-\gamma/2},
\end{equation}
where $r_{\rm EEF}$ is the scale radius and $\gamma$ is the power-law slope at large radii. I.e. a large value of $\gamma$ corresponds to a steep radial distribution, and vice versa. We also compute the core radius, $r_{c}$, with $r_{\rm EEF}$ and $\gamma$ obtained from fitting the EFF profile using Equation (22) in \cite{1987ApJ...323...54E}

\begin{equation}
    r_{c} \approx r_{\rm EEF} \sqrt{2^{2/\gamma}-1}. 
\end{equation}

To fit the RDP in both models, we projected the coordinates of each star onto the plane of the sky, tangential to the celestial sphere at the coordinates of the cluster centers, as recommended by \cite{2006A&A...445..513V} and \cite{2018A&A...612A..70O}. The projected coordinates for each cluster star are calculated as follows:

\begin{equation} \label{proj}
    \begin{aligned}
        X_{proj} & = D \cdot \sin{\left(\alpha-\alpha_{c}\right)} \cdot \cos{\left(\delta \right)} \\
        Y_{proj} & = D \left[\cos{\left(\delta_{c}\right)} \cdot \sin{\left(\delta\right)}- \sin{\left(\delta_{c}\right)} \cdot \cos{\left(\delta\right)} \cdot \cos{\left(\alpha-\alpha_{c}\right)}\right]
    \end{aligned}
\end{equation}
D is the heliocentric distance of the stellar group. We chose the median distance of group members as D. $\alpha_{c}$ and $\delta_{c}$ are the right ascension and declination of the cluster center. 
For open clusters with a distinct core, it is straightforward to determine the center by adopting either the median positional value or the density peak. We utilize the kernel density estimation (KDE) from \texttt{Scikit-learn} to calculate the stellar surface density distribution on a $100 \times 100$ grid. The cluster center is then defined as the average position of the 20 highest-density grid points \citep{2022AJ....164...54Z}.

In contrast, for sparse moving groups lacking a centralized core, the aforementioned averaging method becomes unreliable due to low stellar density. For such cases, we instead adopt the median positional coordinates of group members as the group's center.

\begin{figure*}[!htbp]
    \centering
    \includegraphics[width=0.95\linewidth]{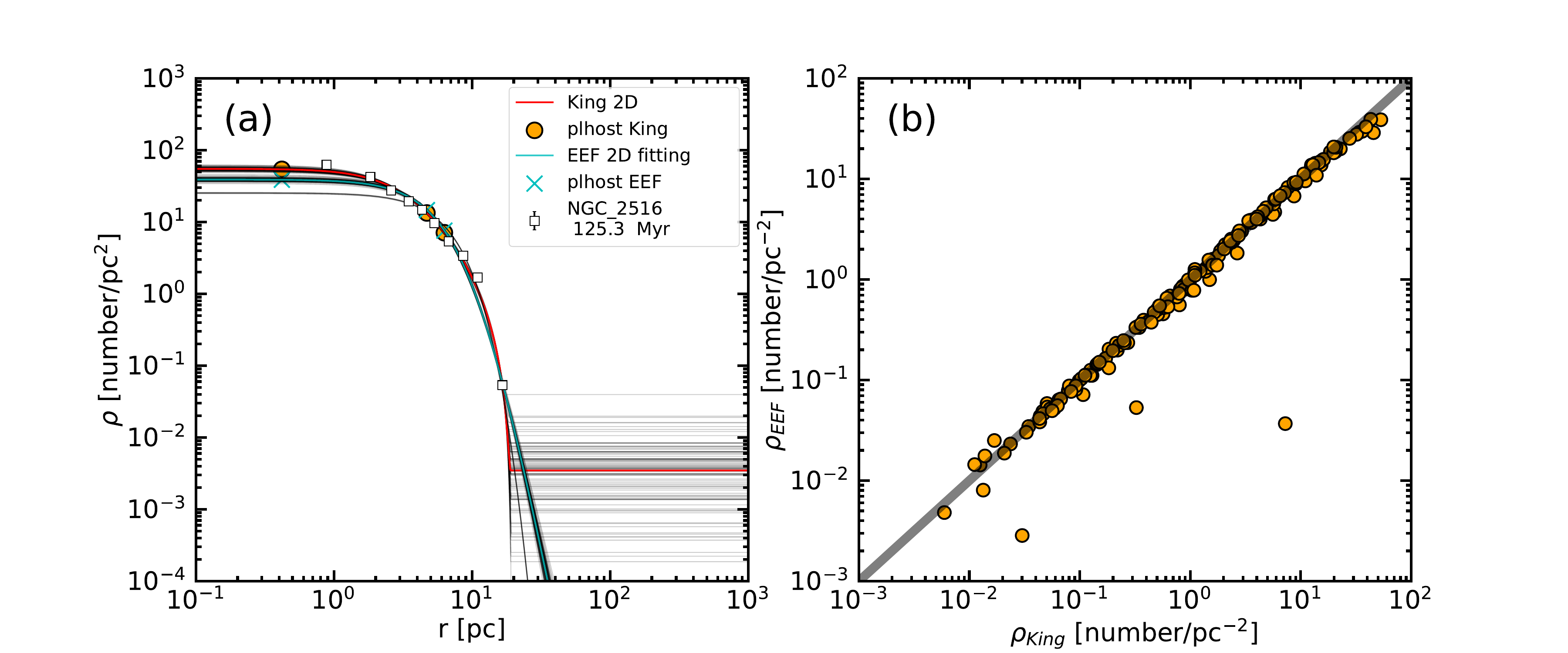}
    \caption{\textbf{Panel (a)}: Radial density profile (RDP) of the typical open cluster NGC~2516, with King (red solid curve) and EFF (blue solid curve) model fits demonstrating excellent agreement with observational data. The projected 2D stellar number density of confirmed and candidate planetary host stars that derived from radial density profile (RDP) analysis are overplotted. I.e. orange circles for King model fit and blue crosses for EEF model fit.  \textbf{Panel (b)}: Comparative density estimates for host stars using King vs. EFF models, with cluster membership data from \cite{2023A&A...673A.114H}. The diagonal unity line (gray dashed) highlights model consistency.}
    \label{RDP}
\end{figure*}

The radial distance ($r$) of each star from the cluster center is calculated as: $r = \sqrt{X_{\rm proj}^2 + Y_{\rm proj}^2}$. Fig.~\ref{RDP} (a) displays the radial density profile (RDP) of the typical open cluster NGC~2516. Both the King profile (red curve) and EFF profile (blue curve) provide good fits to the observational data. The consistency between the two models is further demonstrated in Fig.~\ref{RDP}(b), where surround stellar number density of planet-hosting stars derived from King and EFF model fits show excellent agreement. This indicates no statistically significant differences between the two modeling approaches. Given this equivalence, we primarily adopt the King profile in subsequent analyses for RDP fitting procedures, derivation of morphological parameters (core radius, tidal radius, and etc.), and comparative studies of stellar group structures. For comparison, we also list the fitted parameters via EEF model in Table \ref{plclt}.

\begin{figure*}[!htbp]
    \centering    \includegraphics[width=0.95\linewidth]{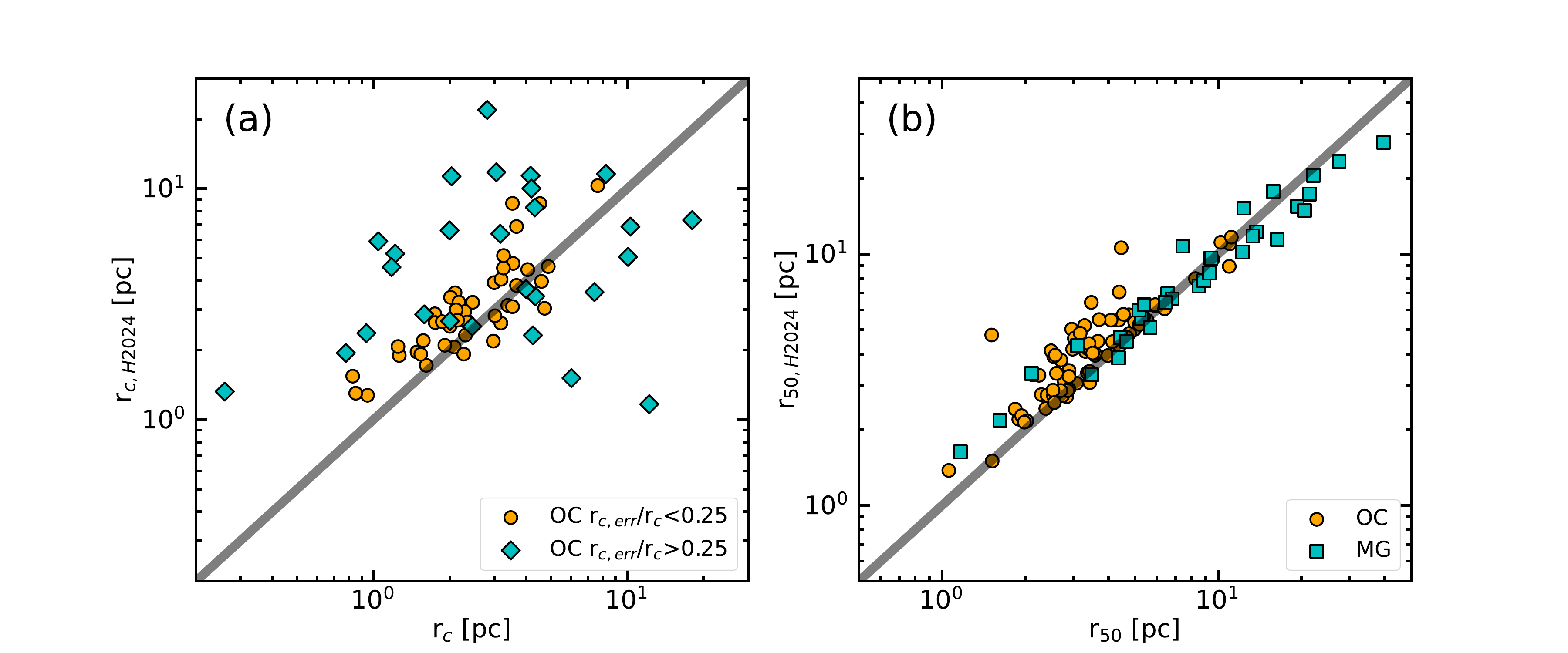}
    \caption{\textbf{Panel (a):} Comparison of the core radius of open clusters between this work and \citetalias{2024A&A...686A..42H}. We divide the open clusters (OCs) into two groups based on the relative uncertainty of the core radius: \( r_{c,\text{err}}/r_{c} < 0.25 \) (orange) and \( r_{c,\text{err}}/r_{c} > 0.25 \) (blue). \textbf{Panel (b):} Comparison of the half-number radius (\( r_{50} \)) of both open clusters (OCs) and moving groups (MGs) between this work and \citetalias{2024A&A...686A..42H}. Note, the labels of OC and MG are directly adopted from \citetalias{2024A&A...686A..42H}.}
    \label{King_EEF_compare}
\end{figure*}

In Fig. \ref{King_EEF_compare}, we compare the cluster morphology parameters derived in our work with those from \citetalias{2024A&A...686A..42H}. Panel (a) shows the comparison of the core radius $r_{c}$. The open clusters classified by \citetalias{2024A&A...686A..42H}  (OCs) are divided into two groups according to the relative uncertainty of the core radius, specifically $\text{r}_{c,\text{err}}/\text{r}_{c} < 0.25$ (orange) and $\text{r}_{c,\text{err}}/\text{r}_{c} > 0.25$ (blue). Here, the relative uncertainty of the core radius is derived from the radial density profile (RDP) fitting using the King model. For the samples with smaller uncertainties, the estimated core radii are consistent with those in \citetalias{2024A&A...686A..42H}. However, the other samples show larger variations in core radius $r_{c}$. This large variation is likely due to poor radial density profile fitting caused by the low number of cluster members. Many of the blue samples have fewer members (with a median value of N${\rm med,b} \sim 200$), whereas most orange samples have higher numbers (N${\rm med,o} \sim 650$). For example, some blue samples —such as Casado 20, SAI 149, and Theia 7—do not exhibit a clear cluster core in their projected 2D distributions.

In panel (b) of Fig.~\ref{King_EEF_compare}, the half-number radius $r_{50}$ shows great consistency for both OCs and Moving groups (MGs) classified in \citetalias{2024A&A...686A..42H}. Since moving groups usually have a very flat density radius profile, in contrast to the typical cluster RDP with a sharp decline, the core radius derived from RDP fitting may be only suitable for characterizing open clusters rather than moving groups. Due to the consistency in the half-number radius $r_{50}$, it's better to use $r_{50}$ to characterize the size of open clusters and moving groups. All the fitted parameters are demonstrated in Table \ref{plclt}. 

\section{Classify stellar groups into OCs and MGs} \label{sec3}
This section outlines the classification scheme for distinguishing between open clusters (OCs) and moving groups (MGs) among the stellar groups hosting planets or candidates. Our sample comprises 112 groups from \citetalias{2023A&A...673A.114H} and 156 from \citetalias{2020AJ....160..279K}. The classification results for these two samples are presented separately in the following sections: Sample 1 (based on \citetalias{2023A&A...673A.114H}) in Section \ref{H2024_classify}, and Sample 2 (based on \citetalias{2020AJ....160..279K}) in Section \ref{K2020_classify}.

\subsection{Stellar Groups from \citetalias{2023A&A...673A.114H}: Sample 1} \label{H2024_classify}
For the stellar groups in \citetalias{2023A&A...673A.114H}, we estimated their morphological parameters as described in Section \ref{RDPfitting}. We then classified them as OCs or MGs based on their Jacobi radius, following a methodology similar to that of \citetalias{2024A&A...686A..42H}.

The Jacobi radius $r_{J}$ is defined as the boundary where the self-gravity of a cluster balances the galactic potential. 
According to \cite{2010ARA&A..48..431P,2024A&A...686A..42H}, the formula for the Jacobi radius is:
\begin{equation} \label{eq_RJ}
    r_{J} = \left(\frac{GM}{4\Omega^{2}-k^{2}}\right)^{\frac{1}{3}}
\end{equation}
\( r_{J} \) is determined by the mass \( M \) of a cluster, the circular frequency \( \Omega \) and the epicyclic frequency \( k \) of the cluster's orbit around the galatic center, assuming the orbit is circular. Stars outside $r_{J}$ can be retained for a period, e.g. the tidal tails of clusters \citep{2021A&A...645A..84M,2022ApJ...931..156P}. Here, we use the \texttt{galpy}\footnote{\url{https://github.com/jobovy/galpy}} package \cite{2015ApJS..216...29B} to calculate the galactic potential, including the circular frequency \( \Omega \) and the epicyclic frequency \( k \).

\begin{figure*}
    \centering    \includegraphics[width=0.95\linewidth]{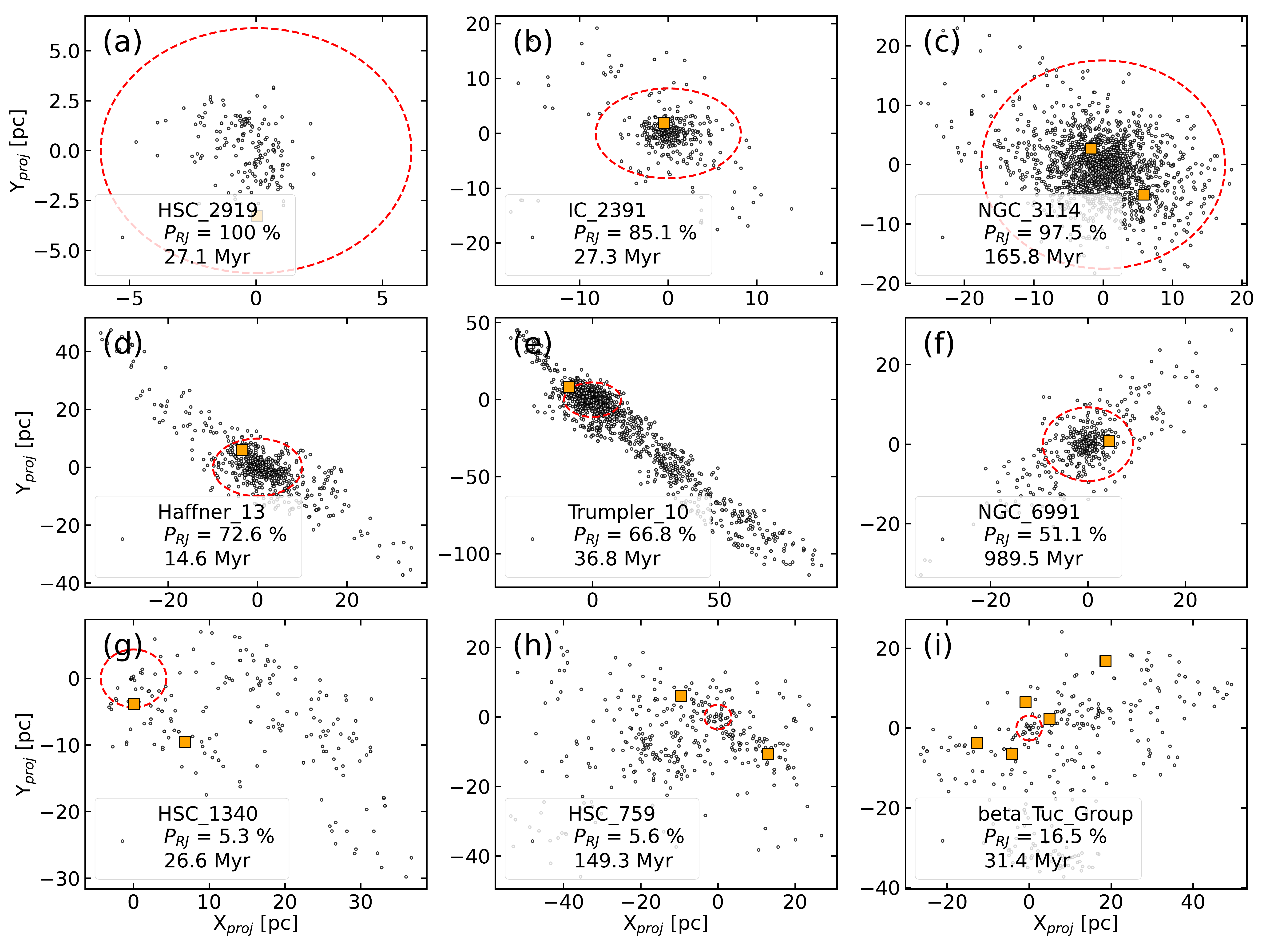}
    \caption{The 2D projection distribution of three typical stellar groups. Red circles: Theoretical Jacobi radii ($r_{\mathrm{J}}$) calculated via Equation~\ref{eq_RJ}. Black dots: members of each stellar groups. Orange squares: host stars of confirmed planets and planet candidates.}   
    \label{OC_MG_classify}
\end{figure*}




We distinguish OCs and MGs as follows:
\begin{itemize}
    \item \textbf{Open Clusters:} The majority of cluster members lie within the Jacobi radius (i.e., \(  P_{RJ} > 50\% \)), or the stellar mass within the Jacobi radius typically exceeds 40 \( M_{\odot} \).
    \item \textbf{Moving Groups:} Most members reside beyond the Jacobi radius (i.e., \(P_{\mathrm{RJ}} < 50\%\)), and the stellar mass within the Jacobi radius (\(M_{\mathrm{RJ}}\)) is typically less than \(40 \, M_{\odot}\).  
\end{itemize}

Compared to the classification criteria described in \citetalias{2024A&A...686A..42H}, we adopt a more relaxed criterion for open clusters (OCs), which includes stellar groups with a stellar mass within the Jacobi radius of less than 40 \( M_{\odot} \), provided that all observed member stars lie within this radius. Since all members are contained within the Jacobi radius, these systems can be regarded as typical open clusters, whose self-gravity is sufficient to balance the tidal potential of the Milky Way. Some stellar groups are challenging to classify, particularly those with \( P_{RJ} < 50\% \) and \( M_{RJ} > 40 \, M_{\odot} \). These transition objects exhibit hybrid characteristics: they retain a core-like structure akin to open clusters, yet most members are located outside the Jacobi radius. Although, we classify them as OCs originally, these stellar groups likely exhibit an evolutionary phase between OCs and MGs (i.e. transitional group). Recent studies \cite{2021ApJ...915L..29G} suggest that some moving groups may be the remnants of disrupted open clusters. 

In Fig.~\ref{OC_MG_classify}, we present the 2D distributions of three typical stellar groups in the projection coordinates: (a--c) OCs lacking tidal structures, (d--f) OCs with significant tidal tails (overdensity extending beyond $r_{\mathrm{J}}$), and (g--i) MGs characterized by sparse kinematic and spatial distributions. 

For example, HSC~2919 exemplifies a typical OC with all members within the Jacobi radius (i.e. $P_{RJ}$=100\%). In contrast, Haffner~13 demonstrates tidal stripping with 28.4\% of its members residing in extended tails. The Beta~Tuc~Group ($\tau = 31.4~\mathrm{Myr}$), a typical young MG, shows a sparse spatial distribution, $\sim 95\%$ of members beyond $r_{\mathrm{J}}$, consistent with unbound kinematics. 

We ultimately classify the 112 stellar groups in \citetalias{2024A&A...686A..42H} into 36 moving groups and 76 open clusters, For comparison, there are 32 MGs and 80 OCs classified by \citetalias{2024A&A...686A..42H}. The difference mainly comes from the transition group. For instance, the MGs of Theia 7 in \citetalias{2024A&A...686A..42H} has a mass (\(M_{\mathrm{RJ}}\))<\(40 \, M_{\odot}\), while \(P_{\mathrm{RJ}} > 50\%\), which is classified as OCs in this paper. 

\subsection{Stellar groups from \citetalias{2020AJ....160..279K}: Sample 2} \label{K2020_classify}
We use the same methods described above to fit the radial density profile and estimate the morphological parameters of the stellar groups in \citetalias{2020AJ....160..279K}. However, we found that some stellar groups in \citetalias{2020AJ....160..279K} exhibit characteristics of multi-components. Similar results are obtained by \cite{2021ApJ...923...20P} in the Vela OB association, which contains five sub-clusters.

Most stellar groups with multiple components are referred to as ``strings'' in \citetalias{2020AJ....160..279K}. They are usually younger than 300 Myr, and some may be more reasonably classified as coeval cluster groups under the scenario of hierarchical star formation \citep{2012MNRAS.426.3008K,2021ApJ...923...20P,2022ApJ...931..156P}. Here, we employ both the \texttt{HDBSCAN} \citep{McInnes2017hdbscanHD} and the Gaussian Mixture Model (GMM) to identify the substructures.

The following steps outline the processes for identifying substructures in stellar groups with multiple components based on the \citetalias{2020AJ....160..279K} dataset:

\begin{itemize}
    \item Use \texttt{HDBSCAN} to identify substructures based on five astrometric parameters of cluster members in \citetalias{2020AJ....160..279K}, including ra, dec, pmra, pmdec, and parallax.
    \item Use GMM to further identify substructures, such as binary clusters that may have been initially classified as one group in \texttt{HDBSCAN}.
    \item Use CMD diagrams, \texttt{ra-dec} diagrams, \texttt{pmra-pmdec} diagrams, and distance distributions to verify group memberships.
\end{itemize}

\begin{figure*}[!htbp]
    \centering
    \includegraphics[width=0.95\linewidth]{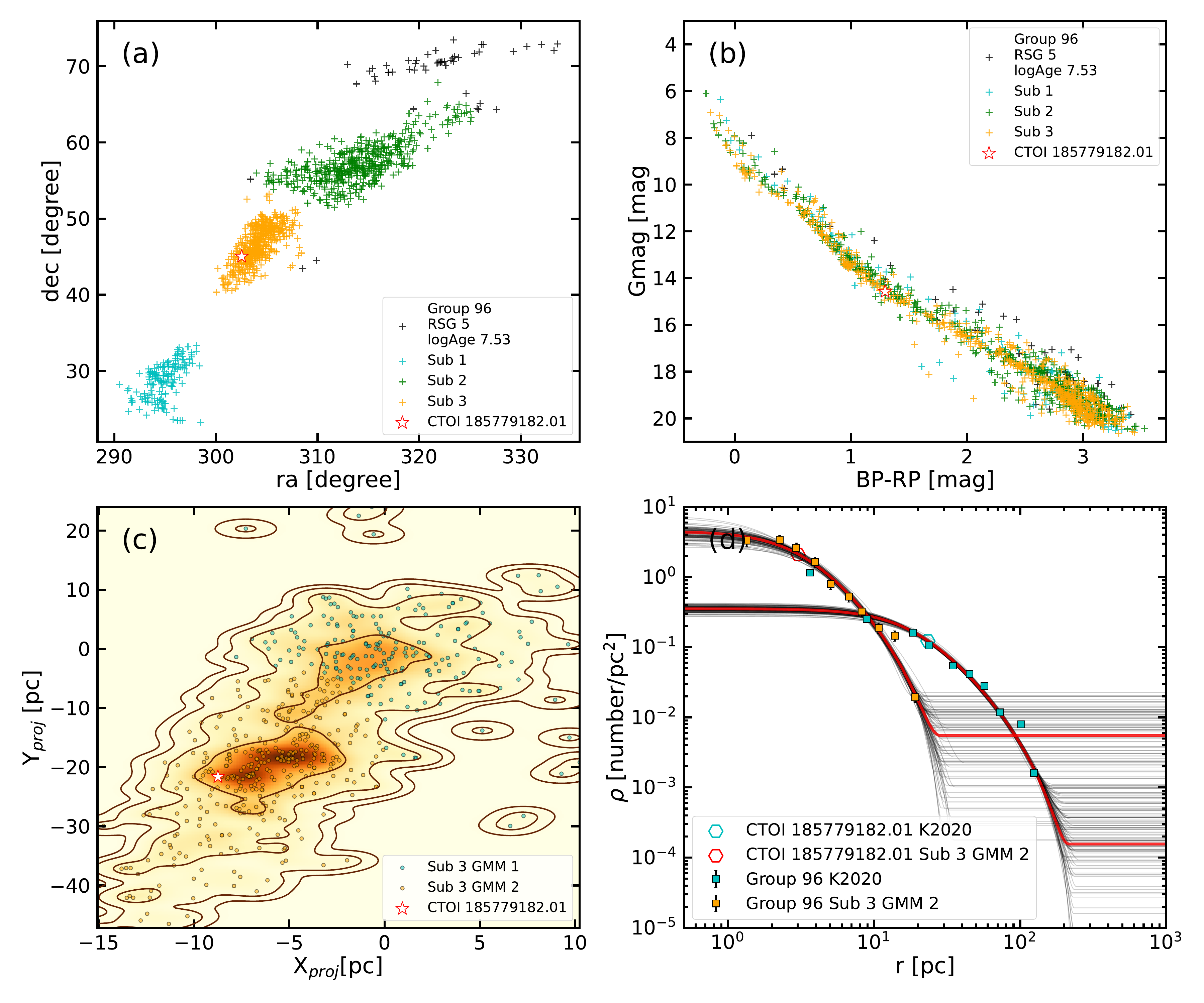}
    \caption{An example of identifying the substructures of Group 96 using \texttt{HDBSCAN} and GMM, along with radial density profile (RDP) fitting. \textbf{Panel (a):} \texttt{HDBSCAN} is used to find substructures, identifying three substructures: Sub 1, Sub 2, and Sub 3 (in blue, green, and orange, respectively). The red star denotes the potential planet candidate CTOI 185779182.01.\textbf{Panel (b):} The color-magnitude diagram of Sub 1, Sub 2, and Sub 3 groups. \textbf{Panel (c):} The density profile of Sub 3 in a 2D projection diagram. We also use the Gaussian Mixture Model (GMM) to further identify two components that may represent potential young binary clusters. \textbf{Panel (d):} The comparison of the RDP fitting before and after substructure identification.}
    \label{Sub_Group96}
\end{figure*}

For instance, we use \texttt{HDBSCAN} to find that Group 96 actually has three main substructures in the \texttt{RA-DEC} diagram, represented by the blue, green, and orange regions (Sub 1--3) in Fig.~\ref{Sub_Group96} (a). Fig.~\ref{Sub_Group96} (b) shows the color-magnitude diagram of Sub 1--3. All three substructures exhibit a clear and similar main sequence in the color-magnitude diagram, suggesting that they are probably coeval. There is one planet candidate, CTOI 185779182.01, located within Sub 3 (orange). Upon enlarging Sub 3, we observe a distinct bimodal distribution in the 2D projection diagram (Fig.~\ref{Sub_Group96} (c)). Interestingly, Sub 3 displays features indicative of binary clusters. A binary cluster is typically defined as two stellar clusters with similar ages, metallicities, and proximity in position \cite{2024arXiv241205376P}. This similarity is reflected in the five-dimensional astrometric data, which may not be suitable for \texttt{HDBSCAN} to classify effectively.

After identifying substructures, we use the Gaussian Mixture Model (GMM) to further distinguish the two components in the 2D projection diagram. After substructure identification, we perform RDP fitting to obtain cluster morphology parameters and the corresponding 2D surrounding stellar density of the host star of the planet candidate CTOI 185779182.01. Fig.~\ref{Sub_Group96} (d) shows the comparison of RDP fitting before and after substructure identification. Using the density of Group 96 from \citetalias{2020AJ....160..279K} as a whole cluster, the fitted RDP is well consistent in the outer region, but underestimated the surface density in the center. Considering the substructures as well as the two components via GMM fitting, the RDP fitted well in the central regions. The different results lead to very different stellar density around the planet host CTOI 185779182.01, i.e. the density from substructure's component is ten times higher than that from \citetalias{2020AJ....160..279K}, which is more realistic as shown in panel (c). Thus, it is crucial to identify the substructures in stellar groups. All the fitted parameters of substructures are listed in Table \ref{plclt}.


\begin{figure}
    \centering    \includegraphics[width=0.95\linewidth]{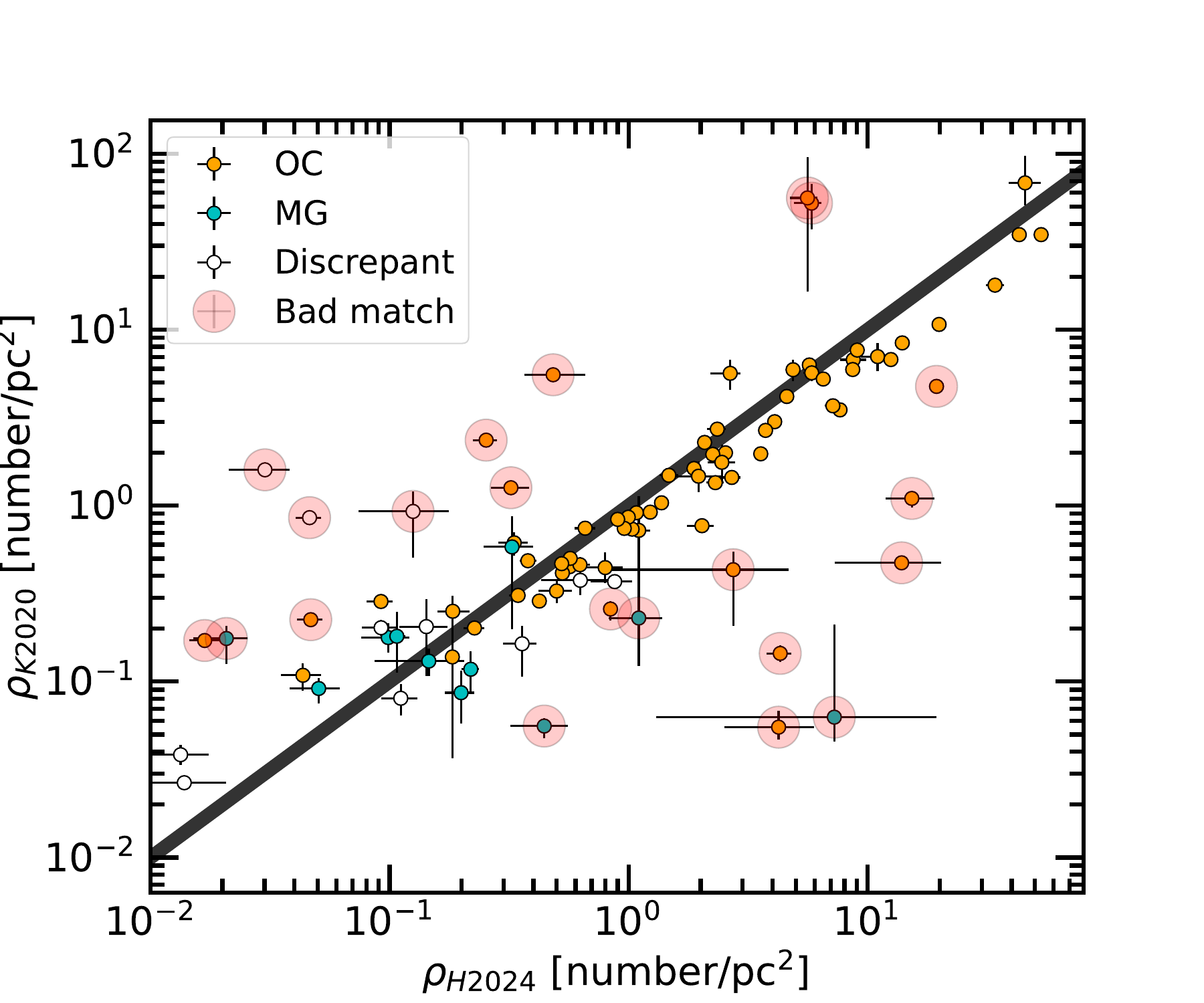}
    \caption{The comparison of the 2D surrounding stellar number density of 94 planet host stars both in \citetalias{2020AJ....160..279K} and \citetalias{2024A&A...686A..42H} dataset. OC(orange) and MG(blue) exhibit that the planet host star's stellar group from \citetalias{2020AJ....160..279K} and \citetalias{2024A&A...686A..42H} show consistent classification, while white circles(11) show inconsistent classification. Large red circles(21) present the density difference is larger than 0.5 index in log scale.}
    \label{Density_compare}
\end{figure}

To test the robustness of the 2D stellar number density density \( \rho \), we compare the \( \rho \) values of planet-hosting stars derived from both the \citetalias{2020AJ....160..279K} and \citetalias{2024A&A...686A..42H} datasets (Fig.~\ref{Density_compare}). Among the stellar groups of 94 planet-host stars, most(\(\sim 80\%\)) of the their classification show great consistency. However, 21 stars show significant discrepancies in density estimates (marked as large red circles), defined as a logarithmic-scale difference exceeding 0.5 dex. We further compare the spatial distributions of the stellar groups associated with these 21 stars in the \citetalias{2020AJ....160..279K} and \citetalias{2024A&A...686A..42H} catalogs. Given that these planetary host stellar groups exhibit a discrepancy exceeding 30\% in either relative size or member fraction within the Jacobi radius (P$_{RJ}$), the differences observed in  \( \rho \) can likely be attributed to inconsistencies in the membership criteria of the host groups across the two datasets. Further works that unify both OCs and MGs can help to improve the \( \rho \) estimation, e.g. \cite{2023MNRAS.526.4107P} unify different catalogues of OCs to get more completed star numbers.



After identification the substructures in \citetalias{2020AJ....160..279K}, we classify the 156 stellar groups in \citetalias{2020AJ....160..279K} into 168 OCs and MGs, including 85 open clusters and 83 moving groups. There are 10 groups have substructures, and 22 substructures are newly founded.

\subsection{Comparison of OCs and MGs in two samples} \label{OC&MG}

\begin{figure*}
\begin{center} 
\includegraphics[width=.45\linewidth]{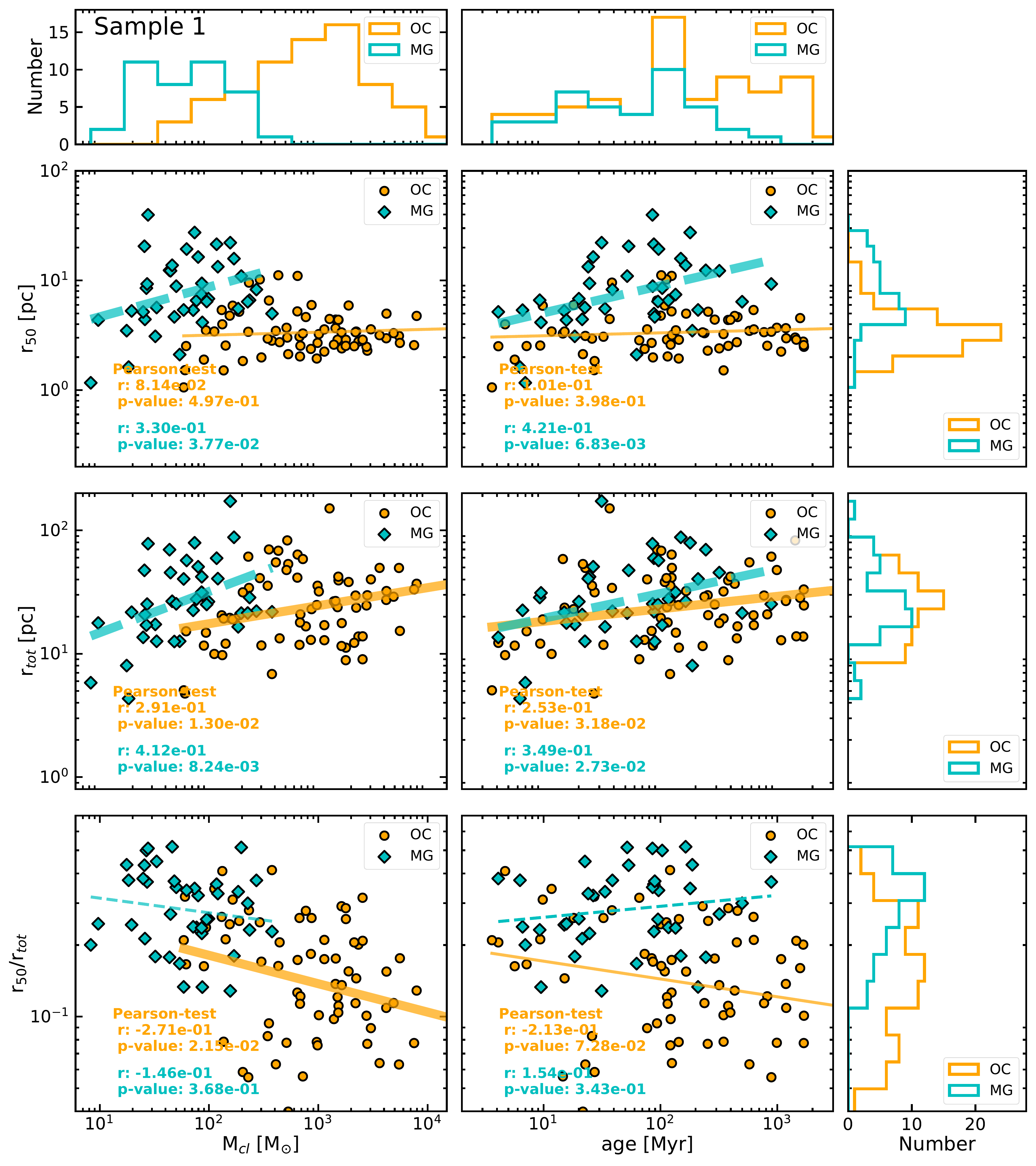}\quad\includegraphics[width=.45\linewidth]{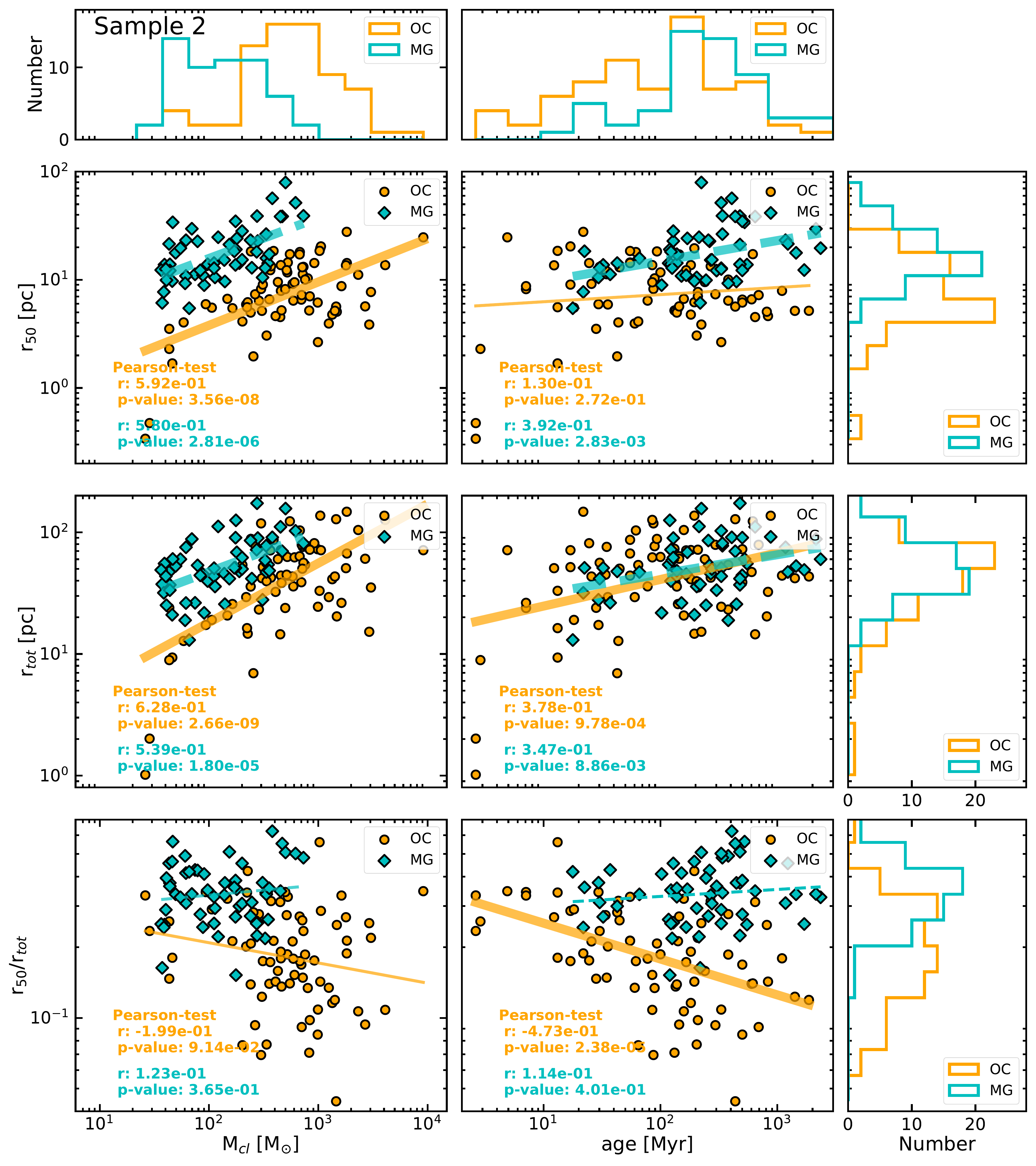}
\end{center}
\caption{The mass-radius and age-radius distributions for planet-hosting stellar groups are shown. The top, middle, and bottom panels show these relationships for the i.e. \(r_{50}\), \(r_{\rm tot}\), and \(r_{50}\ /r_{\rm tot}\) radii, respectively. Open clusters (orange circles) and moving groups (cyan squares) from Sample 1 (left panel) and Sample 2 (right panel) are plotted. The orange solid lines and cyan dashed lines represent linear fits to their corresponding data. The heavy lines indicate a significant correlation (p-value $<$ 0.05), whereas the light lines show no significant trend (p-value $>$ 0.05).}
\label{OCMG_comparison}
\end{figure*}
Here we presents the mass-radius and age-radius distributions for planet-hosting stellar groups. As shown in Fig.~\ref{OCMG_comparison}, the 2D scatter plots display the relationships between the radii (\(r_{50}\), \(r_{\rm tot}\), and \(r_{50}\/r_{\rm tot}\) from top to bottom) and the cluster masses/ages. The left and right panels show the distributions from Sample 1 and Sample 2.

In the top panels, we find that Moving Groups (MGs) in both Sample 1 and 2 exhibit a positive \(M_{\rm cl}-r_{50}\) correlation , with Pearson correlation coefficients $>$ 0 and p-values $<$ 0.05. In contrast, Open Clusters (OCs) do not show a consistent correlation between the two samples. Specifically, OCs in Sample 1 show no significant trend, while those in Sample 2 show a positive correlation. The division of OCs and MGs for stellar groups with low masses may lead to the inconsistency. 

Furthermore, we find that MGs exhibit a positive \(age-r_{50}\) correlation, whereas OCs show no obvious trend. This difference in the \(age-r_{50}\) correlation may be due to their different cluster masses \(M_{\rm cl}\). OCs are usually more massive, which allows them to balance the Milky Way's gravitational potential, especially for members in the cluster core. In contrast, lower-mass MGs are more likely to be dissolved with increading ages.

In the middle panels, we find that both MGs and OCs in Sample 1 and 2 exhibit positive correlations for both \(M_{\rm cl}-r_{tot}\) and \(age-r_{tot}\), which is naturally \citetalias{2024A&A...686A..42H}

In the bottom panels, the dimensionless radius\(r_{50} /r_{tot}\) can represent the stellar number density profile. A smaller \(r_{50} /r_{tot}\) in OCs indicates a more concentrated density distribution compared to MGs. We find that MGs and OCs exhibit different relationships for  \( age - r_{50} /r_{tot}\). Specifically, the dimensionless radius \( r_{50} /r_{tot}\) of MGs shows no trend with either the mass of the cluster \( M_{\rm cl}\) or age, whereas OCs exhibit obvious negative correlations with ages. Since the MGs is more sparse, closely resembling that of field stars, and therefore remain approximately constant. In OCs, however, mass segregation can cause massive stars to migrate inward, increasing the central stellar number density over time and creating an anti-correlation between age and \( r_{50} /r_{tot}\). The result is also consistent with \citetalias{2024A&A...686A..42H}.

\begin{figure*}
    \centering
    \includegraphics[width=0.95\linewidth]{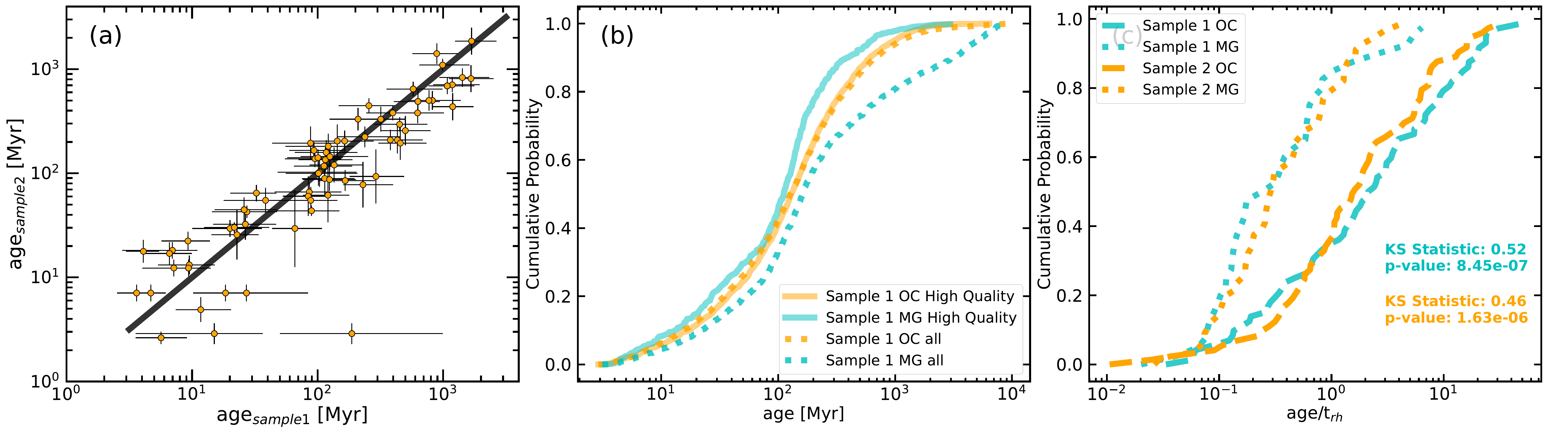}
    \caption{A comparison of the ages of planet-hosting stars in open clusters (OCs) and moving groups (MGs). (a) Cross-matched planet-hosting stars from Sample 1 and Sample 2. (b) Cumulative age distribution of the stars in Sample 1. Solid and dotted lines represent high-quality and total member stars, respectively. (c) Cumulative distribution of the dynamical timescale \(age/t_{rh}\), where the half-mass relaxation timescale \(t_{rh}\) is calculated using Eq. 3 from \cite{2019A&A...624A.110F}} 
    \label{H2024_K2020_age_compare}
\end{figure*} 
Intriguingly, we find that moving groups (MGs) in Sample 1 are systematically younger than those in Sample 2. This age difference may suggest two distinct origins: the younger MGs in Sample 1 likely formed from the dissolution of stellar groups following gas expulsion, whereas the older MGs in Sample 2 could be remnants of open clusters (OCs) that have undergone galaxy tides (see Section 4 for further discussion of the density–age correlation).

Before comparinging the age distribution in the two samples, systematic errors and observational biases should be checked. Firstly, to evaluate systematic errors in age measurements, we cross-matched planet-hosting stars from both samples. As shown in Fig.~\ref{H2024_K2020_age_compare}(a), the ages of these stars align closely along the 1:1 line, indicating consistent age determination between the two samples. Thus, systematic measurement errors are unlikely to be the cause of the age difference. 
Secondly, we considered observational biases. For instance, if Sample 2 contains more low-quality older groups—where ``low-quality'' refers to a high false-positive probability, i.e., chance alignments of field stars—this could bias the age distribution. To test this, we selected high-quality OCs and MGs from Sample 1 using the criteria from \citetalias{2024A&A...686A..42H} (median CMD class $>$ 0.5 and astrometric S/N CST $>$ 5$\sigma$). Fig.~\ref{H2024_K2020_age_compare}(b) shows the cumulative age distribution of planet-hosting stars in Sample 1, where solid and dotted lines represent high-quality and total members, respectively. We find that high-quality MGs are even younger than the full sample, whereas high-quality OCs show no significant age shift. This suggests that observational biases may contribute to the younger apparent age of MGs in Sample 1. 

Additionally, the brightness of planet-hosting stars in MGs in Sample 1 is systematically higher than that in Sample 2 (see Fig.~\ref{Plhost_CDF}). Since older MGs are more likely to dissipate under Galactic tidal forces and become indistinguishable from field stars, this apparent brightness bias, which often correlates with proximity, could further explain the over-representation of young MGs in the solar neighborhood. The high-quality MGs in \citetalias{2024A&A...686A..42H} are likely closer to the Solar System, making them easier to detect in transit surveys.

However, we cannot apply the same quality criteria to Sample 2 for a direct comparison. Thus, while observational bias may partially explain the age difference, it is unlikely to account for it entirely.

While the absolute ages of moving groups (MGs) may vary between samples due to the observational biases and potential physical differences discussed above, their classification as MGs implies shared physical characteristics that distinguish them from open clusters (OCs). To uncover this commonality, we turn to the concept of dynamical age, i.e. \( \tau_{\text{dyn}} = \text{age}/t_{rh} \), which measures a system's evolutionary stage relative to its internal relaxation timescale \( t_{rh} \) (calculated from Eq. 3 in \cite{2019A&A...624A.110F}). As shown in Fig.~\ref{H2024_K2020_age_compare}(c), the cumulative distributions of \( \tau_{\text{dyn}}\) reveal a striking consensus: despite their chronological age differences, all MGs are consistently and significantly dynamically younger than OCs. I.e.  the p-value is less than 0.05 via KS test in each sample. Thus, within the unified framework of dynamical age, MGs emerge as a distinct population characterized by a less dynamically evolved state, clearly setting them apart from the more relaxed OCs.

\section{Properties of planet hosts} \label{Planethosts}
\subsection{Planets host stars in OCs and MGs} \label{4.1}
\begin{figure*} 
    \centering    \includegraphics[width=1\linewidth]{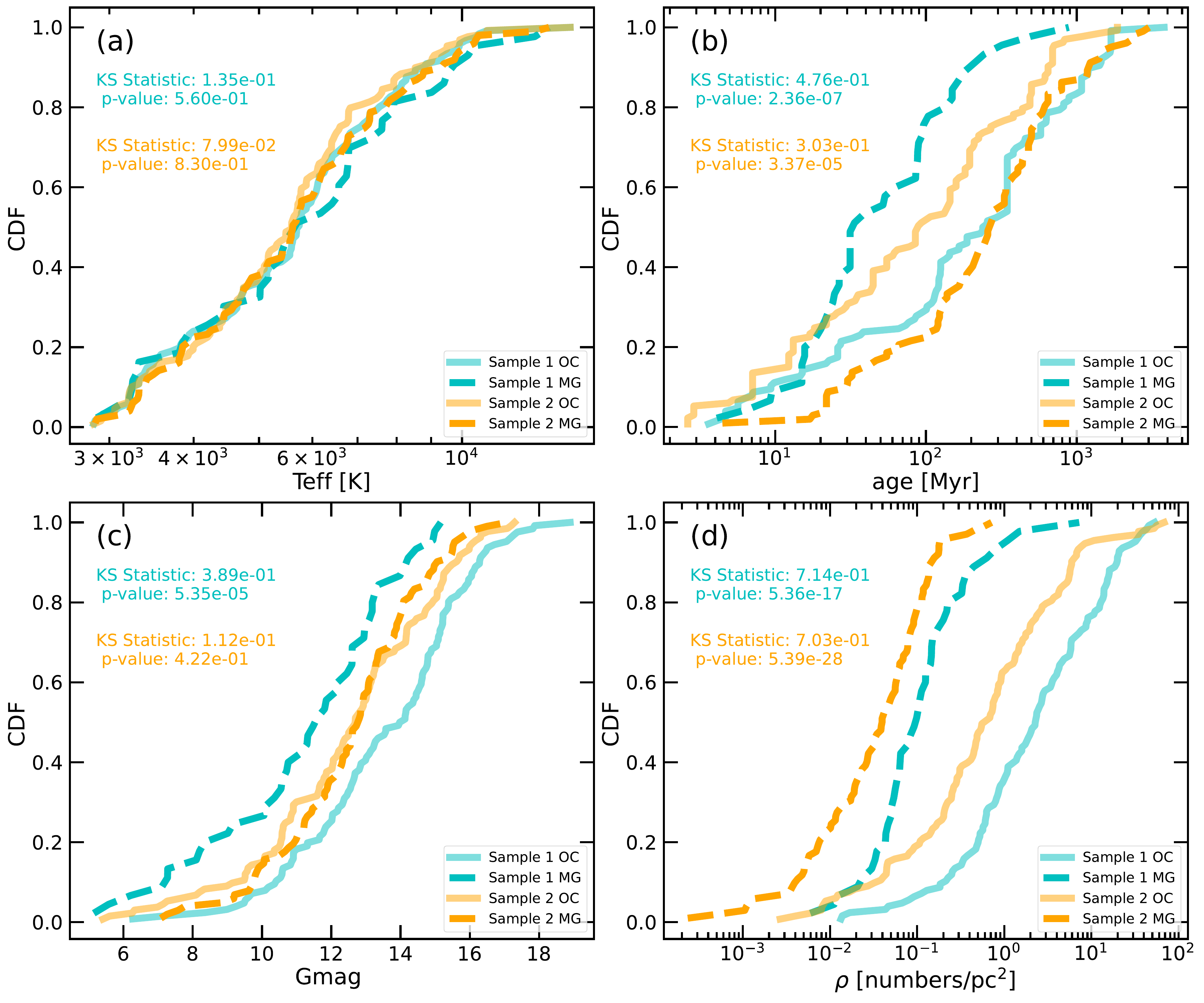}
    \caption{Cumulative distribution functions (CDFs) for parameters of planet-hosting stars in OCs and MGs: (a) stellar effective temperature (\(T_{\rm eff}\)), (b) stellar age, (c) G magnitude, and (d) stellar number density (\(\rho\)).}
    \label{Plhost_CDF}
\end{figure*}

\begin{figure*}
    \centering    \includegraphics[width=1\linewidth]{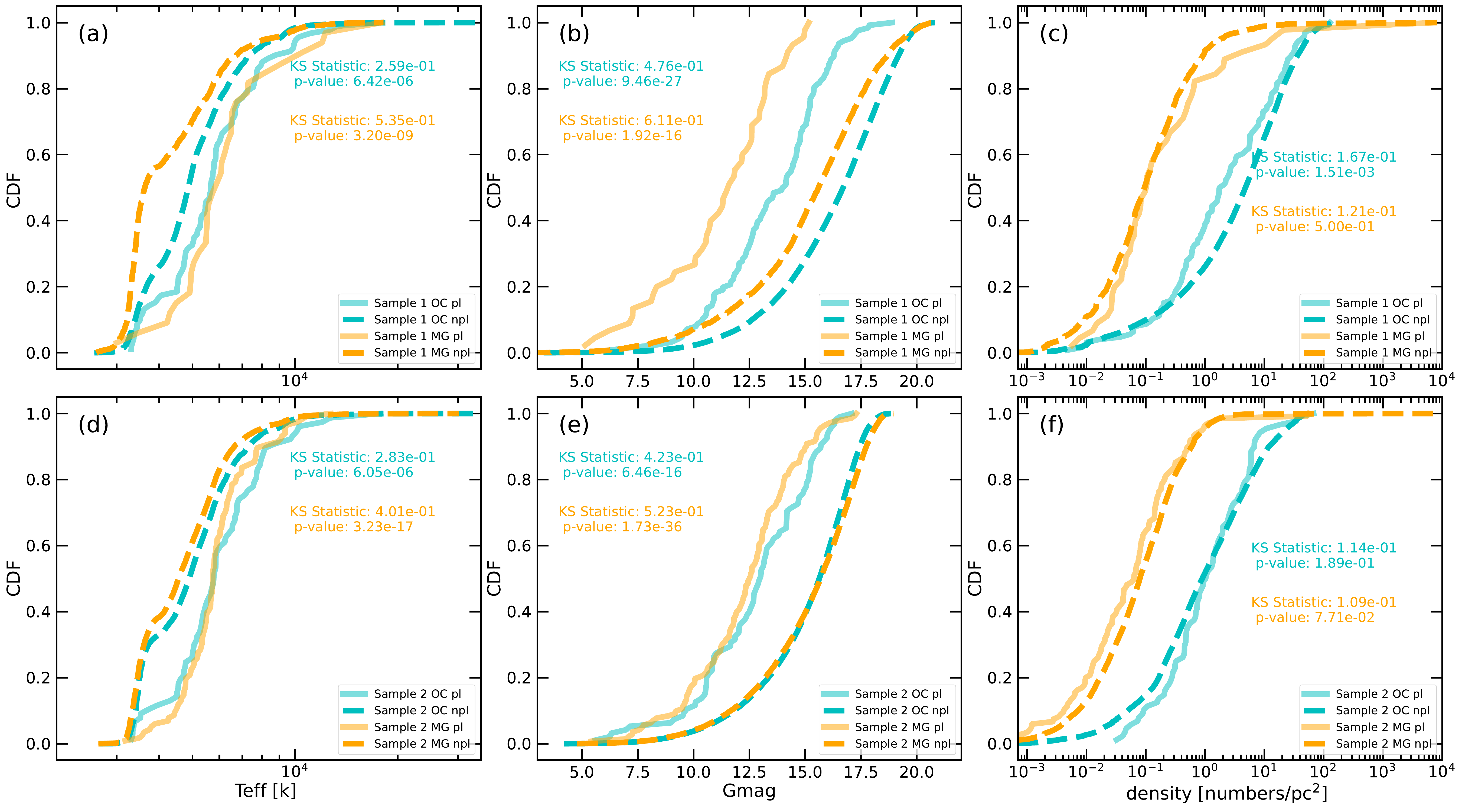}
    \caption{Cumulative distribution functions (CDFs) for parameters of planet-hosting stars (pl) and none-planet-hosting stars (npl). We compare the stellar effective temperature (\(T_{\rm eff}\)), G magnitude, and stellar number density (\(\rho\)) from left to right respectively. The results of sample1 and sample 2 are displayed from top to bottom, respectively.}    
    \label{Plhost_nplhost_CDF}
\end{figure*}

In this section, we investigate the characteristics of planet-hosting stars in open clusters (OCs) and moving groups (MGs). We present the cumulative distribution functions (CDFs) for key stellar parameters including effective temperature (\(T_{\rm eff}\)), age, G magnitude, and stellar number density (\(\rho\)). We also compare these CDFs with those of ``non-planet-hosting stars''—a term referring to stars with none detected exoplanets.

Fig.~\ref{Plhost_CDF} shows that planet-hosting stars in open clusters (OCs) and moving groups (MGs) have similar effective temperature \(T_{\rm eff}\) distribution. This is confirmed by Kolmogorov–Smirnov (KS) two-sample tests, which yield p-values above 0.05, failing to reject the null hypothesis of identical distributions.
However, CDF of age in panel (b) reveals a significant difference. In Sample 1, planet-hosting stars in MGs are younger than those in OCs, while the opposite is true in Sample 2. We discuss the origin of this age difference and the further analysis of its correlation with stellar density in Section \ref{density-age}.


Additionally, we find that in Sample 1, planet-hosting stars in moving groups (MGs) are typically brighter (in G magnitude) than those in open clusters (OCs). In contrast, both groups share a similar G magnitude distribution in Sample 2.

The brighter appearance of MG stars in Sample 1 can be attributed to their closer average distance. For instance, the median distance for the total sample of MGs in \citetalias{2024A&A...686A..42H} is 597 pc, compared to 2259 pc for OCs. Similarly, for the planet-hosting stars in Sample 1, the median distance for OCs (844 pc) is greater than that for MGs (274 pc). This proximity of MGs may partially account for the observed age difference between the samples, as mentioned in Section \ref{OC&MG}.

For the stellar number density \(\rho\), planet-hosting stars in moving groups (MGs) exhibit lower densities in both samples. Furthermore, the \(\rho\) distribution for these stars is higher in Sample 1 than in Sample 2. This discrepancy is likely related to the differing stellar group identification methods used by \citetalias{2020AJ....160..279K} and \citetalias{2023A&A...673A.114H}.
In addition to comparing the distribution of planet-hosting stars in open clusters (OCs) and MGs, we also compare the parameters of planet-hosting stars and field stars (i.e., ``non-planet-hosting stars''). Fig.~\ref{Plhost_nplhost_CDF} shows the cumulative distribution functions (CDFs) for stellar effective temperature(\(T_{\rm eff}\)), G magnitude, and stellar number density (\(\rho\)). Panels a, b, and c represent Sample 1, while panels d, e, and f represent Sample 2. We find that field stars are generally hotter and brighter than planet-hosting stars. In contrast, the stellar number density distributions of field stars and planet-hosting stars are similar, with the exception of OC members in Sample 1.

\subsection{Density-age correlation} \label{density-age}
\begin{figure*}
    \centering
    \includegraphics[width=1\linewidth]{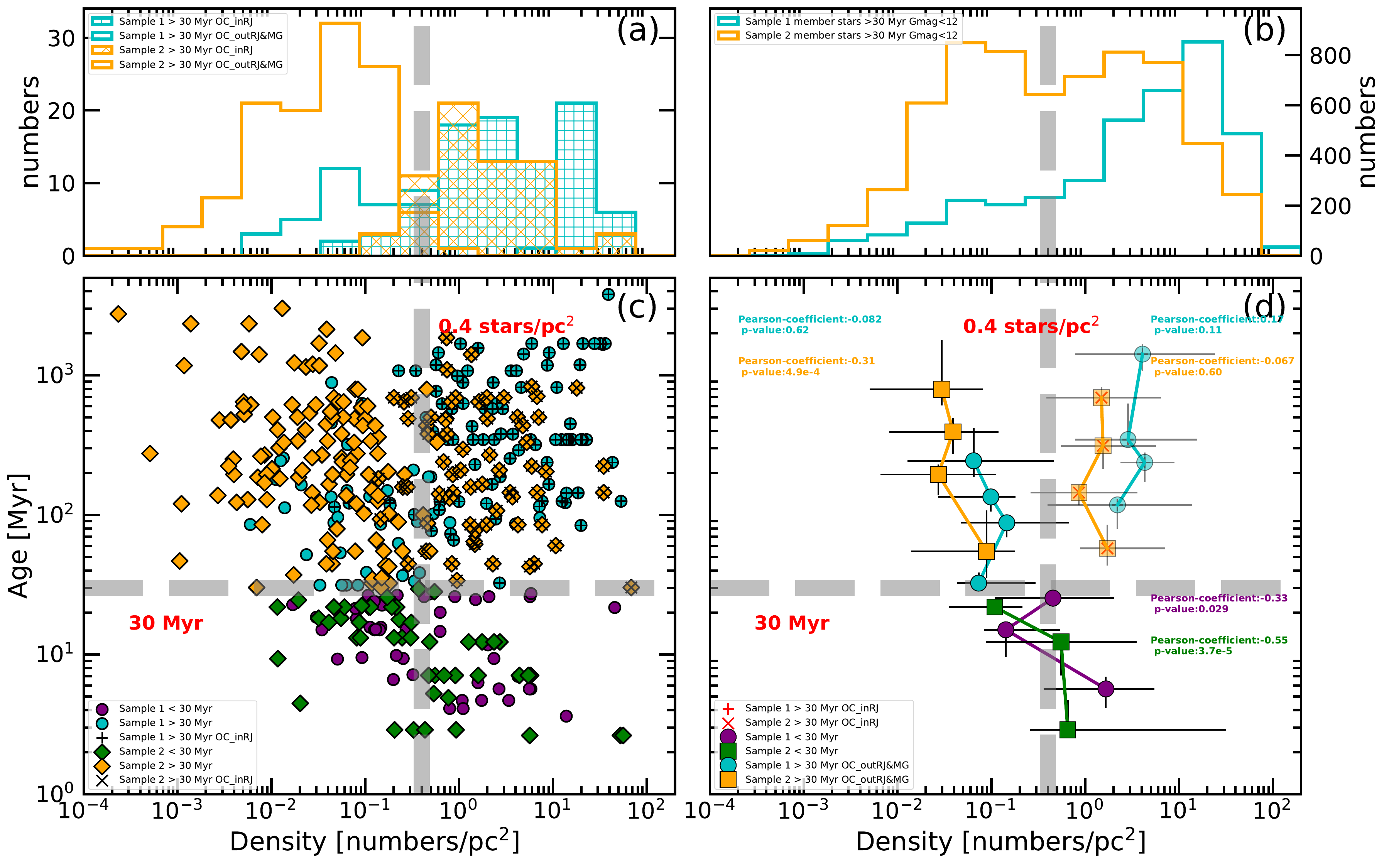}
    \caption{The density-age relation, illustrating the distribution of surrounding stellar number density for planet-hosting stars alongside their ages. \textbf{Panel (a)}: Histogram of stellar number density of planet-hosting stars older than 30 Myr. For both samples, hatched histograms (``+'' for sample 1, ``x'' for sample 2) show stars in OC members within the Jacobi radius. Solid histograms (cyan for sample 1, orange for sample 2) show stars in OC members beyond the Jacobi radius or in MG members. \textbf{Panlel (b)}: Histogram of stellar number density of member stars in planet hosting stellar groups older than 30 Myr. \textbf{Panlel (c)}: Scatter plot of the density-age relation. Young stars (<30 Myr) are shown as purple circles (sample 1) and green squares (sample 2). Older stars (>30 Myr) are shown as cyan circles (sample 1) and orange squares (sample 2), with ``+'' or ``x'' symbols indicating OC members outside the Jacobi radius. \textbf{Panlel (d)}: Binned density-age relation. Results of the Pearson correlation coefficient test for different groups with associated p-values are shown in corresponding regions. The gray dashed lines highlight the typical age of 30 Myr (horizontal) and stellar number density of 0.4 stars/pc$^{2}$ (vertical)}    \label{Density-age}
\end{figure*}

In Fig.~\ref{Plhost_CDF}-(d), we find that the average surrounding stellar number density (\( \rho \)) of planet-hosting stars in open clusters (OCs) is an order of magnitude higher than that of those in moving groups (MGs). To understand the origin of this difference, we investigate the time evolution of stellar density.

Fig.~\ref{Density-age} shows the density-age relation, illustrating the distribution of surrounding stellar number density for planet-hosting stars as a function of their ages. We adopt 30 Myr as a threshold to distinguish stellar groups that have experienced gas expulsion, consistent with previous work (e.g., \cite{2022AJ....164...54Z, 2023A&A...673A.114H}), which shows cluster expansion typically occurs within the first 40 Myr. For young groups (<30 Myr), the density-age relation shows a clear anti-correlation in both samples (Panels c and d). This is consistent with cluster evolution driven by supernova explosions, which expel gas, weaken the gravitational potential, and induce outward bulk motion of stars, thereby increasing the cluster's size. We determine the critical age for this anti-correlation by identifying the age threshold that yields the minimum Pearson correlation coefficient; this is \(\sim\)20 Myr for Sample 1 and \(\sim\)70 Myr for Sample 2.

For systems older than 30 Myr, a different pattern emerges. Fig~\ref{Density-age} (a) reveals a bimodal distribution in stellar density for planet-hosting stars, particularly in Sample 2, with a pronounced gap around 0.4 stars/pc$^{2}$. We adopt this value as a density threshold, likely corresponding to the Jacobi radius. To test the robustness of this bimodality, we analyzed all member stars (Gmag < 12) in the planet-hosting groups. Fig~\ref{Density-age} (b) confirms the bimodal distribution in Sample 2, which contains both low-density MGs and high-density OCs, while the OC-dominated Sample 1 shows only a single peak. 

Crucially, a comparison between Fig.~\ref{Density-age} (a) and (b) shows that a larger proportion ($\sim$0.65\%) of known planet-hosting stars are located in the lower-density peak, while only $\sim$49\% of all member stars are located in the same peak. This suggests that planets are either more frequently detected or more likely to survive in low-density environments, such as MGs and the sparse regions of OCs. This higher peak at lower densities could result from observational biases or intrinsic planet evolution processes. We will explore the planet fraction in OCs and MGs in more detail in Section~\ref{sec5.1}.

This aligns with theoretical work describing the divergent fates of clusters after gas expulsion. Some groups would revirialize while others dissolve, leaving some sparse groups as cluster remnants \citep{2020ApJ...900L...4P, 2021ApJ...915L..29G}. The prevalence of planet hosts in low-density environments may thus reflect this dynamical evolution.



\section{Results of planets/candidates in OCs and MGs} \label{results}
Having detailed the properties of planet-hosting stars in Open Clusters (OCs) and Moving Groups (MGs) in Section \ref{Planethosts}, we now investigate the properties of the planets themselves to understand planetary formation and evolution. In Section \ref{sec5.1}, we compare the fraction of planets and candidates in OCs versus MGs. Section \ref{sec5.2} presents a statistical analysis of the planetary radius-age relation, and Section \ref{sec5.3} discusses evidence related to the formation of the Hot Neptune desert.

\subsection{Fraction of planets/candidates in OCs and MGs} \label{sec5.1}


A common approach for estimating planetary occurrence rates from a survey of \( N_{*} \) stars is the inverse detection efficiency method (IDEM) \cite{2021ARA&A..59..291Z}. The occurrence rate, \( f_{p}^{\rm IDEM} \), is defined as:

\begin{equation}
    f_{p}^{\rm IDEM}= \frac{1}{N_{*}} \sum_{i=1}^{N_{p}} \frac{1}{p_{i}} = \frac{N_{p}}{N_{*}} \left\langle \frac{1}{p_{i}} \right\rangle
\end{equation}

Here, \( p_{i} \) is the survey detection efficiency for the \( i \)-th of the \( N_{p} \) detected planets, and \( \langle \cdots \rangle \) denotes the average efficiency over all detected planets.

Although the correction of detection efficiency is beyond the scope of this paper, it is crucial to reveal the difference between different stellar groups. Key observational bias, e.g. variations in membership completeness and photometric data completeness, differ between stellar groups. These biases are often linked to stellar population and number density distribution. 
A critical bias arises from the relationship between stellar mass and planet occurrence. Studies of \emph{Kepler} data show that the occurrence rate of Kepler-like planets (orbital periods \( P < 400 \) days, radii \( 1 < R_p < 4 R_\oplus \)) is higher around lower-mass stars \cite{2015ApJ...798..112M,2020AJ....159..164Y}. Since membership completeness affects the inclusion of these more common, fainter low-mass stars, it can indirectly bias the derived occurrence rate. A group with higher completeness may include more observable low-mass stars, potentially inflating its observed planet count.


To enable a fair comparison across OCs and MGs, we should therefore control for completeness. This involves establishing the relationship between membership completeness and magnitude for each group. By applying group-specific magnitude cuts tailored to their completeness limits, we can construct a uniformly completeness-limited stellar sample. The occurrence rate for the \( k \)-th OC or MG within this sample is then:

\begin{equation}
\begin{split}
    f_{p,k}(T,C)^{\mathrm{IDEM}}
    &= \frac{1}{N_{*,k}(T,C)} \sum_{i=1}^{N_{p,k}(T,C)} \frac{1}{p_{i}} \\
    &= \frac{N_{p,k}(T,C)}{N_{*,k}(T,C)} \left\langle \frac{1}{p_{i,k}} \right\rangle
\end{split}
\end{equation}

Here, \( p_{i,k} \) is the detection efficiency for the \( i \)-th planet among the \( N_{p,k} \) detected in the \( k \)-th group. \( N_{*,k}(T,C) \) and \( N_{p,k}(T,C) \) are the numbers of member stars and planets under specific TESS magnitude (\( T \)) and membership completeness (\( C \)) conditions. While \( T \) and \( C \) are related, we include both to clarify that a magnitude-limited sample holds \( T \) constant, while a completeness-limited sample holds \( C \) constant.

However, even when observational biases are addressed, calculating precise occurrence rates for individual OCs and MGs remains challenging due to statistics noise of limited small sample. This limitation is clearly demonstrated by Praesepe for instance, one of the best-studied open clusters for exoplanet research, which consists of the largest number of confirmed planets and planet candidates among such groups.

According to \cite{2024A&A...686A..42H}, Praesepe contains 1,314 confirmed members and hosts 10 confirmed planets with 8 additional candidates (based on our catalog in Appendix A). To test how different selection criteria affect the derived planet fraction (i.e. $\frac{N_{p,k}(T,C)}{N_{*,k}(T,C)}$), we calculated values under various Gaia magnitude and membership probability cuts—a straightforward method to partially correct for observational selection effects, as brighter stars typically yield more complete photometric information (Note, the uncertainty is calculated via poisson noise):

\begin{itemize}
    \item With a Gaia \( G < 18 \) cut: \( f = \frac{18}{984} = 1.8^{+0.5}_{-0.4}\% \)
    \item With a stricter \( G < 12 \) cut: \( f = \frac{5}{196} = 2.5^{+1.2}_{-1.0}\% \)
    \item Using membership probability \( p > 0.9 \): \( f = \frac{5}{150} = 3.2^{+1.7}_{-1.3}\% \)
    \item Using \( p > 0.5 \): \( f = \frac{16}{793} = 2.0^{+0.5}_{-0.5}\% \)
\end{itemize}

Adopting different sample cuts, the estimated planetary fractions all lie within 1\(\sigma\) of each other, showing no statistically significant differences. This illustrates the challenge posed by small sample sizes, the statistics uncertainties make it difficult to reveal a precise correlation to the stellar magnitude or member completeness.

To enlarge the planet sample, it's reasonable to compute the planet fraction using all planets in different clusters. I.e. the planet fraction $F$ should be

\begin{equation}
    F = \frac{\sum_{k} N_{k,p}}{\sum_{k} N_{k,s}}
\end{equation}

where \( N_{k,p} \) is the number of planets and candidates, and \( N_{k,s} \) is the number of member stars in the \( k \)-th OC or MG. Note the observational bias for each cluster is not corrected, and we try to find if the planet fractions before correction are obvious differences for different groups. 

Using the estimated fraction F shown in Table~\ref{Fraction}, we find that the fraction of planets and candidates in MGs is significantly higher than in OCs across all tested criteria, with significances of approximately 3.3\(\sigma\) for sample 1 and 4.9\(\sigma\) for sample 2 on average. This result is robust when splitting the sample by detection method (“Transit” vs. “Not Transit”) and remains after applying cuts in Gaia G-magnitude and parallax to control for brightness and distance biases (Fig.~\ref{Plhost_CDF}). The trend also persists within subgroups from different missions (“KOI”, “K2”, “TOI”, “CTOI”), which helps mitigate mission-specific pipeline effects.





Additionally, we split the sample by different age ranges or only select star having TESS light curves and the trend persists. This indicates that MGs may have a higher intrinsic planet fraction from the outset, which may be correlated with the different initial environments of OCs and MGs. OCs usually host massive stars that have strong UV radiation; this can evaporate nearby protoplanetary disks, making it difficult for planets—especially giant planets—to form. Recent papers show that the dust mass in protoplanetary disks in massive clusters like the Orion Nebula Cluster is systematically lower than in isolated regions like Taurus and Chamaeleon I, e.g., \cite{Tobin2020}. This is widely attributed to external photoevaporation driven by massive stars. Thus, we speculate that the initial radiation environment may account for the different planet fraction between OCs and MGs, at least partially.

However, until we can obtain a more quantitative planetary fraction or occurrence rate by correcting for observational biases, it is difficult to draw robust conclusions. Different missions like KOI and TOI have different observational biases; for instance, KOIs generally have longer observational baselines and are fainter than TOIs. Furthermore, observational completeness varies significantly between individual OCs and MGs. The higher stellar density in OCs presents a particular challenge for transit detection, especially with TESS's large pixel scale, as analyzing light curves in crowded fields is notoriously difficult. While methods exist to address this \cite{2019ApJS..245...13B,2019MNRAS.490.3806N}, a full correction is beyond the scope of this paper and will be carefully addressed in future work.

\begin{deluxetable*}{ccccc} \label{plft}
\tablenum{2}
\tablecaption{Fraction of planets+candidates in OCs and MGs}
\tablewidth{0pt}
\tablehead{
\colhead{Criteria} & \colhead{Sample 1 OCs} & \colhead{Sample 1 MGs} & \colhead{Sample 2 OCs} & \colhead{Sample 2 MGs}
}
\startdata
Transit & 0.12$\pm$0.01\% & 0.66$\pm$0.11\% & 0.11$\pm$0.01\% & 0.54$\pm$0.05\% \\
Not Transit & 0.11$\pm$0.02\% & 1.20$\pm$0.30\% & 0.13$\pm$0.02\% & 0.54$\pm$0.15\% \\
Total & 0.11$\pm$0.01\% & 0.76$\pm$0.10\% & 0.11$\pm$0.01\% & 0.54$\pm$0.05\% \\
Gmag$<$10 & 1.7$\pm$0.5\% & 7.1$\pm$2.1\% & 0.16$\pm$0.04\% & 1.4$\pm$0.8\% \\
10$<$Gmag$<$14 & 0.13$\pm$0.02\% & 0.82$\pm$0.14\% & 0.14$\pm$0.03\% & 0.57$\pm$0.12\% \\
Plx$>$1 & 0.11$\pm$0.1\% & 0.69$\pm$0.10\% & 0.10$\pm$0.01\% & 0.47$\pm$0.06\% \\
CP & 0.12$\pm$0.02\% & 0.79$\pm$0.18\% & 0.15$\pm$0.02\% & 0.65$\pm$0.12\% \\
KOI & 0.10$\pm$0.03\% & 1.6$\pm$1.1\% & 0.13$\pm$0.03\% & 1.1$\pm$0.2\% \\
K2 & 0.10$\pm$0.02\% & 0.62$\pm$0.28\% & 0.12$\pm$0.05\% & 0.75$\pm$0.28\% \\
TOI & 0.12$\pm$0.03\% & 0.81$\pm$0.19\% & 0.12$\pm$0.02\% & 0.50$\pm$0.07\% \\
CTOI & 0.13$\pm$0.02\% & 0.70$\pm$0.14\% & 0.10$\pm$0.01\% & 0.45$\pm$0.07\% \\
TOI\&10$<$Gmag$<$14 & 0.12$\pm$0.03\% & 1.2$\pm$0.3\% & 0.11$\pm$0.02\% & 0.76$\pm$0.14\% \\
CTOI\&10$<$Gmag$<$14 & 0.14$\pm$0.03\% & 0.71$\pm$0.17\% & 0.10$\pm$0.02\% & 0.56$\pm$0.11\% \\
age<100 Myr & 0.26$\pm$0.04\% & 0.75$\pm$0.11\% & 0.07$\pm$0.01\% & 0.28$\pm$0.04\% \\
age<100 Myr\&CP & 0.47$\pm$0.16\% & 0.84$\pm$0.20\% & 0.09$\pm$0.03\% & 0.28$\pm$0.06\% \\
100<age<1000 Myr & 0.10$\pm$0.01\% & 0.80$\pm$0.24\% & 0.13$\pm$0.02\% & 0.33$\pm$0.03\% \\
100<age<1000 Myr\&CP & 0.11$\pm$0.02\% & 0.50$\pm$0.35\% & 0.14$\pm$0.04\% & 0.33$\pm$0.07\% \\
Having TESS light curve & 0.12$\pm$0.01\% & 1.01$\pm$0.17\% & \nodata & \nodata \\
\enddata
\tablecomments{Fraction of planets + candidates in OCs and MGs with different selection criteria. The uncertainty is estimated via Poisson noise}
\label{Fraction}
\end{deluxetable*}

\subsection{Planet radius-age relation in OCs and MGs} \label{sec5.2}
\begin{figure*}
    \centering    
    \includegraphics[width=1\linewidth]{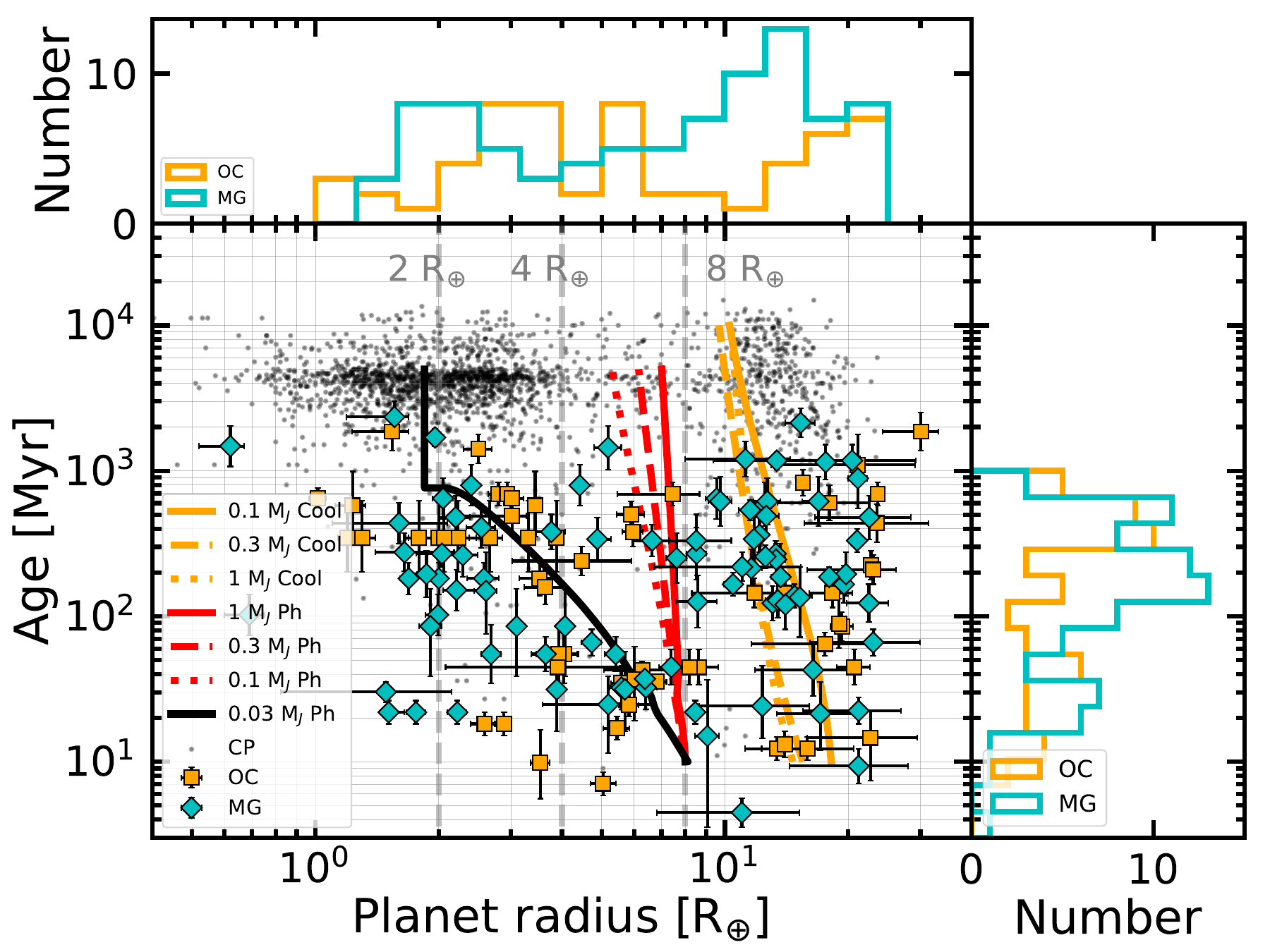}
    \caption{The relationship between planetary radius and age for short-period planets (P < 20 days). The black dots are confirmed planets with measured ages. Orange squares and blue diamonds denote planetary candidates in open clusters and moving groups, respectively. The theoretical tracks show radius evolution for hot planets of different masses, with orange lines representing cooling contraction and red/black lines showing photoevaporation effects. The line styles (solid, dashed, dotted) correspond to different planetary masses or model assumptions.}\label{Rp_age}
\end{figure*}

\begin{figure*}
    \centering    \includegraphics[width=1\linewidth]{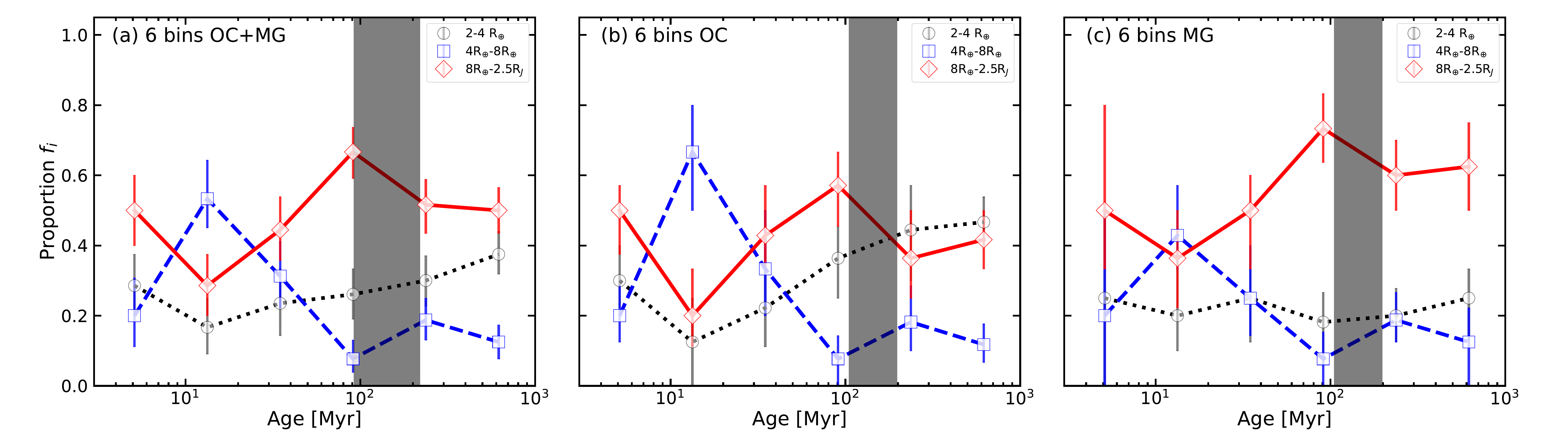}
    \caption{The probability distribution function of the planet radius in different age range. We selected main-sequence single stars to mitigate the potential observation bias correlated to stellar properties.}
    \label{Planet_ratio}
\end{figure*}
In this paper, we present an updated view of the planet radius-age distribution using the updated catalog of planets and candidates within open clusters (OCs) and moving groups (MGs) we compiled above. Following the methodology of our previous work \citep{2023AJ....166..219D}, we selected 41 confirmed planets (CPs) and 106 planet candidates (PCs) with orbital periods shorter than 20 days, excluding potential binary systems with RUWE values greater than 1.4 and those with large radius error (\( \sigma_{\rm R_{p}}/R_{\rm p} > 0.5 \)). This selection yields 23 CPs and 41 PCs in OCs, and 18 CPs and 65 PCs in MGs. For simplicity, we classify planets into three size groups: Sub-Neptunes (\(2 \, R_{\oplus} < R_{p} < 4 \, R_{\oplus}\)), Sub-Jupiters(\(4 \, R_{\oplus} < R_{p} < 8 \, R_{\oplus}\)), and Jovian planets (\(8 \, R_{\oplus} < R_{p} < 2.5 \, R_{\text{J}}\)). 

Fig.~\ref{Rp_age} reveals several key features: dozens of hot Jupiters are clustered around 100 Myr, and a gap for sub-Jupiters is observed between 100 and 200 Myr, consistent with our earlier findings in UPiC-I. Interestingly, these trends persist in both OCs and MGs. A similar correlation can be seen in Fig.~\ref{Planet_ratio}, which shows the relationship between cluster age and the relative proportion of planets of different sizes. The relative proportion for each group is defined as:

\begin{equation}
    f_{\rm i} = \frac{N_{\rm i}}{N_{\rm SubN} + N_{\rm SubJ} + N_{\rm J}},
\end{equation}

where \( N_{i} \) is the number of planets in star clusters of different sizes, specifically \( N_{\rm SubN} \), \( N_{\rm SubJ} \), and \( N_{\rm J} \) for Sub-Neptunes, Sub-Jupiters, and Jovian planets, respectively. 

The histograms in Fig.~\ref{Rp_age} further indicate that Jovian-sized planets predominantly orbit stars in MGs. While the number of sub-Jupiters is larger in OCs, the overall number of planets and candidates is dominant in MGs (88 versus 58 in OCs). This is consistent with the distribution of planet host stars shown in Fig.~\ref{Density-age} and Table~\ref{Fraction}.

\begin{figure} 
    \centering    
    \includegraphics[width=1\linewidth]{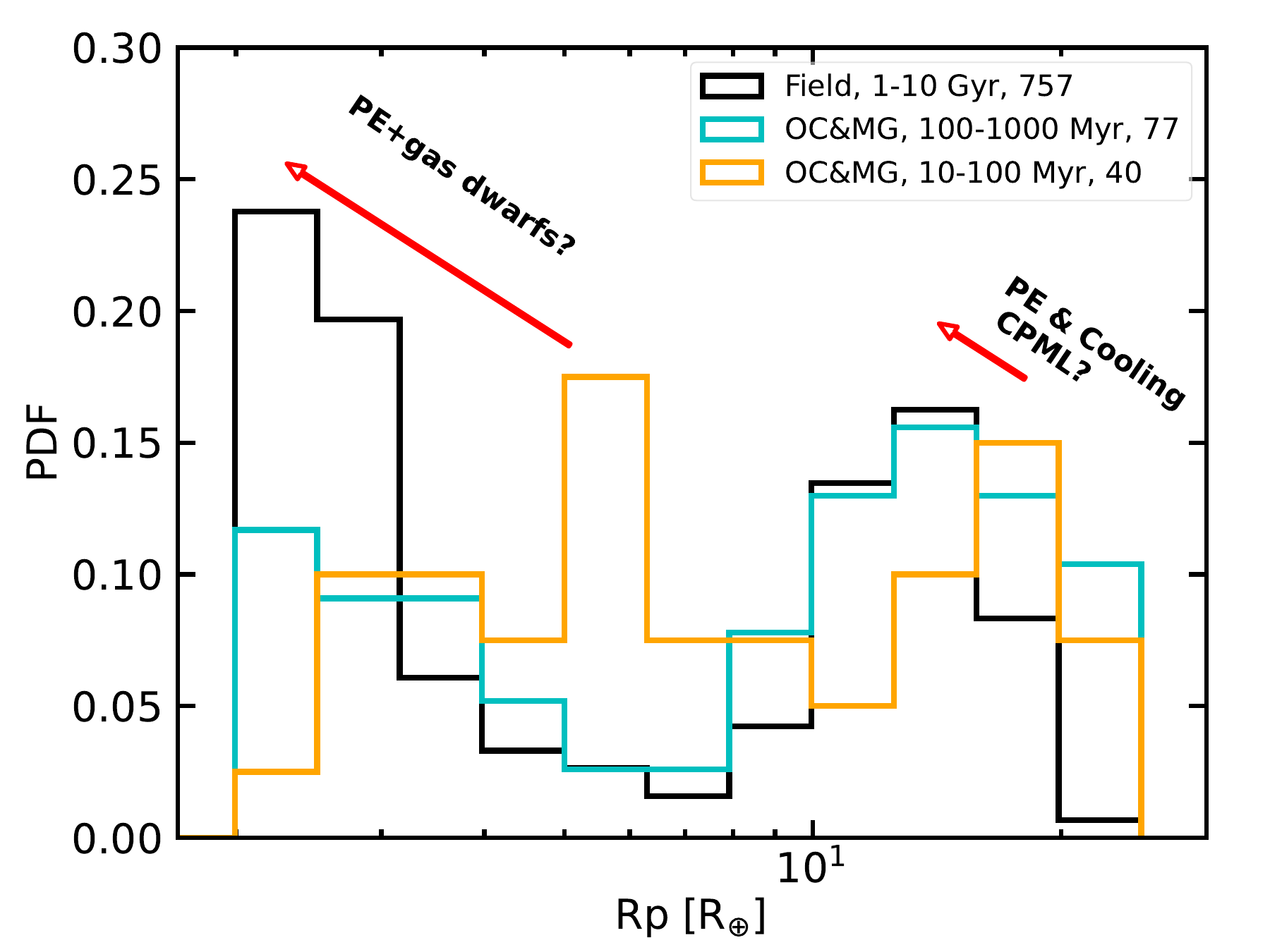}
    \caption{The size distribution of planets with different ages: \textcolor{black}{757 planets around field stars (black)}, \textcolor{blue}{77 confirmed planets and planet candidates with ages between 100 and 1000 Myr in open clusters (OCs) and moving groups (MGs) (blue)}, and \textcolor{orange}{40 confirmed planets and planet candidates with ages between 10 and 100 Myr in OCs and MGs (orange)}. The selection criteria are as follows: 1) Orbital periods $P < 20$\,days; 2) Relative error of planet radius $\delta R_{\rm p}/R_{\rm p} < 0.5$; 3) RUWE $< 1.4$. CPML: core-powered mass loss \citep{2018MNRAS.476..759G}, PE: photoevaporation \citep{2013ApJ...775..105O,2017ApJ...847...29O}}
    \label{Rp_PDF}
\end{figure}

To clarify the planet radius–age relation, we also combine those relatively old planets around field stars with age measurements. These planet are select with the same criterion as those in OCs and MGs (P$<$20 days, RUWE$<$1.4, and \(\sigma_{\rm R_{p}}/R_{\rm p}\) $<$ 0.5). We then split this sample into three age groups: 
\begin{itemize}
    \item Young (10–100 Myr): 40 confirmed planets and candidates in OCs and MGs
    \item Intermediate (100–1000 Myr): 77 confirmed planets and candidates in OCs and MGs.
    \item Old (1–10 Gyr): 757 confirmed planets from the NASA Exoplanet Archive.
\end{itemize} 
Fig.~\ref{Rp_PDF} shows how the planet radius distribution evolves with age, where the y-axis represents the ratio of planets in each radius bin to the total number. In the young group (10–100 Myr), a peak of Sub-Jupiters ($5-6 R_{\oplus}$) is present. This peak evolves into a gap in the intermediate group (100–1000 Myr), where Sub-Neptunes and Jovian-sized planets dominate, creating a clear bimodal distribution. This bimodality, or the corresponding Sub-Jupiter gap, persists into the older sample. 

Previous works, including \cite{2023AJ....166..219D}, interpret this evolution as the formation of the ``hot Neptune desert,'' potentially driven by photoevaporation and flyby-induced high-eccentricity migration. Young stars emit strong UV radiation, which can drive significant atmospheric mass loss for close-in planets, particularly puffy gas dwarfs. A recent study \cite{2025MNRAS.tmp..591R} supports this, finding that these Sub-Jupiters are likely puffy gas dwarfs undergoing photoevaporation (see the black radius evolution tracks in Fig. \ref{Rp_age}). 

We also find suggestive evidence that the radii of giant planets may shrink slightly within the first 100 Myr. Although the sample size is limited, the peak of the giant planet distribution appears to move inwards. This evolution is consistent with traditional cooling models and photoevaporation (see the orange and red tracks in Fig. \ref{Rp_age}).

In UPiC-I, we explained the pile-up of Jovian-sized planets around 100–200 Myr using flyby-induced high-eccentricity tidal migration. This model suggests that a single effective flyby event can trigger migration for an initial Jupiter-like system with an outer companion. Consequently, we did not previously emphasize the role of stellar density. However, other studies show that planetary systems in high-density environments experience stronger dynamical instabilities due to increased flyby frequency and perturbations \citep{2022MNRAS.509.5253W}. Dense star clusters should thus trigger more high-eccentricity migration, leading to a higher hot Jupiter occurrence rate compared to sparse regions \cite{2020ApJ...905..136W}. This predicts more hot Jupiters in OCs than in MGs, yet our sample shows the opposite.

This discrepancy may be due to observational biases. For instance, many ``hot Jupiters'' in our sample are Jovian-sized candidates without mass measurements and could be extremely puffy Neptunian-mass planets. Furthermore, TESS's relatively large pixels make it difficult to extract clean light curves in crowded fields like open clusters, giving it a preference for detecting targets in sparse regions. This likely contributes to the higher number of hot Jupiter candidates detected in MGs.

In Fig.~\ref{Rp_age}, we overlay radius evolution tracks for hot sub-Jupiters and hot Jupiters, incorporating both cooling contraction and photoevaporation. The cooling contraction tracks (orange line), generated with the Python package \texttt{PlanetSynth}\footnote{\url{https://github.com/tiny-hippo/planetsynth}} \citep{2021MNRAS.507.2094M}, simulate planets with initial masses of 0.1 M$_{J}$, 0.3 M$_{J}$, and 1 M$_{J}$, with a fixed heavy element fraction of 0.1. The orange tracks show that cooling contraction alone does not cause significant radius shrinkage for hot Jupiters, indicating it cannot explain the observed gap around 100 Myr for Neptune-sized planets. 

We also used \texttt{Photoevolver}\footnote{\url{https://github.com/jorgefz/photoevolver}} \citep{2023MNRAS.522.4251F} to simulate radius evolution under photoevaporation. Young stars ($<$100 Myr) are highly active, emitting strong XUV radiation that heats the atmospheres of close-orbiting planets and drives escape. By integrating \texttt{Mors}\footnote{\url{https://github.com/Go-Spect/MORS}} \citep{2021A&A...649A..96J} to estimate XUV fluxes from young stars, we modeled radius evolution for planets of different masses, assuming a 1 M$_{\oplus}$ and a 3 days orbital period. For hot Jupiters, photoevaporation does not cause substantial atmospheric loss, leading only to a gradual radius decrease. This aligns with theories suggesting that massive planets (with \( M_{p} > 0.5 \, M_{J} \)) can resist photoevaporation even at short orbital periods \cite{2012MNRAS.425.2931O,2015ApJ...808..173T,2016ApJ...816...34O}. In contrast, lower-mass, inflated planets
(e.g. 10 M$_{\oplus}$) are susceptible to rapid radius shrinkage under strong XUV irradiation, as shown by the black track in our models.

Our findings are consistent with previous studies reporting larger radii for planets younger than 100 Myr, such as those in AU Mic \citep{2020Natur.582..497P}, V1298 Tau \citep{2019ApJ...885L..12D}, and IRAS 04125+2902 \citep{2024Natur.635..574B}. The elevated fraction of hot sub-Jupiters also aligns with recent occurrence rate statistics for young planets \cite{2024AJ....167..210V}. We propose that cooling contraction alone cannot fully explain the high fraction of young sub-Jupiters ($<$100 Myr). Instead, photoevaporation-driven mass loss likely plays a critical role, especially for lower-density planets. Future detection of young planets, coupled with precise mass measurements, will help test this hypothesis.

The radius evolution of Super-Earths and Sub-Neptunes ($<$4 R\(_{\oplus}\)) is more ambiguous than that of their larger counterparts. We changed the bin size of the planet radius distribution; however, we found no obvious signature of the radius valley \cite{2017AJ....154..109F} because of the small sample size. Including more data may help us further understand the radius evolution of these small planets and distinguish between the mechanisms of photoevaporation \citep{2012MNRAS.425.2931O,2013ApJ...775..105O} and core-powered mass loss \cite{2018MNRAS.476..759G}. 


\subsection{Young planets in Hot Neptune Desert, Ridge, and Savanna} \label{sec5.3}
Recent studies \cite{2025AJ....169..117V} reveal that host stars of planets in both the Neptune desert and the ridge exhibit similar metal-rich properties, which contrasts with their counterparts in the longer-period savanna region. Here we try to find the planet fraction of planet in different regions evolve with ages. We also check the differences for planets in the savanna in OCs and MGs, to see if the stellar environments can induce these differences.

\begin{figure}
    \centering    \includegraphics[width=1\linewidth]{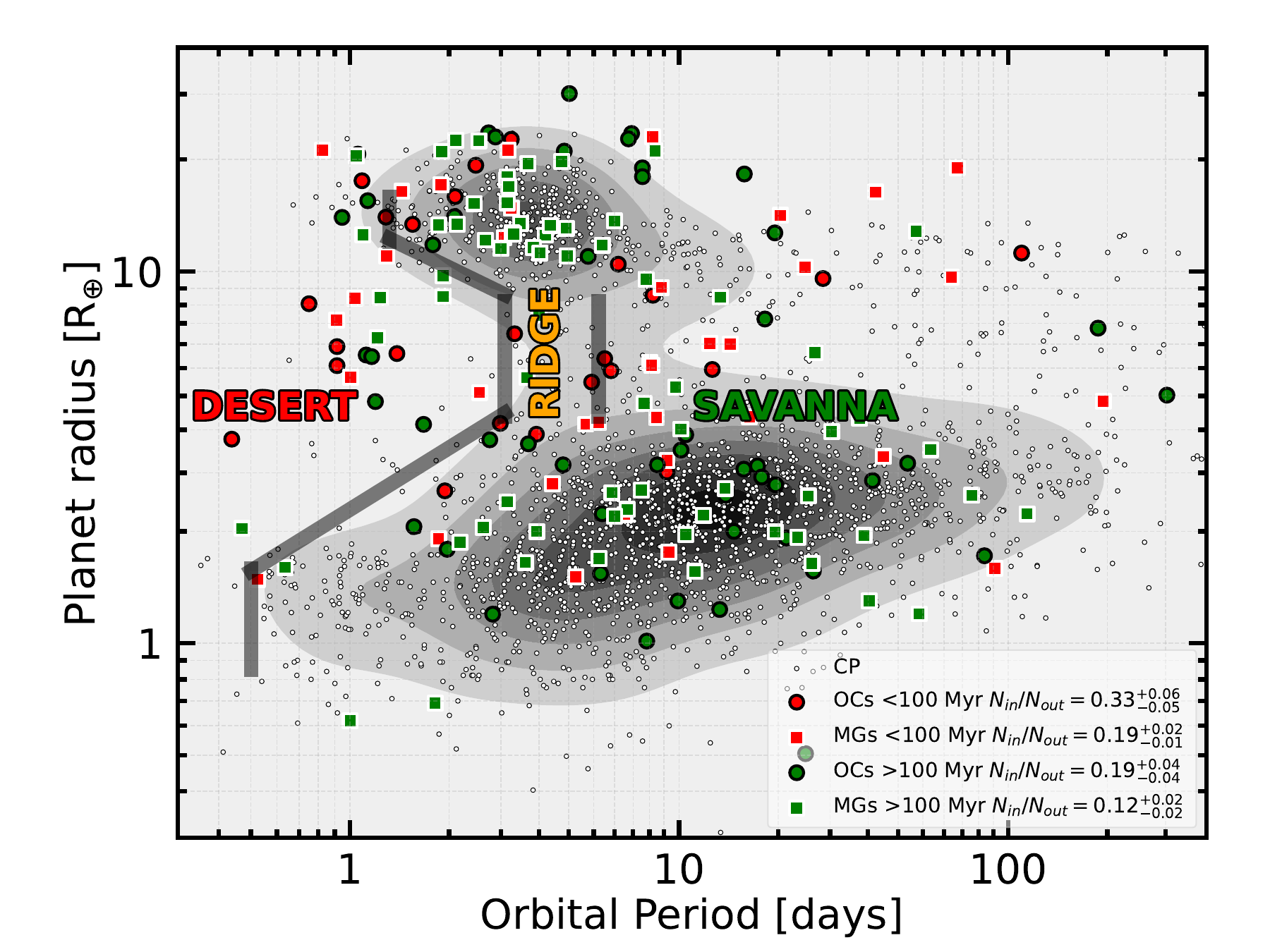}
    \caption{The planet radius--period diagram. The black dots are confirmed planets with age measurements. We use the boundary of the hot-Neptune desert according to the recent paper \cite{2024A&A...689A.250C}.} 
    \label{RP_P}
\end{figure}

In UPiC-I \cite{2023AJ....166..219D}, we applied the Hot-Neptune desert boundary from \cite{2016A&A...589A..75M} and found a higher fraction of young planets ($<$100 Myr) within the desert compared to older systems ($>$100 Myr). Here, we update this analysis using the revised desert definition from \cite{2024A&A...689A.250C} (Fig.~\ref{RP_P}) by calculating the ratio \( N_{in}/N_{out}\)--the number of planets and candidates with P $<$ 20 days inside versus outside the desert.

Consistent with UPiC-I, young systems ($<$100 Myr) show a significantly higher fraction of desert planet (\(N_{in}/N_{out} \) = \( 0.39^{-0.06}_{+0.07} \)) compared to \( 0.18^{-0.02}_{+0.02} \) for old systems. This trend persists when separating OCs and MGs: both young OCs (\( 0.33^{-0.06}_{+0.05}\)) and young MGs (\(0.19^{-0.01}_{+0.01}\)) exceed their older counterpart (OCs: \(0.19^{-0.04}_{+0.04} \); MGs \( 0.12^{-0.02}_{+0.02}\). Notably, OC planets exhibit slightly higher desert occupancy than their MG counterparts at different ages. Furthermore, the fraction of desert planets in OCs decreases more than those in MGs, therefore, we thought more original dersert planets are evolving outside this region after 100 Myrs than planets in MGs.

Our catalog also provides insights into the ``ridge'' of Neptune-sized planets (4.1-8.3 \(R_{\odot}\), 3.2\(<\)P\(<\)5.7 days \cite{2024A&A...689A.250C}). There are one planetary candidate TOI 4373.01 and one confirmed planets TOI 434398831 c in young OCs in the ridge, while two from old moving groups (MGs) in this region. Although the planet number is limited, it also hint that planet in OCs prefer to stay in young ridge population, similar with the desert region. 

We performed a similar analysis for the ``savanna'' region (\(P \approx \)5.7-20  days, 4.1-8.3 \(R_{\odot}\) ). Consistent with the Hot-Neptune desert result, young systems (<100 Myr) show a higher fraction (\(N_{in}/N_{out}\) = \( 0.174^{-0.025}_{+0.022} \)) than old systems (\( 0.033^{-0.001}_{+0.011}\)). This indicates that sub-Jupiters are predominantly young, regardless of their original orbital period. Number of planets in all three regions (the desert, ridge, or savanna) will decrease with age.

This trend also holds when separating OCs and MGs: young OCs \( 0.130^{-0.05}_{+0.44} \) and young MGs \( 0.208^{-0.041}_{+0.032} \) exceed their older counterparts (OCs: \( 0.025^{-0.001}_{+0.001} \); MGs: \( 0.037^{-0.001}_{+0.019} \). In this case, however, MG planets exhibit slightly higher savanna occupancy than OC counterparts.

If HEM is the primary mechanism for forming both ridge Neptunes and hot Jupiters (HJs), we would expect ridge planets to be older due to longer migration timescales. A comparison of confirmed planets with identical periods (\(P \approx 3.2-5.7 \) days) shows that HJs are indeed younger on average (\(\tau_{HJ}= 4.0^{+3.4}_{-2.1}\) Gyr) than ridge planets (\(\tau_{ridge}= 4.5^{+2.7}_{-1.3}\) Gyr) which is consistent with HEM predictions. However, the presence of several young ridge planets in OCs presents an anomaly. For instance, the young ridge planet K2-33 b (10 Myr; \cite{2016Natur.534..658D}) conflicts with HEM timescales that typically exceed the disk dispersal phase. This necessitates alternative formation mechanisms, with \cite{2016Natur.534..658D} favoring in-situ formation or early disk-driven migration.

Environmental differences between OCs and MGs may play a role. Divergent initial disk conditions (e.g., mass, viscosity) in dense OCs could accelerate Type I/II migration, enhancing primordial occupancy of the desert and ridge. While this could explain the higher number of young ridge planets in OCs (4 systems vs 0 in MGs), the limited sample size prevents robust conclusions. Detection biases and stochastic migration outcomes remain plausible alternatives. Resolving these questions demands an expanded demographic census of young (<100 Myr) transiting planets, particularly in low-mass open clusters and moving groups.

\section{Discussion} \label{discussion}
\subsection{Stellar Group Classification} \label{sec6.1}
\begin{figure}
    \centering    \includegraphics[width=1\linewidth]{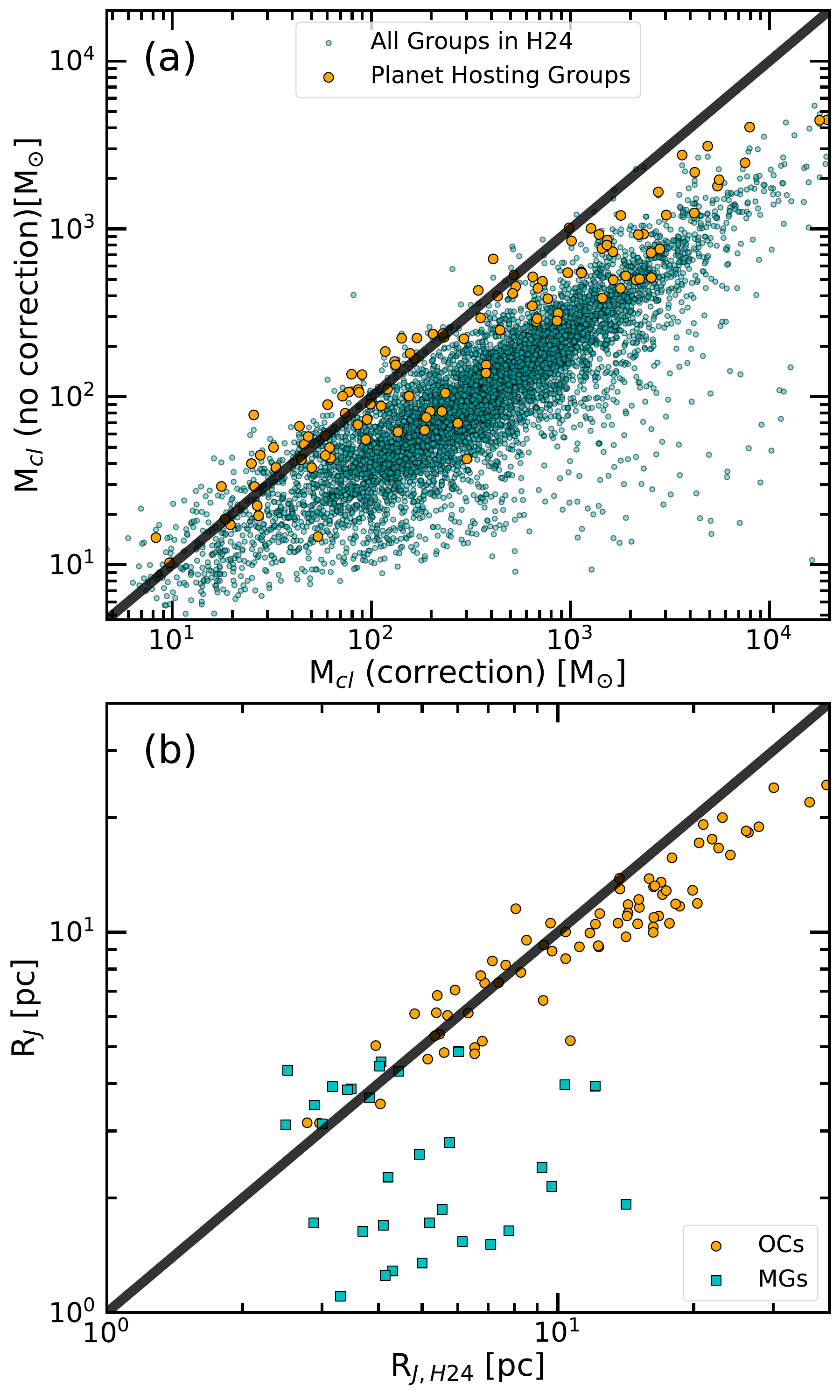}
    \caption{A comparison of the total mass \(M_{cl}\) and Jacobi radius \(R_{J}\) with or without the correction. Panel (a) compares \(M_{cl}\) for all stellar groups from \citetalias{2024A&A...686A..42H} (cyan circles) and for the planet-hosting stellar groups (orange circles). Panel (b) shows the same comparison for \(R_{J}\), but only for the planet-hosting open clusters (OCs) and moving groups (MGs) in \citetalias{2024A&A...686A..42H}.}
    \label{mass_comparison}
\end{figure}

In Section \ref{H2024_classify}, we describe a method to classify stellar groups as open clusters (OCs) or moving groups (MGs) using the Jacobi radius \(R_{J}\). The calculation of \(R_{J}\) depends critically on the total mass of the stellar group \(M_{cl}\), making an accurate mass determination essential. While our initial mass analysis was not exhaustive, this section discusses key factors that influence \(M_{cl}\) and outlines how we plan to address them in future work. The primary factors are the binary fraction and selection effects in the data


binary fraction:
Our current analysis does not account for unresolved binary stars, which leads to an underestimation of \(M_{cl}\) by approximately 10–30\%. Directly measuring the binary fraction in each stellar group is challenging with Gaia data alone, particularly for low-mass-ratio binaries. In \citetalias{2024A&A...686A..42H}, this issue was addressed by simulating a binary star population for each cluster using field star data from \cite{2017ApJS..230...15M}. The simulated binaries were then used to correct the cluster mass function based on their unresolved mass. A similar approach could be applied to our sample in future work.


Selection Effects:
Selection effects in the Gaia data can influence stellar group membership determination and, consequently, the derived \(M_{cl}\). \citetalias{2024A&A...686A..42H} corrected for these effects using the \texttt{gaiaunlimited} Python package. While we could apply a similar correction, our current dataset lacks the intermediate data from the group identification process. Therefore, correcting for selection effects is beyond the scope of this paper but remains a goal for future analysis.

Impact on Mass and Classification:
Fig.~\ref{mass_comparison}(a) compares our total mass estimates, \(M_{cl}\), with and without the corrections applied in \citetalias{2024A&A...686A..42H}. For planet-hosting stellar groups (orange circles), the lack of correction leads to an average mass underestimation of ~26\%, which is consistent with the expected effect of the binary fraction. The underestimation for the full sample of groups in \citetalias{2024A&A...686A..42H} is significantly larger. We attribute this difference to the different distance distributions; our planet-hosting groups are significantly closer (median ~550 pc) than the full sample (median ~2000 pc). At ~550 pc, the Gaia data is more complete, making the uncorrected mass estimate less severely affected than for more distant groups.

As shown in Fig.~\ref{mass_comparison}(b), our values of \(R_{J}\) for OCs are systematically lower than those from \citetalias{2024A&A...686A..42H} by about 18\%. Despite this systematic offset in the absolute value of \(R_{J}\), our classification of groups into OCs and MGs is broadly consistent with \citetalias{2024A&A...686A..42H}. This is because the stellar number density in OCs is centrally concentrated. Therefore, even with an underestimated \(R_{J}\), the relative population of member stars within this radius remains consistent, which is the essential criterion for classification.

\subsection{The inhomogeneity of age measurements}
In Fig.~\ref{H2024_K2020_age_compare}, we compare the age measurements of planet-hosting stars from both \citetalias{2024A&A...686A..42H} and \citetalias{2020AJ....160..279K}, finding them to be broadly consistent. However, for planet-hosting stars that appear in only one of the two catalogs, we did not verify the robustness of their age measurements.

\begin{figure}
    \centering    \includegraphics[width=1\linewidth]{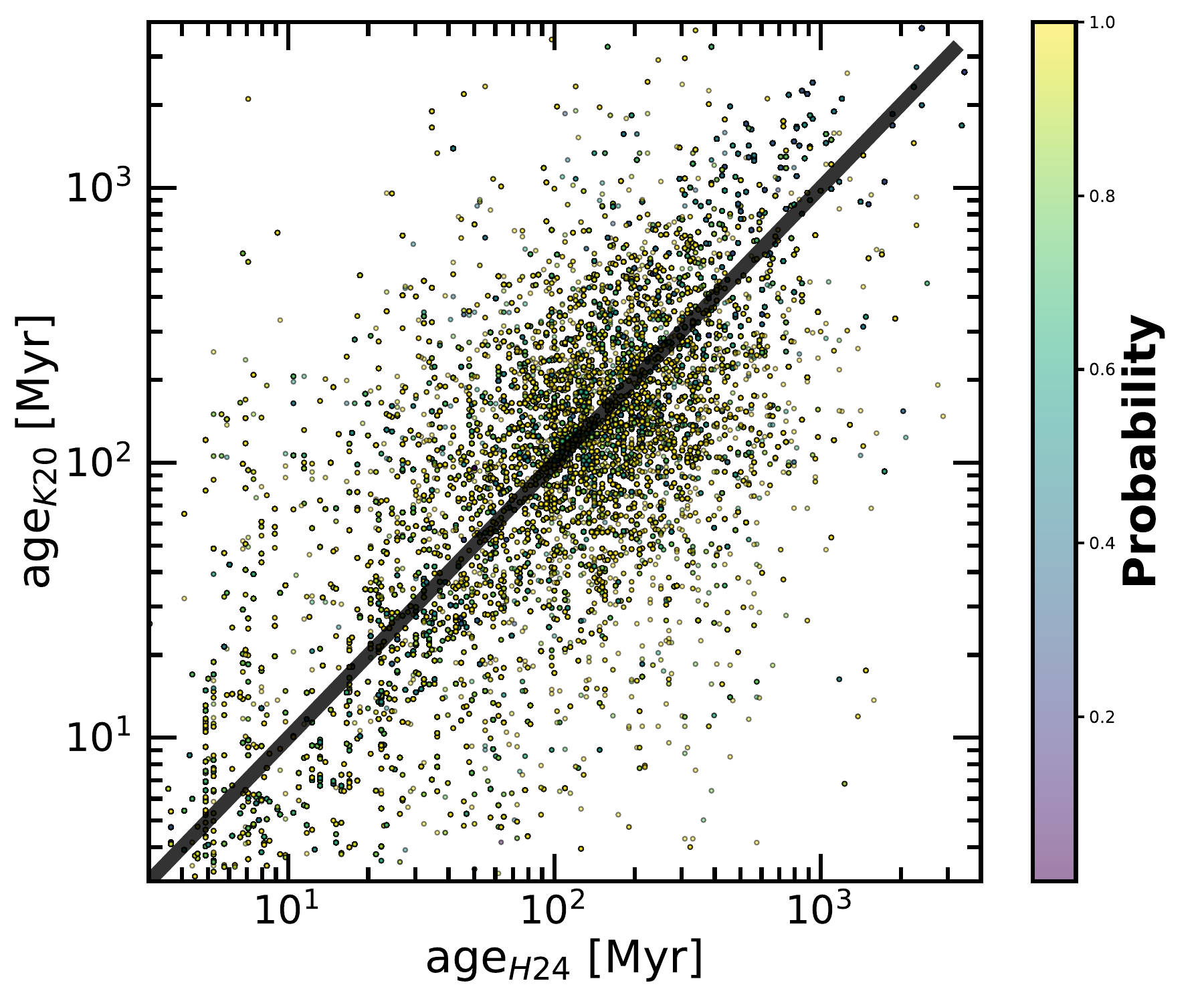}
    \caption{The age comparison of cross-matched member stars from \citetalias{2020AJ....160..279K} and \citetalias{2024A&A...686A..42H}. The color represents the membership probability of stars in \citetalias{2024A&A...686A..42H}.}
    \label{age_comparison}
\end{figure}

Fig.~\ref{age_comparison} shows an age comparison for the cross-matched stars common to both \citetalias{2020AJ....160..279K} and \citetalias{2024A&A...686A..42H}. The age measurements show broad consistency, with 61\% of stars having a relative age difference smaller than 1\(\sigma\). Nevertheless, significant scatter is evident. If we assume that the planet-hosting stars found in only one catalog have a similar distribution, we can extrapolate that approximately 61\% of them (111 of 182 stars) would also have consistent ages. Combining this with the 89 cross-matched stars with consistent ages, we estimate that roughly 73\% of all planet-hosting stars in the combined sample have consistent age measurements between the two catalogs.

We also split the cross-matched stars from Fig.~\ref{age_comparison} into three age groups based on their \(age_{H24}\): $<$10 Myr, 10–100 Myr, and 100–1000 Myr. The consistency of the age measurements varies with age, with a higher probability of agreement for older stars. Specifically, 46\% of stars with \(age_{H24}<\) 10 Myr have measurements consistent within 1 \(\sigma\), compared to 58\% for stars between 10–100 Myr, and 70\% for stars between 100–1000 Myr. Although all these stars are younger than 1 Gyr, we should be cautious in the statistical analysis of planet-hosting stars within 100 Myr due to their relatively large age measurement uncertainties. Nonetheless, this inhomogeneity in age measurements has little influence on our analysis of environmental influences, such as the comparison of planets within open clusters and moving groups.

Several factors may contribute to the discrepancies in age measurements between the two studies. First, their different scientific objectives led to different membership definitions for the same stellar groups. Specifically, \citetalias{2020AJ....160..279K} focused on extended structures (including more moving groups), while \citetalias{2024A&A...686A..42H} concentrated on open clusters.

The more inclusive approach to extended structures in \citetalias{2020AJ....160..279K} inevitably introduces more field stars that coincidentally share similar spatial and kinematic properties with genuine group members. Conversely, this approach enables the discovery of more moving groups.

One solution to this problem would be to establish a comprehensive catalog of both open clusters and moving groups and then apply a uniform method for age estimation. This approach would mitigate systematic errors and benefit subsequent statistical analyses. Another solution is to use complementary methods to constrain and obtain accurate ages, such as lithium abundance, gyrochronology, and isochrone fitting. For gyrochronology, obtaining rotation periods for other member stars via photometric data can provide more robust results when applied to a group of stars.

\section{summary and conclusion} \label{summary}
Young planets, particularly those younger than 100 Myr, are crucial for understanding planetary formation and evolution. They represent a temporal bridge between protoplanetary disks ($<$10 Myr) and mature planetary systems ($>$1 Gyr). However, only about 30 such young planets are currently listed in the NASA exoplanet archive. To enlarge this sample, we compiled a more comprehensive catalog than UPiC-I by cross-matching catalogs of confirmed planets and candidates with catalogs of stellar groups.

In Section \ref{method}, we derived Sample-1 and Sample-2 by cross-matching the catalogs from \citetalias{2024A&A...686A..42H} and \citetalias{2020AJ....160..279K}, respectively. Combining these two samples, we obtained a full catalog of stellar groups containing planet/candidate hosts. We also updated the parameters of the stellar groups, such as total mass\(M_{cl}\), core radius(\(R_{c}\)), and tidal radius(\(R_{t}\))(See tables in Appendix \ref{catalogs}).

In Section \ref{sec3}, after identifying substructures using the \texttt{HDBSCAN} clustering algorithm and a Gaussian Mixture Model (GMM), we classified the stellar groups as open clusters (OCs) or moving groups (MGs) based on their Jacobi radius (\(R_{J}\)). This process identified \text{122 planets/candidates in OCs and 152 in MGs} (see Appendix \ref{pl} and \ref{plcl}).

In Section \ref{OC&MG}, we compare the properties of OCs and MGs. OCs possess distinct core regions, while MGs are more sparsely distributed (Fig.~\ref{OC_MG_classify}). Fig.~\ref{OCMG_comparison} reveals that OCs are generally more massive, more concentrated, and smaller in size than MGs. MGs show positive \(M_{cl}\)--\(r_{50}\) and age--\(r_{50}\) correlations, whereas OCs show no obvious trend. For the total radius (\(r_{tot}\)), however, both MGs and OCs exhibit positive correlations with both \(M_{cl}\) and age. Interestingly, we find that MGs in Sample 1 are typically older than those in Sample 2, although their dynamical ages (age/\(r_{th}\)) show similar distributions (Fig.~\ref{H2024_K2020_age_compare}a). We attribute this difference to the distinct origins of the moving groups in each sample, i.e., primordial young MGs versus older MGs that have dissolved from OCs (a more detailed analysis is provided in Section \ref{density-age}).

Beyond comparing the properties of OCs and MGs, we also compare the parameters of planet-hosting stars within them (Section \ref{4.1}). We find that planet-hosting stars in OCs and MGs have similar \texttt{T\_eff} distributions, but those in OCs reside in significantly higher surrounding stellar number densities \(\rho\) than those in MGs. We also compare planet-hosting stars with non-planet-hosting stars. The latter are generally hotter (higher \texttt{T\_eff}) and fainter (larger Gmag), yet their \(\rho\)  distributions are nearly identical. 

In Section \ref{density-age}, Fig.~\ref{Density-age} shows the distribution of surrounding stellar number density for planet-hosting stars as a function of age. This reveals primordial young MGs and OCs that dissolve quickly within the first 30 Myr during the gas expulsion phase, as well as older MGs that dissolve via tidal interactions between OCs and the Milky Way—effectively the remnants of disrupted clusters.

We also performed preliminary statistical analyses in Section \ref{results}:

First, we find that MGs have a significantly higher planet occurrence rate than OCs, especially for Jovian-sized planets (Table \ref{Fraction} and Fig.~\ref{Rp_age}). This difference may be due to different primordial radiation environments. OCs often host massive stars with strong UV radiation, which can evaporate nearby protoplanetary disks and inhibit planet formation, particularly for giant planets. However, this finding requires further validation to account for observational biases in crowded regions.

Second, we explore the planet radius–age relation in Section \ref{sec5.2}. We find a pile-up of Hot Jupiters around 100 Myr, consistent with our previous work, which explained this peak through flyby-induced high-eccentricity tidal migration (HEM). This mechanism predicts a higher occurrence of Hot Jupiters in open clusters. Contrary to this prediction, we found that Hot Jupiters predominantly reside in MGs. While we cannot rule out the early arrival of some Hot Jupiters via HEM, our results suggest that in-situ formation and disk migration also play important roles. Furthermore, we find the formation of the Hot-Neptune desert around 100 Myr. We use models of photoevaporation and planetary cooling contraction to constrain the radius evolution of sub-Jupiters. Simulations suggest these young sub-Jupiters are likely ``puffy'', and that Neptune-mass planets undergo strong photoevaporation from their host stars within the first 100 Myr (Fig.~\ref{Rp_age} and Fig.~\ref{Rp_PDF}). These young, puffy Neptune-mass planets are also consistent with the findings of \cite{2025MNRAS.tmp..591R}.

Third, in Fig.~\ref{RP_P}, we find that sub-Jupiters are predominantly young, regardless of their orbital period. The number of planets in all three regions—the desert, ridge, and savanna—decreases with age. We also find that OC planets exhibit higher desert and ridge occupancy than their MG counterparts, while MGs have higher savanna occupancy. However, more data are needed to validate this phenomenon and explore the underlying mechanism.

In the future, with the extended TESS mission, the upcoming Earth 2.0 mission (ET2.0; \citep{2022arXiv220606693G}), the Chinese Space Station Telescope (CSST; \citep{2011SSPMA..41.1441Z, 2019ApJ...883..203G}), PLATO \citep{2014ExA....38..249R}, and the forthcoming Gaia DR4, we expect to detect more young planets in both open clusters and moving groups. This will deepen our understanding of environmental influences on planet formation and evolution.

\acknowledgments
We thank Dr. Emily Hunt, Prof. Fei Dai, and Prof. Chengyuan Li for their helpful recommendations to improve the paper.

This work is supported by the National Key R\&D Program of China (Grant No.~\texttt{2024YFA1611801}) and the National Natural Science Foundation of China (Grant No.~\texttt{11973028}). We also acknowledge the science research grants from the China Manned Space Project (No.~\texttt{CMSCSST-2025-A16}) and the Civil Aerospace Technology Research Project (No.~\texttt{D010102}). Xiaoying Pang acknowledges the financial support of the National Natural Science Foundation of China through Grants~\texttt{12573036} and~\texttt{12233013}.

This research has made use of the Exoplanet Follow-up Observation Program (\href{https://exofop.ipac.caltech.edu/}{ExoFOP}), which is operated by the California Institute of Technology, under contract with the National Aeronautics and the NASA Exoplanet Archive, which is operated by the California Institute of Technology, under contract with the National Aeronautics and Space Administration under the Exoplanet Exploration Program.

This work has made use of data from the European Space Agency (ESA) mission \emph{Gaia} (\url{https://www.cosmos.esa.int/gaia}), processed by the Gaia Data Processing and Analysis Consortium (DPAC, \url{https://www.cosmos.esa.int/web/gaia/dpac/consortium}). Funding for the DPAC has been provided by national institutions, in particular the institutions participating in the Gaia Multilateral Agreement (MLA).

\vspace{5mm}
\software{astropy \citep{2013A&A...558A..33A}, 
          matplotlib \citep{Hunter:2007},
          pandas \citep{mckinney-proc-scipy-2010},
          galpy \citep{2015ApJS..216...29B},
          PlanetSynth \citep{2021MNRAS.507.2094M},
          Photoevolver \citep{2023MNRAS.522.4251F},
          Mors \citep{2021A&A...649A..96J}
          }



\clearpage

\appendix 

\section{Catalogs} \label{catalogs}
This paper primarily describes four catalogs:
\begin{itemize}
    \item The catalog of planets and candidates in open clusters and moving groups (Table \ref{plt} in Appendix \ref{pl}).
    \item The catalog of planet-hosting stars in open clusters and moving groups (Table \ref{plht} in Appendix \ref{plh}).
    \item The catalog of open clusters and moving groups with planets/candidates (Table \ref{plclt} in Appendix \ref{plcl}).
    \item The catalog of member stars in planet-hosting open clusters and moving groups (Table \ref{plclmt} in Appendix \ref{plclm}).
\end{itemize}

\subsection{Catalog Description} \label{description}
This subsection provides a combined description of all columns from the four catalogs, presented in Table \ref{combine}.

\begin{deluxetable}{lllll}
\tablewidth{0pt}
\tablecaption{Columns of the four catalogs\label{combine}}
\tablenum{A1} 
\tablehead{
\colhead{Column} & \colhead{Name} & \colhead{Format} & \colhead{Unit} & \colhead{Description}
}
\startdata
\multicolumn{5}{c}{Columns for Planets/Candidates in OCs and MGs (Table \ref{plt})} \\
\hline
1 & \texttt{Planet\_name} & String &  & The name of planet \\
2 & \texttt{P} & Double & days & Planetary orbital periods\\
3 & \texttt{P\_err1} & Double & days & The lower error of planetary orbital periods\\
4 & \texttt{P\_err2} & Double & days & The upper error of planetary orbital periods\\
5 & \texttt{R\_p} & Double & \(R_{\oplus}\) & Planetary radius\\
6 & \texttt{R\_p\_err1} & Double & \(R_{\oplus}\) & The lower error of planetary radius\\
7 & \texttt{R\_p\_err2} & Double & \(R_{\oplus}\) & The upper error of planetary radius\\
8 & \texttt{M\_p} & Double & \(M_{\oplus}\) & Planetary mass\\
9 & \texttt{M\_p\_err1} & Double & \(M_{\oplus}\) & The lower error of planetary mass\\
10 & \texttt{M\_p\_err2} & Double & \(M_{\oplus}\) & The upper error of planetary mass\\
11 & \texttt{ecc} & Double &  & Planetary orbital eccentricity\\
12 & \texttt{ecc\_err1} & Double &  & The lower error of planetary orbital eccentricity\\
13 & \texttt{ecc\_err2} & Double &  & The upper error of planetary orbital eccentricity\\
14 & \texttt{CP\_flag} & Integer &  & Flag presents whether it is a confirmed planet (1: CP, 0: not CP) \\
15 & \texttt{KOI\_flag} & Integer &  & Flag presents whether it is a Kepler Objects of Interest(KOI) \\
16 & \texttt{TOI\_flag} & Integer &  & Flag presents whether it is a TESS Objects of Interest(TOI) \\
17 & \texttt{CTOI\_flag} & Integer & & Flag presents whether it is a Community TESS Objects of Interest(CTOI) \\
18 & \texttt{PATHOS\_flag} & Integer & & Flag presents whether it is in PATHOS catalog \\
19 & \texttt{Source\_ID} & Long & & Gaia DR3 source id \\
20 & \texttt{Disposition} & String & & Whether this target is a Confirmed Planet(CP) or a Planetary Candidates(PC) or a False Positive(FP). \\
21 & \texttt{Comments} & String & & The additional notes about TOIs and CTOIs \\
\hline
\multicolumn{5}{c}{Columns of Planet host stars in OCs and MGs (Table \ref{plht})}  \\
\hline
1 & \texttt{Source\_ID} & Long & & Gaia DR3 source id \\
2 & \texttt{Cluster\_Name} & String & & Cluster names in H23\&H24 \\
3 & \texttt{Group} & Integer & & Group number in K20 \\
4 & \texttt{Newgroup} & Integer & & New group number after substructure identification from original sample in K20 \\
5 & \texttt{X\_proj} & Double & pc & X-coordinate of the projection plane \\
6 & \texttt{Y\_proj} & Double & pc & Y-coordinate of the projection plane \\
7 & \texttt{reference} & String & & The reference paper \\
8 & \texttt{age\_H24} & Double & Myr & The median age of stellar groups in H24 \\
9 & \texttt{age\_H24\_err1} & Double & Myr & The lower error of age of stellar groups in H24 \\
10 & \texttt{age\_H24\_err2} & Double & Myr & The upper error of age of stellar groups in H24 \\
11 & \texttt{age\_K20} & Double & Myr & The median age of stellar groups in K20 \\
12 & \texttt{age\_K20\_err1} & Double & Myr & The lower error of age of stellar groups in K20 \\
13 & \texttt{age\_K20\_err1} & Double & Myr & The upper error of age of stellar groups in K20 \\
14 & \texttt{Ms\_H24} & Double & \(M_{\odot}\) & The median mass of stars in H24 \\
15 & \texttt{Ms\_H24\_err1} & Double & \(M_{\odot}\) & The lower error of mass of stars in H24 \\
16 & \texttt{Ms\_H24\_err2} & Double & \(M_{\odot}\) & The upper error of mass of stars in H24 \\
17 & \texttt{Ms\_iso\_H24} & Double & \(M_{\odot}\) & The median mass of stars in H24 (This work) \\
18 & \texttt{Ms\_iso\_K20} & Double & \(M_{\odot}\) & The median mass of stars in K20 (This work) \\
19 & \texttt{\(\rho\)\_King} & Double & (\#/pc\(^{2}\)) & The surrounding stellar number density of the star (King model) \\
20 & \texttt{\(\rho\)\_King\_err1} & Double & (\#/pc\(^{2}\)) & The lower error of the surrounding stellar number density (King model) \\
21 & \texttt{\(\rho\)\_King\_err2} & Double & (\#/pc\(^{2}\)) & The upper error of the surrounding stellar number density (King model) \\
22 & \texttt{\(\rho\)\_EEF} & Double & (\#/pc\(^{2}\)) & The surrounding stellar number density of the star fitting by EEF model \\
23 & \texttt{\(\rho\)\_EEF\_err1} & Double & (\#/pc\(^{2}\)) & The lower error of the surrounding stellar number density (EEF model) \\
24 & \texttt{\(\rho\)\_EEF\_err2} & Double & (\#/pc\(^{2}\)) & The upper error of the surrounding stellar number density (EEF model) \\
25 & \texttt{within\_R\_J} & Boolean & & The flag shows whether the planet hosting star is within the Jacobi radius \(R_{J}\) \\
26 & \texttt{Probability} & Double & & The membership probability of planet hosting stars in H24\\
27 & \texttt{Comments} & String & & Notes about whether this star is high-probability(HP) or low-probability(LP) member stars. \\
\hline
\multicolumn{5}{c}{Columns of Planet Hosting Groups (Table \ref{plclt})}\\
\hline
1 & \texttt{Cluster\_Name} & String & & The name of planet hosting groups in sample 1 \\
2 & \texttt{Group} & Int64 & & Original group number in K20(before the substructure extraction) \\
3 & \texttt{Newgroup} & Int64 & & The number of groups in sample 2(after the substructure extraction) \\
4 & \texttt{M\_cluster} & Double & & The total stellar mass of the stellar groups (no correction, this work) \\
5 & \texttt{M\_cluster\_H24} & Double & & The total stellar mass of the stellar groups (no correction, H24) \\
6 & \texttt{M\_cluster\_H24\_cor} & Double & & The total stellar mass of the stellar groups (with correction, H24) \\
7 & \(\rho\)\texttt{\_c,King} & Double & (\#/pc\(^{2}\)) & The central stellar number density of a stellar group (King model) \\
8 & \(\rho\)\texttt{\_c,King\_err1} & Double & (\#/pc\(^{2}\)) & The lower error of the central stellar number density of a stellar group (King model) \\
9 & \(\rho\)\texttt{\_c,King\_err2} & Double & (\#/pc\(^{2}\)) & The upper error of the central stellar number density of a stellar group (King model) \\
10 & \texttt{R\_c,King} & Double & (pc) & The core radius of a stellar group (King model) \\
11 & \texttt{R\_c,King\_err1} & Double & (pc) & The lower error of core radius of a stellar group (King model) \\
12 & \texttt{R\_c,King\_err2} & Double & (pc) & The upper error of core radius of a stellar group (King model) \\
13 & \texttt{R\_t,King} & Double & (pc) & The tidal radius of a stellar group (King model) \\
14 & \texttt{R\_t,King\_err1} & Double & (pc) & The lower error of tidal radius of a stellar group (King model) \\
15 & \texttt{R\_t,King\_err2} & Double & (pc) & The upper error of tidal radius of a stellar group (King model) \\
\enddata
\end{deluxetable}

\begin{deluxetable}{lllll}
\tablewidth{0pt}
\tablecaption{Continue}
\tablenum{A1} 
\tablehead{
\colhead{Column} & \colhead{Name} & \colhead{Format} & \colhead{Unit} & \colhead{Description}
}
\startdata
16 & \texttt{C} & Double & (\#/pc\(^{2}\)) & The constant (King model) \\ 
17 & \texttt{C\_err1} & Double & (\#/pc\(^{2}\)) & The lower error of the constant (King model) \\
18 & \texttt{C\_err2} & Double & (\#/pc\(^{2}\)) & The upper error of the constant (King model) \\
19 & \(\rho\)\texttt{\_c,EEF} & Double & (\#/pc\(^{2}\)) & The central stellar number density of a stellar group (EEF model) \\
20 & \(\rho\)\texttt{\_c,EEF\_err1} & Double & (\#/pc\(^{2}\)) & The lower error of the central stellar number density of a stellar group (EEF model) \\
21 & \(\rho\)\texttt{\_c,EEF\_err2} & Double & (\#/pc\(^{2}\)) & The upper error of the central stellar number density of a stellar group (EEF model) \\
22 & \texttt{R\_c,EEF} & Double & (pc) & The core radius of a stellar group (EEF model) \\
23 & \texttt{R\_c,EEF\_err1} & Double & (pc) & The lower error of core radius of a stellar group (EEF model) \\
24 & \texttt{R\_c,EEF\_err2} & Double & (pc) & The upper error of core radius of a stellar group (EEF model) \\
25 & \texttt{R\_a,EEF} & Double & (pc) & The typical radius of a stellar group (EEF model) \\
26 & \texttt{R\_a,EEF\_err1} & Double & (pc) & The lower error of typical radius of a stellar group (EEF model) \\
27 & \texttt{R\_a,EEF\_err2} & Double & (pc) & The upper error of typical radius of a stellar group (EEF model) \\
28 & \(\gamma\) & Double &  & The slope of the stellar number density distribution(EEF model) \\
29 & \(\gamma\)\texttt{\_err1} & Double &  & The lower error of the slope of the stellar number density distribution(EEF model) \\
30 & \(\gamma\)\texttt{\_err2} & Double &  & The upper error of the slope of the stellar number density distribution(EEF model) \\
31 & \texttt{R\_c\_H24} & Double & (pc) & The core radius of the stellar group (H24)\\ 
32 & \texttt{R\_50} & Double & (pc) & The half number radius of the stellar group \\ 
33 & \texttt{R\_50\_H24} & Double & (pc) & The half number radius of the stellar group (H24)\\ 
34 & \texttt{R\_tot} & Double & (pc) & The total radius of the stellar group \\
35 & \texttt{R\_tot\_H24} & Double & (pc) & The total radius of the stellar group (H24)\\
36 & \texttt{R\_J} & Double & (pc) & The Jacobi radius of the stellar group \\
37 & \texttt{R\_J\_H24} & Double & (pc) & The Jacobi radius of the stellar group (H24) \\
38 & \texttt{M\_J} & Double & \(M_{\odot}\) & The stellar mass within the Jacobi radius\\
39 & \texttt{P\_RJ} & Double & \(M_{\odot}\) & The fraction of the number of stars within the Jacobi radius\\
40 & \texttt{kind} & String & & The classification of stellar groups. `o' means open clusters and `m' means moving groups \\
41 & \texttt{kind\_2024} & String & & The classification of stellar groups in H24. `o' means open clusters and `m' means moving groups \\
42 & \texttt{nstars} & Int64 & & The number of member stars in a stellar group \\
43 & \texttt{comments} & String & & Notes whether it is a high-quality(HQ) or low-quality(LQ) stellar group. \\
\hline
\multicolumn{5}{c}{Columns of Member stars in OCs and MGs hosting planets (Table \ref{plclmt})}  \\
\hline
1 & \texttt{Source\_ID} & Long & & Gaia DR3 source id \\
2 & \texttt{Cluster\_Name} & String & & Cluster names in H23\&H24 \\
3 & \texttt{Group} & Integer & & Group number in K20 \\
4 & \texttt{Newgroup} & Integer & & New group number after substructure identification from original sample in K20 \\
5 & \texttt{X\_proj} & Double & pc & X-coordinate of the projection plane \\
6 & \texttt{Y\_proj} & Double & pc & Y-coordinate of the projection plane \\
7 & \texttt{ra} & Double & Degree & The right ascension  \\
8 & \texttt{ra\_error} & Double & Degree & The standard error of the right ascension \\
9 & \texttt{dec} & Double &  Degree &The declination \\
10 & \texttt{dec\_error} & Double &  Degree &The standard error of the declination \\
11 & \texttt{l} & Double & Degree & The galactic longitude \\
12 & \texttt{b} & Double & Degree & The galactic latitude \\
13 & \texttt{pmra} & Double & mas/yr & Proper motion in ra (multiplied with cos (dec)) \\
14 & \texttt{pmra\_error} & Double & mas/yr & The standard error of proper motion in ra (multiplied with cos (dec)) \\
15 & \texttt{pmdec} & Double & mas/yr & Proper motion in dec \\
16 & \texttt{pmdec\_error} & Double & mas/yr & The standard error of proper motion in dec\\
17 & \texttt{ruwe} & Double &  & Renormalised Unit Weight Error\\ 
18 & \texttt{phot\_g\_mean\_mag} & Double & mag & G-band mean magnitude \\
19 & \texttt{phot\_bp\_mean\_mag} & Double & mag & Integrated RP mean magnitude, Optical B band between 400 and 500 nm\\
20 & \texttt{phot\_rp\_mean\_mag} & Double & mag & Integrated RP mean magnitude, Optical R band between 600 and 750 nm  \\
21 & \texttt{bp\_rp} & Double & mag & \texttt{bp\_rp=phot\_bp\_mean\_mag-phot\_rp\_mean\_mag}\\
22 & \texttt{bp\_g} & Double & mag & \texttt{bp\_g=phot\_bp\_mean\_mag-phot\_g\_mean\_mag}\\
23 & \texttt{g\_rp} & Double & mag & \texttt{g\_rp=phot\_g\_mean\_mag-phot\_rp\_mean\_mag}\\
24 & \texttt{radial\_velocity} & Double & km/s & Radial velocity derived from a wavelength shift using the optical convention\\
25 & \texttt{radial\_velocity\_error} & Double & km/s & Radial velocity error \\
26 & \texttt{mass\_50} & Double & \(M_{\odot}\) & Median value of stellar mass \\
27 & \texttt{T\_eff} & Double & K & Effective temperature from GSP-Phot Aeneas best library using BP/RP spectra \\
28 & \texttt{parallax} & Double & mas & Parallax\\
29 & \texttt{parallax\_error} & Double & mas & Parallax error\\
30 & \texttt{r\_med\_geo} & Double & pc & Median of the geometric distance posterior\\ 
31 & \texttt{r\_lo\_geo} & Double & pc & 16th percentile of the geometric distance posterior\\ 
32 & \texttt{r\_hi\_geo} & Double & pc & 84th percentile of the geometric distance posterior\\ 
33 & \texttt{r\_med\_photogeo} & Double & pc & Median of the photogeometric distance posterior\\ 
34 & \texttt{r\_lo\_photogeo} & Double & pc & 16th percentile of the photogeometric distance posterior\\ 
35 & \texttt{r\_hi\_photogeo} & Double & pc & 84th percentile of the geometric distance posterior\\
\enddata
\tablecomments{Additional notes about the data}
\end{deluxetable}
\subsection{Catalog of Planets in OCs and MGs} \label{pl}
\begin{deluxetable*}{lrrrrrrrrrrrrr}
    \rotate
    \tabletypesize{\tiny} 
    \tablecaption{Catalog of Planets and Candidates in OCs and MGs\label{plt}}
    \tablewidth{0pt}
    \tablenum{A2}
    \tablehead{
        \colhead{\texttt{Pl\_name}} & \colhead{\texttt{P}} & \colhead{\texttt{Rp}} & \colhead{\texttt{Mp}} & \colhead{\texttt{ecc}} & \colhead{\texttt{CP\_flag}} & \colhead{\texttt{KOI\_flag}} & \colhead{\texttt{K2\_flag}} & \colhead{\texttt{TOI\_flag}} & \colhead{\texttt{CTOI\_flag}} & \colhead{\texttt{PATHOS\_flag}} & \colhead{\texttt{Source\_ID}} & \colhead{\texttt{Disposition}} & \colhead{\texttt{Comments}} \\
        \colhead{} & \colhead{(days)} & \colhead{(R$_{\oplus}$)} & \colhead{(M$_{\oplus}$)} & \colhead{} & \colhead{} & \colhead{} & \colhead{} & \colhead{} & \colhead{} & \colhead{} & \colhead{} & \colhead{} & \colhead{} \\
    }
    \colnumbers
    \startdata
    1RXS J160929.1-210524 b & -- & -- & 3000.00 & -- & 1 & 0 & 0 & 0 & 0 & 0 & 6243841249531772800 & CP & -- \\
    2MASS J02192210-3925225 b & -- & 16.14 & 4417.84 & -- & 1 & 0 & 0 & 0 & 0 & 0 & 4963614887043956096 & CP & -- \\
    51 Eri b & -- & -- & 635.66 & -- & 1 & 0 & 0 & 0 & 0 & 0 & 3205095125321700480 & CP & -- \\
    AB Aur b & -- & -- & 2860.46 & 0.400 & 1 & 0 & 0 & 0 & 0 & 0 & 156917493449670656 & CP & -- \\
    AF Lep b & 8030.000 & -- & 1017.05 & 0.240 & 1 & 0 & 0 & 0 & 0 & 0 & 3009908378049913216 & CP & -- \\
    BD-13 2130 b & 714.300 & -- & -- & 0.210 & 1 & 0 & 0 & 0 & 0 & 0 & 3030262468592291072 & CP & -- \\
    CHXR 73 b & -- & -- & 3994.49 & -- & 1 & 0 & 0 & 0 & 0 & 0 & 5201175987817179136 & CP & -- \\
    CI Tau c & 25.200 & -- & -- & 0.580 & 1 & 0 & 0 & 0 & 0 & 0 & 145203159127518336 & CP & -- \\
    CT Cha b & -- & 24.66 & 5403.00 & -- & 1 & 0 & 0 & 0 & 0 & 0 & 5201360671411974912 & CP & -- \\
    DH Tau b & -- & -- & 3496.00 & -- & 1 & 0 & 0 & 0 & 0 & 0 & 151374202498079872 & CP & -- \\
    DS Tuc A b & 8.138 & 5.70 & -- & 0.000 & 1 & 0 & 0 & 1 & 0 & 0 & 6387058411482257536 & CP & DS Tuc A b; ... \\
    EPIC 211822797 b & 21.170 & 1.92 & -- & -- & 1 & 0 & 1 & 0 & 0 & 0 & 658607950370340608 & CP & -- \\
    FU Tau b & -- & -- & 5085.28 & -- & 1 & 0 & 0 & 0 & 0 & 0 & 149629483705467008 & CP & -- \\
    GSC 06214-00210 b & -- & -- & 5000.00 & -- & 1 & 0 & 0 & 0 & 0 & 0 & 6244440552088537600 & CP & -- \\
    HD 97048 b & -- & -- & 794.58 & -- & 1 & 0 & 0 & 0 & 0 & 0 & 5201128124701636864 & CP & -- \\
    HIP 21152 b & 22000.000 & -- & 7627.88 & 0.360 & 1 & 0 & 0 & 0 & 0 & 0 & 3285426613077584384 & CP & -- \\
    IC 4651 9122 b & 734.000 & -- & -- & 0.180 & 1 & 0 & 0 & 0 & 0 & 0 & 5949553973093167104 & CP & -- \\
    K2-100 b & 1.674 & 3.88 & 21.80 & 0.000 & 1 & 0 & 1 & 1 & 0 & 0 & 664337230586013312 & CP & K2-100 b \\
    K2-101 b & 14.676 & 2.00 & -- & -- & 1 & 0 & 1 & 0 & 0 & 0 & 661004400386393984 & CP & -- \\
    K2-102 b & 9.916 & 1.30 & -- & 0.100 & 1 & 0 & 1 & 0 & 0 & 0 & 661310756111835392 & CP & -- \\
     \ldots & \ldots & \ldots & \ldots & \ldots & \ldots & \ldots & \ldots & \ldots & \ldots & \ldots & \ldots & \ldots & \ldots \\
    CTOI 81353413.01 & 2.087 & 15.90 & -- & -- & 0 & 0 & 0 & 0 & 1 & 1 & 5519320261438248704 & PC & new CTOI lik... \\
    CTOI 93275779.01 & 0.914 & 6.28 & -- & -- & 0 & 0 & 0 & 0 & 1 & 0 & 5321581959996214016 & PC & Cand super-N... \\
    CTOI 94589619.01 & 12.083 & 30.40 & -- & -- & 0 & 0 & 0 & 0 & 1 & 1 & 3029589773632754688 & FP & new CTOI lik... \\
    CTOI 110718787.01 & 3.092 & 22.67 & -- & -- & 0 & 0 & 0 & 0 & 1 & 0 & 5599752663752776192 & PC & Cand HJ in H... \\
    CTOI 125414447.01 & 3.707 & 25.10 & -- & -- & 0 & 0 & 0 & 0 & 1 & 1 & 3051560489855207936 & FP & new CTOI lik... \\
    CTOI 144995073.01 & 20.021 & 36.00 & -- & -- & 0 & 0 & 0 & 0 & 1 & 1 & 5318327336862657408 & FP & new CTOI lik... \\
    CTOI 147069011.01 & 7.328 & 41.73 & -- & -- & 0 & 0 & 0 & 0 & 1 & 1 & 5889360246986113024 & FP & Likely EB, A... \\
    CTOI 147426828.01 & 1.462 & 35.90 & -- & -- & 0 & 0 & 0 & 0 & 1 & 1 & 2949493547003693056 & FP & new CTOI lik... \\
    CTOI 148158540.01 & 1.849 & -- & -- & -- & 0 & 0 & 0 & 0 & 1 & 0 & 6005849243285423104 & FP & Consistent w... \\
    CTOI 150070085.01 & 10.475 & 3.64 & -- & -- & 0 & 0 & 0 & 0 & 1 & 0 & 986757333119241216 & PC & Identified a... \\
    CTOI 153734545.01 & 3.220 & 53.70 & -- & -- & 0 & 0 & 0 & 0 & 1 & 1 & 5597873804538618496 & FP & new CTOI lik... \\
    CTOI 153735144.01 & 3.779 & 29.00 & -- & -- & 0 & 0 & 0 & 0 & 1 & 1 & 5597846003199406976 & FP & new CTOI lik... \\
    CTOI 159059181.01 & 5.217 & 27.00 & -- & -- & 0 & 0 & 0 & 0 & 1 & 1 & 3218880217994699136 & FP & new CTOI lik... \\
    CTOI 181602717.01 & 2.456 & 40.20 & -- & -- & 0 & 0 & 0 & 0 & 1 & 1 & 5524972919429439232 & FP & new CTOI lik... \\
    CTOI 198390707.01 & 0.141 & 5.50 & -- & -- & 0 & 0 & 0 & 0 & 1 & 0 & 1434127655020848768 & PC & seems to be ... \\
    CTOI 206544316.01 & 0.324 & 4.06 & -- & -- & 0 & 0 & 0 & 0 & 1 & 0 & 4909009394396454144 & PC & new CTOI \\
    CTOI 206544316.02 & 0.325 & 13.54 & -- & -- & 0 & 0 & 0 & 0 & 1 & 0 & 4909009394396454144 & PC & new CTOI \\
    CTOI 206544316.03 & 0.321 & 7.19 & -- & -- & 0 & 0 & 0 & 0 & 1 & 0 & 4909009394396454144 & PC & new CTOI \\
    CTOI 206544316.04 & 0.326 & 13.67 & -- & -- & 0 & 0 & 0 & 0 & 1 & 0 & 4909009394396454144 & PC & new CTOI \\
    CTOI 234284556.01 & 1.106 & -- & -- & -- & 0 & 0 & 0 & 0 & 1 & 0 & 6405089921141776128 & FP & Planet Candi... \\
    \enddata
    \tablecomments{This table is published in its entirety in machine-readable format. We just show 40 rows with key parameters(the entire parameters with errors are shown in \ref{combine}).}
\end{deluxetable*}

We checked the provided orbital periods of CTOI 198390707.01, CTOI 206544316.01, CTOI 206544316.02, CTOI 206544316.03, and CTOI 206544316.04, but did not detect obvious transit signals. For instance, the rotation period of TIC 206544316 is close to the orbital periods of the associated CTOIs. Therefore, we classify them as False Positives (FP). Among the remaining objects, there are 106 confirmed planets, 168 planet candidates, 3 targets with single transit events, and one confirmed brown dwarf, TOI-5090.01 (also known as AD 3116 b, a 54 M$_{\mathrm{Jup}}$ brown dwarf; \cite{2017ApJ...849...11G}).

\subsection{Catalog of Planet Hosting Stars in OCs and MGs} \label{plh}
Table \ref{plht} comprises 406 stars, of which 171 are planet-hosting stars from Sample 1 and 235 from Sample 2. A further breakdown shows that, within Sample 1, 126 reside in open clusters (OCs) and 45 in moving groups (MGs); the corresponding numbers for Sample 2 are 95 in OCs and 140 in MGs.

\begin{deluxetable*}{lrrrrrrrrrrrrrrr}
    \rotate
    \tabletypesize{\tiny} 
    \tablecaption{Catalog of Planet Host Stars in Open Clusters and Moving Groups\label{plht}}
    \tablewidth{0pt}
    \tablenum{A3}
    \tablehead{
        \colhead{\texttt{Source\_ID}} & \colhead{\texttt{Cluster\_name}} & \colhead{\texttt{Group}} & \colhead{\texttt{Newgroup}} & \colhead{\texttt{X\_proj}} & \colhead{\texttt{Y\_proj}} & \colhead{\texttt{reference}} & \colhead{\texttt{age\_H24}} & \colhead{\texttt{age\_K20}} & \colhead{\texttt{Ms\_H24}} & \colhead{\texttt{Ms\_K20}} & \colhead{\texttt{\(\rho\)\_King}} & \colhead{\texttt{\(\rho\)\_EEF}} & \colhead{\texttt{within\_R\_J}} & \colhead{\texttt{Probability}} & \colhead{\texttt{Comments}} \\
        \colhead{} & \colhead{} & \colhead{} & \colhead{} & \colhead{(pc)} & \colhead{(pc)} & \colhead{} & \colhead{(Myr)} & \colhead{(Myr)} & \colhead{(M$_{\odot}$)} & \colhead{(M$_{\odot}$)} & \colhead{(stars/pc$^2$)} & \colhead{(stars/pc$^2$)} & \colhead{} & \colhead{} & \colhead{} \\
    }
    \colnumbers
    \startdata
    6243841249531772800 & \texttt{OCSN\_96} & -- & -- & 3.50 & 3.11 & H24 & 4.7 & -- & 0.9 & -- & 1.1 & 0.8 & True & 0.791 & HP \\
    4963614887043956096 & \texttt{beta\_Tuc\_Group} & -- & -- & -6.88 & 16.08 & H24 & 31.4 & -- & 0.2 & -- & 0.1 & 0.1 & False & 0.633 & HP \\
    3205095125321700480 & \texttt{FSR\_1017} & -- & -- & -5.83 & 0.27 & H24 & 15.7 & -- & 1.8 & -- & 0.1 & 0.1 & False & 1.000 & HP \\
    156917493449670656 & \texttt{Theia\_54} & -- & -- & -1.76 & 1.23 & H24 & 4.1 & -- & 2.6 & -- & 1.1 & 1.1 & True & 1.000 & HP \\
    3009908378049913216 & \texttt{FSR\_1017} & -- & -- & 0.85 & -4.81 & H24 & 15.7 & -- & 1.5 & -- & 0.1 & 0.1 & False & 1.000 & HP \\
    3030262468592291072 & \texttt{NGC\_2423} & -- & -- & -0.26 & -0.70 & H24 & 1182.4 & -- & 2.2 & -- & 8.7 & 6.8 & True & 1.000 & HP \\
    5201175987817179136 & \texttt{Chamaleon\_I} & -- & -- & -0.45 & -1.64 & H24 & 5.7 & -- & 0.3 & -- & 5.8 & 4.7 & True & 0.972 & HP \\
    145203159127518336 & \texttt{CWNU\_1129} & -- & -- & 0.40 & -3.88 & H24 & 15.0 & -- & 1.3 & -- & 0.1 & 0.1 & True & 1.000 & HP \\
    5201360671411974912 & \texttt{Chamaleon\_I} & -- & -- & -0.94 & 2.21 & H24 & 5.7 & -- & 1.6 & -- & 2.7 & 1.8 & True & 1.000 & HP \\
    151374202498079872 & \texttt{Theia\_7} & -- & -- & -2.23 & 4.53 & H24 & 187.3 & -- & 1.2 & -- & 0.5 & 0.5 & False & 1.000 & HP \\
    6387058411482257536 & \texttt{beta\_Tuc\_Group} & -- & -- & -14.30 & -16.94 & H24 & 31.4 & -- & 1.1 & -- & 0.1 & 0.1 & False & 1.000 & HP \\
    658607950370340608 & \texttt{NGC\_2632} & -- & -- & 1.30 & -6.11 & H24 & 346.2 & -- & 0.6 & -- & 2.4 & 2.3 & True & 0.628 & HP \\
    149629483705467008 & \texttt{Theia\_7} & -- & -- & -5.39 & 1.18 & H24 & 187.3 & -- & 0.7 & -- & 0.5 & 0.5 & False & 1.000 & HP \\
    6244440552088537600 & \texttt{CWNU\_1143} & -- & -- & 11.50 & 7.26 & H24 & 18.5 & -- & 0.9 & -- & 0.0 & 0.0 & False & 0.735 & HP \\
    5201128124701636864 & \texttt{Chamaleon\_I} & -- & -- & -0.17 & -1.73 & H24 & 5.7 & -- & 4.8 & -- & 5.6 & 4.4 & True & 0.886 & HP \\
    3285426613077584384 & \texttt{Melotte\_25} & -- & -- & 0.99 & -8.55 & H24 & 576.8 & -- & 1.6 & -- & 0.8 & 0.8 & True & 0.400 & LP \\
    5949553973093167104 & \texttt{IC\_4651} & -- & -- & -0.04 & -0.55 & H24 & 1665.1 & -- & 1.9 & -- & 34.1 & 29.2 & True & 0.927 & HP \\
    664337230586013312 & \texttt{NGC\_2632} & -- & -- & -1.15 & 1.79 & H24 & 346.2 & -- & 1.2 & -- & 15.1 & 14.9 & True & 0.955 & HP \\
    661004400386393984 & \texttt{NGC\_2632} & -- & -- & 1.09 & -1.97 & H24 & 346.2 & -- & 0.8 & -- & 13.9 & 14.0 & True & 1.000 & HP \\
    661310756111835392 & \texttt{NGC\_2632} & -- & -- & 0.21 & 0.74 & H24 & 346.2 & -- & 0.8 & -- & 22.8 & 19.9 & True & 0.984 & HP \\
    664292459846946560 & \texttt{NGC\_2632} & -- & -- & -1.05 & 0.72 & H24 & 346.2 & -- & 0.6 & -- & 20.6 & 18.6 & True & 0.806 & HP \\
    145916050683920128 & \texttt{Melotte\_25} & -- & -- & 0.45 & 5.78 & H24 & 576.8 & -- & 0.7 & -- & 1.9 & 1.9 & True & 0.482 & LP \\
    4184182737768311296 & \texttt{Ruprecht\_147} & -- & -- & 0.10 & 2.34 & H24 & 888.7 & -- & 1.0 & -- & 2.3 & 2.4 & True & 1.000 & HP \\
    3311804515502788352 & \texttt{Melotte\_25} & -- & -- & -2.81 & -0.44 & H24 & 576.8 & -- & 0.4 & -- & 2.9 & 3.0 & True & 0.775 & HP \\
    661167785238757376 & \texttt{NGC\_2632} & -- & -- & 4.14 & 0.50 & H24 & 346.2 & -- & 0.5 & -- & 5.8 & 6.3 & True & 0.746 & HP \\
    6245758900889486720 & \texttt{OCSN\_100} & -- & -- & -0.21 & 0.02 & H24 & 3.6 & -- & 0.5 & -- & 13.9 & 10.9 & True & 1.000 & HP \\
    659744295638254336 & \texttt{NGC\_2632} & -- & -- & -1.88 & -1.83 & H24 & 346.2 & -- & 0.5 & -- & 12.3 & 12.7 & True & 0.761 & HP \\
    2081880913079644288 & \texttt{NGC\_6866} & -- & -- & 4.76 & 1.83 & H24 & 629.2 & -- & 1.1 & -- & 1.1 & 1.3 & True & 0.872 & HP \\
    2052827207364859264 & \texttt{HSC\_572} & -- & -- & -0.94 & -7.77 & H24 & 103.5 & -- & 1.0 & -- & 0.1 & 0.1 & False & 1.000 & HP \\
    2128198836827103616 & \texttt{NGC\_6811} & -- & -- & -4.92 & 6.17 & H24 & 1077.6 & -- & 1.1 & -- & 0.3 & 0.3 & True & 0.405 & LP \\
    2128112181565948800 & \texttt{NGC\_6811} & -- & -- & -2.66 & -3.94 & H24 & 1077.6 & -- & 0.9 & -- & 1.2 & 1.2 & True & 0.510 & HP \\
    144936836795636864 & \texttt{CWNU\_1129} & -- & -- & 3.89 & -5.22 & H24 & 15.0 & -- & 1.4 & -- & 0.1 & 0.1 & True & 1.000 & HP \\
    604903468852379904 & \texttt{NGC\_2682} & -- & -- & 2.44 & -1.90 & H24 & 1688.3 & -- & 1.5 & -- & 12.9 & 13.8 & True & 0.666 & HP \\
    604911375882674560 & \texttt{NGC\_2682} & -- & -- & -0.31 & -0.93 & H24 & 1688.3 & -- & 1.9 & -- & 36.1 & 29.8 & True & 1.000 & HP \\
    604914949295282816 & \texttt{NGC\_2682} & -- & -- & -1.31 & -0.08 & H24 & 1688.3 & -- & 1.1 & -- & 32.1 & 27.8 & True & 0.908 & HP \\
    604922096120852864 & \texttt{NGC\_2682} & -- & -- & -1.31 & 0.97 & H24 & 1688.3 & -- & 1.0 & -- & 27.7 & 25.3 & True & 0.723 & HP \\
    604909657895767680 & \texttt{NGC\_2682} & -- & -- & -0.22 & -2.18 & H24 & 1688.3 & -- & 1.2 & -- & 20.8 & 20.6 & True & 0.819 & HP \\
    6655168686921108864 & \texttt{HSC\_2846} & -- & -- & -3.65 & -12.81 & H24 & 53.5 & -- & 1.2 & -- & 0.1 & 0.1 & False & 1.000 & HP \\
    661222279785743616 & \texttt{NGC\_2632} & -- & -- & 1.70 & -0.87 & H24 & 346.2 & -- & 0.9 & -- & 15.8 & 15.4 & True & 0.838 & HP \\
    6048935358761628288 & \texttt{HSC\_2919} & -- & -- & 0.29 & -3.37 & H24 & 27.1 & -- & 1.1 & -- & 0.5 & 0.5 & True & 0.944 & HP \\
    \enddata
    \tablecomments{This table is published in its entirety in machine-readable format. We just show 40 rows with key parameters(the entire parameters with errors are shown in \ref{combine}).}
\end{deluxetable*}

\subsection{Catalog of OCs and MGs with Planets/Candidates} 
There are 76 open clusters (OCs) and 36 moving groups (MGs) from Sample~1, and 56 OCs and 112 MGs from Sample~2. There are 122 planets/candidates in OCs and 152 in MGs. Due to space limitations, we only list the key parameters without uncertainties. The full machine-readable table with corresponding columns as in Table~\ref{combine} is published in the online material. Additionally, we list the stellar groups separately from Sample~1 and Sample~2; specifically, the first 112 stellar groups are from Sample~1 and the next 168 stellar groups are from Sample~2. We do not combine duplicate OCs or MGs that appear in both samples.

\label{plcl}
\begin{deluxetable*}{lrrrrrrrrrrrrrrrrrrrrrrr}
    \rotate
    \tabletypesize{\tiny} 
    \tablecaption{Catalog of Structural Parameters for Open Clusters and Moving Groups\label{plclt}}
    \tablewidth{0pt}
    \tablenum{A4}
    \tablehead{
        \colhead{\texttt{Cluster\_Name}} & \colhead{\texttt{Group}} & \colhead{\texttt{Newgroup}} & \colhead{\texttt{rho\_c\_King}} & \colhead{\texttt{R\_c\_King}} & \colhead{\texttt{R\_t\_King}} & \colhead{\texttt{C}} & \colhead{\texttt{rho\_c\_EEF}} & \colhead{\texttt{R\_c\_EEF}} & \colhead{\texttt{R\_a\_EEF}} & \colhead{\texttt{gamma}} & \colhead{\texttt{R\_c\_H24}} & \colhead{\texttt{R\_50}} & \colhead{\texttt{R\_50\_H24}} & \colhead{\texttt{R\_tot}} & \colhead{\texttt{R\_tot\_H24}} & \colhead{\texttt{R\_J}} & \colhead{\texttt{R\_J\_H24}} & \colhead{\texttt{M\_J}} & \colhead{\texttt{P\_RJ}} & \colhead{\texttt{kind}} & \colhead{\texttt{kind\_H2024}} & \colhead{\texttt{nstars}} & \colhead{\texttt{comments}} \\
        \colhead{} & \colhead{} & \colhead{} & \colhead{(stars/pc$^3$)} & \colhead{(pc)} & \colhead{(pc)} & \colhead{} & \colhead{(stars/pc$^3$)} & \colhead{(pc)} & \colhead{(pc)} & \colhead{} & \colhead{(pc)} & \colhead{(pc)} & \colhead{(pc)} & \colhead{(pc)} & \colhead{(pc)} & \colhead{(pc)} & \colhead{(pc)} & \colhead{(M$_{\odot}$)} & \colhead{} & \colhead{} & \colhead{} & \colhead{} & \colhead{} \\
    }
    \colnumbers
    \startdata
    \texttt{ASCC\_13} & -- & -- & 4.1 & 1.75 & 23.80 & 0.00 & 0.7 & 5.33 & 24.20 & 28.48 & 5.91 & 5.01 & 4.98 & 21.45 & 21.87 & 3.97 & 10.36 & 24.1 & 0.247 & m & o & 117 & HQ \\
    \texttt{ASCC\_85} & -- & -- & 3.9 & 3.57 & 18.14 & 0.00 & 2.8 & 2.73 & 3.88 & 3.46 & 2.32 & 4.91 & 5.22 & 41.45 & 41.37 & 8.51 & 10.40 & 285.9 & 0.766 & o & o & 201 & HQ \\
    \texttt{ASCC\_88} & -- & -- & 44.6 & 9.10 & 6.78 & 0.07 & 2.3 & 2.80 & 13.62 & 31.81 & 6.85 & 2.70 & 2.83 & 6.89 & 6.85 & 6.61 & 9.27 & 135.8 & 0.979 & o & o & 96 & HQ \\
    \texttt{ASCC\_114} & -- & -- & 14.9 & 1.72 & 12.96 & 0.03 & 9.8 & 1.76 & 2.66 & 3.81 & 2.87 & 2.92 & 3.07 & 11.90 & 11.81 & 9.14 & 12.32 & 274.1 & 0.964 & o & o & 249 & HQ \\
    \texttt{Alessi\_3} & -- & -- & 3.6 & 4.10 & 18.81 & 0.02 & 1.9 & 3.84 & 6.75 & 4.93 & 11.36 & 5.06 & 5.38 & 17.83 & 20.49 & 6.81 & 5.40 & 116.5 & 0.651 & o & o & 186 & HQ \\
    \texttt{Alessi\_8} & -- & -- & 3.9 & 1.08 & 410.73 & 0.01 & 3.1 & 1.28 & 1.33 & 2.09 & 5.23 & 4.84 & 4.78 & 18.88 & 19.52 & 5.39 & 5.46 & 67.6 & 0.548 & o & o & 93 & HQ \\
    \texttt{Alessi\_13} & -- & -- & 2.0 & 2.37 & 77.31 & 0.00 & 1.1 & 3.58 & 4.40 & 2.72 & 9.38 & 9.31 & 9.42 & 39.26 & 42.07 & 4.56 & 4.06 & 35.3 & 0.305 & m & m & 167 & HQ \\
    \texttt{Alessi\_37} & -- & -- & 2.5 & 3.94 & 56.50 & 0.00 & 1.8 & 4.41 & 6.06 & 3.27 & 6.59 & 10.77 & 11.02 & 63.28 & 63.64 & 8.91 & 9.71 & 253.7 & 0.452 & o & o & 396 & HQ \\
    \texttt{Alessi\_84} & -- & -- & 1.7 & 2.40 & 81.41 & 0.00 & 1.3 & 2.94 & 3.49 & 2.60 & 5.53 & 6.95 & 6.57 & 27.72 & 27.70 & 5.03 & 3.95 & 45.5 & 0.391 & o & m & 128 & HQ \\
    \texttt{CWNU\_129} & -- & -- & -- & -- & -- & -- & -- & -- & -- & -- & 11.67 & 5.47 & 5.53 & 15.36 & 16.52 & 1.51 & 7.09 & 1.1 & 0.024 & m & o & 42 & HQ \\
    \texttt{CWNU\_136} & -- & -- & 189.4 & 44.57 & 14.69 & 0.01 & 0.5 & 6.22 & 20.61 & 15.16 & 12.74 & 6.76 & 6.61 & 25.47 & 28.55 & 1.54 & 6.14 & 1.3 & 0.009 & m & o & 107 & HQ \\
    \texttt{CWNU\_1129} & -- & -- & -- & -- & -- & -- & -- & -- & -- & -- & 4.58 & 5.54 & 5.28 & 17.59 & 21.70 & 3.13 & 3.00 & 1.1 & 0.029 & m & m & 34 & HQ \\
    \texttt{CWNU\_1143} & -- & -- & 39.3 & 8.82 & 8.79 & 0.02 & 0.8 & 3.45 & 15.50 & 27.94 & 6.60 & 4.33 & 3.09 & 13.77 & 17.29 & 3.92 & 3.16 & 23.2 & 0.438 & m & m & 73 & HQ \\
    \texttt{Casado\_20} & -- & -- & -- & -- & -- & -- & -- & -- & -- & -- & 1.32 & 4.61 & 1.51 & 18.07 & 19.29 & 5.17 & 6.80 & 42.0 & 0.538 & o & o & 39 & LQ \\
    \texttt{Chamaleon\_I} & -- & -- & 592.3 & 6.03 & 3.56 & 0.07 & 12.8 & 1.40 & 7.44 & 39.14 & 1.51 & 1.82 & 1.90 & 10.40 & 11.64 & 7.04 & 5.91 & 133.4 & 0.969 & o & o & 192 & HQ \\
    \texttt{FSR\_1017} & -- & -- & -- & -- & -- & -- & -- & -- & -- & -- & 9.83 & 3.88 & 4.36 & 15.45 & 17.73 & 1.72 & 5.19 & 1.9 & 0.294 & m & m & 17 & LQ \\
    \texttt{Gulliver\_49} & -- & -- & 7.8 & 2.41 & 21.42 & 0.01 & 5.1 & 2.65 & 4.11 & 3.98 & 2.66 & 4.51 & 4.40 & 24.25 & 25.05 & 10.31 & 16.27 & 340.5 & 0.858 & o & o & 296 & LQ \\
    \texttt{HSC\_453} & -- & -- & 0.2 & 6.72 & 326.08 & 0.00 & 0.1 & 11.44 & 17.06 & 3.78 & 24.19 & 15.79 & 12.39 & 43.15 & 69.71 & 2.60 & 4.93 & 6.6 & 0.045 & m & m & 110 & LQ \\
    \texttt{HSC\_572} & -- & -- & -- & -- & -- & -- & -- & -- & -- & -- & 17.06 & 7.50 & 8.52 & 17.59 & 17.06 & 1.35 & 5.00 & 0.6 & 0.031 & m & m & 32 & HQ \\
    \texttt{HSC\_590} & -- & -- & 84.2 & 77.49 & 21.62 & 0.00 & 0.1 & 9.43 & 44.78 & 31.47 & 26.72 & 10.27 & 12.27 & 34.14 & 45.46 & 1.72 & 2.88 & 1.9 & 0.070 & m & m & 71 & HQ \\
     \ldots & \ldots & \ldots & \ldots & \ldots & \ldots & \ldots & \ldots & \ldots & \ldots & \ldots & \ldots & \ldots & \ldots & \ldots & \ldots & \ldots & \ldots & \ldots & \ldots & \ldots & \ldots & \ldots & \ldots \\
    \texttt{Hyades} & 1004 & 1 & 20.0 & 6.55 & 11.90 & 0.06 & 5.3 & 4.18 & 20.21 & 32.87 & -- & 4.53 & -- & 14.51 & -- & 10.56 & -- & 435.6 & 0.961 & o & -- & 492 & HQ \\
     & 1008 & 2 & 0.5 & 13.71 & 61.45 & 0.00 & 0.2 & 14.91 & 35.57 & 8.61 & -- & 17.81 & -- & 68.78 & -- & 3.44 & -- & 15.1 & 0.062 & m & -- & 305 & LQ \\
     & 1009 & 3 & -- & -- & -- & -- & -- & -- & -- & -- & -- & 13.81 & -- & 55.29 & -- & 1.38 & -- & 0.9 & 0.021 & m & -- & 48 & HQ \\
    \texttt{Alessi\_3} & 1011 & 4 & 3.5 & 5.56 & 21.63 & 0.01 & 1.4 & 5.63 & 14.18 & 9.47 & -- & 6.41 & -- & 23.18 & -- & 8.08 & -- & 194.3 & 0.641 & o & -- & 270 & HQ \\
     & 1019 & 5 & 2.1 & 6.07 & 29.52 & 0.01 & 0.2 & 13.95 & 16.63 & 2.61 & -- & 26.46 & -- & 77.52 & -- & 1.49 & -- & 1.5 & 0.007 & m & -- & 307 & LQ \\
    \texttt{NGC\_2527} & 1047 & 6 & 8.1 & 2.56 & 39.63 & 0.00 & 5.4 & 3.08 & 4.23 & 3.25 & -- & 6.56 & -- & 38.85 & -- & 11.32 & -- & 503.3 & 0.679 & o & -- & 546 & HQ \\
     & 1080 & 7 & 10.4 & 81.82 & 35.26 & 0.00 & 0.1 & 14.22 & 47.88 & 15.57 & -- & 16.71 & -- & 87.75 & -- & 2.18 & -- & 1.1 & 0.010 & m & -- & 100 & HQ \\
    \texttt{NGC\_2215} & 1087 & 8 & 2.9 & 2.23 & 59.32 & 0.00 & 2.1 & 2.69 & 3.50 & 2.98 & -- & 7.35 & -- & 78.87 & -- & 7.68 & -- & 141.9 & 0.502 & o & -- & 201 & HQ \\
    \texttt{UBC\_165} & 108 & 9 & 2.2 & 3.32 & 36.48 & 0.00 & 1.6 & 3.56 & 4.88 & 3.24 & -- & 5.91 & -- & 29.34 & -- & 6.97 & -- & 123.0 & 0.527 & o & -- & 182 & HQ \\
     & 1098 & 10 & 62.3 & 78.58 & 22.94 & 0.00 & 0.1 & 14.17 & 19.13 & 3.21 & -- & 22.37 & -- & 51.78 & -- & 1.18 & -- & 0.7 & 0.005 & m & -- & 185 & LQ \\
     & 1103 & 11 & 39.2 & 77.15 & 21.69 & 0.00 & 0.0 & 12.32 & 50.92 & 22.94 & -- & 13.90 & -- & 36.96 & -- & 1.50 & -- & 1.3 & 0.038 & m & -- & 52 & HQ \\
     & 1114 & 12 & -- & -- & -- & -- & 0.0 & 49.43 & 87.18 & 4.73 & -- & 56.24 & -- & 103.94 & -- & 1.17 & -- & 0.7 & 0.003 & m & -- & 314 & LQ \\
    \texttt{Alessi\_62} & 1131 & 13 & 8.7 & 1.64 & 57.30 & 0.00 & 6.3 & 2.03 & 2.44 & 2.63 & -- & 6.60 & -- & 47.17 & -- & 7.93 & -- & 205.1 & 0.536 & o & -- & 351 & HQ \\
     & 1141 & 14 & 0.4 & 31.00 & 50.89 & 0.00 & 0.1 & 18.81 & 64.50 & 16.66 & -- & 20.53 & -- & 70.38 & -- & 1.72 & -- & 2.0 & 0.012 & m & -- & 169 & LQ \\
    \texttt{IC\_2391} & 114 & 15 & 34.4 & 1.35 & 7.88 & 0.14 & 18.4 & 1.42 & 2.39 & 4.48 & -- & 1.96 & -- & 6.95 & -- & 8.80 & -- & 254.8 & 1.000 & o & -- & 253 & HQ \\
     & 1153 & 16 & 0.2 & 7.14 & 214.20 & 0.00 & 0.1 & 14.89 & 61.18 & 22.99 & -- & 15.03 & -- & 36.07 & -- & 3.45 & -- & 1.4 & 0.013 & m & -- & 77 & LQ \\
     & 116 & 17 & 1.5 & 5.70 & 248.07 & 0.00 & 1.3 & 5.98 & 6.46 & 2.24 & -- & 14.45 & -- & 44.03 & -- & 6.86 & -- & 116.3 & 0.214 & m & -- & 552 & LQ \\
     & 1172 & 18 & 12.4 & 81.43 & 24.73 & 0.00 & 0.0 & 14.04 & 20.36 & 3.59 & -- & 23.23 & -- & 75.28 & -- & 2.20 & -- & 4.2 & 0.054 & m & -- & 56 & LQ \\
    \texttt{NGC\_2423} & 1176 & 19 & 7.9 & 2.65 & 82.29 & 0.00 & 5.8 & 3.35 & 4.33 & 2.96 & -- & 8.77 & -- & 99.77 & -- & 12.10 & -- & 569.2 & 0.597 & o & -- & 755 & HQ \\
     & 1179 & 20 & 1.0 & 2.78 & 445.18 & 0.01 & 0.4 & 6.12 & 7.40 & 2.74 & -- & 9.65 & -- & 25.65 & -- & 4.75 & -- & 39.0 & 0.254 & m & -- & 118 & LQ \\
    \enddata
    \tablecomments{This table is published in its entirety in machine-readable format. We just show 40 rows with key parameters(the entire parameters with errors are shown in \ref{combine}).}
\end{deluxetable*}

\subsection{Catalog of members of OCs and MGs with Planets/Candidates} \label{plclm}
There are 76 OCs and 36 MGs from sample 1 and 56 OCs and 112 MGs from sample 2. We combine all of the member stars(134,351) in sample 1(58,128) and 2(76,223) into the catalog of members of OCs and MGs with planets/candidates.

\begin{deluxetable*}{lrrrrrrrrrrrrrrr}
    \rotate
    \tabletypesize{\tiny} 
    \tablecaption{Catalog of members of OCs and MGs with Planets/Candidates\label{plclmt}}
    \tablewidth{0pt}
    \tablenum{A5}
    \tablehead{
        \colhead{\texttt{Source\_ID}} & \colhead{\texttt{Cluster\_Name}} & \colhead{\texttt{Group}} & \colhead{\texttt{Newgroup}} & \colhead{\texttt{X\_proj}} & \colhead{\texttt{Y\_proj}} & \colhead{\texttt{ra}} & \colhead{\texttt{dec}} & \colhead{\texttt{pmra}} & \colhead{\texttt{pmdec}} & \colhead{\texttt{phot\_g\_mean\_mag}} & \colhead{\texttt{bp\_rp}} & \colhead{\texttt{mass\_50}} & \colhead{\texttt{T\_eff}} & \colhead{\texttt{r\_med\_geo}} & \colhead{\texttt{r\_med\_photogeo}} \\
        \colhead{} & \colhead{} & \colhead{} & \colhead{} & \colhead{(pc)} & \colhead{(pc)} & \colhead{(deg)} & \colhead{(deg)} & \colhead{(mas/yr)} & \colhead{(mas/yr)} & \colhead{(mag)} & \colhead{(mag)} & \colhead{(M$_{\odot}$)} & \colhead{(K)} & \colhead{(pc)} & \colhead{(pc)} \\
    }
    \colnumbers
    \startdata
    202372193199103360 & \texttt{ASCC\_13} & -- & -- & -3.33 & -9.82 & 77.9610 & 43.8772 & -0.58 & -1.74 & 11.604 & 0.209 & 2.685 & 7997 & 1011.5 & 1016.2 \\
    202379825360029824 & \texttt{ASCC\_13} & -- & -- & -1.48 & -5.39 & 78.0971 & 44.1132 & -0.45 & -1.88 & 14.775 & 1.073 & 1.228 & 5759 & 949.0 & 960.2 \\
    202559488134232704 & \texttt{ASCC\_13} & -- & -- & -5.17 & -7.50 & 77.8244 & 44.0007 & -0.55 & -1.68 & 14.688 & 0.944 & 1.248 & 6036 & 1101.4 & 1099.9 \\
    202560007828803328 & \texttt{ASCC\_13} & -- & -- & -6.65 & -7.68 & 77.7149 & 43.9907 & -0.57 & -1.77 & 13.932 & 0.809 & 1.446 & 7005 & 1099.3 & 1102.9 \\
    202568975720365568 & \texttt{ASCC\_13} & -- & -- & -1.53 & -3.13 & 78.0931 & 44.2339 & -0.47 & -1.72 & 17.940 & 1.915 & 0.706 & 3688 & 1072.8 & 1000.6 \\
    202570212670995712 & \texttt{ASCC\_13} & -- & -- & -3.39 & -4.14 & 77.9556 & 44.1796 & -0.61 & -1.85 & 17.095 & 1.519 & 0.805 & 5641 & 1135.2 & 1186.4 \\
    202572166877684224 & \texttt{ASCC\_13} & -- & -- & -2.52 & -2.61 & 78.0195 & 44.2611 & -0.57 & -1.91 & 11.112 & 0.219 & 3.136 & -- & 1050.0 & 1046.9 \\
    202572407396443136 & \texttt{ASCC\_13} & -- & -- & -1.08 & -2.16 & 78.1270 & 44.2855 & -0.54 & -1.78 & 16.350 & 1.331 & 0.917 & 4809 & 1112.5 & 1140.8 \\
    202572617852595968 & \texttt{ASCC\_13} & -- & -- & -0.77 & -1.84 & 78.1499 & 44.3023 & -0.53 & -2.09 & 11.080 & 0.260 & 3.167 & 7696 & 1061.2 & 1065.7 \\
    202572716633496832 & \texttt{ASCC\_13} & -- & -- & -1.43 & -1.78 & 78.1007 & 44.3055 & -0.64 & -1.72 & 10.571 & 0.172 & 3.673 & 12351 & 1094.6 & 1089.8 \\
    202573850505695232 & \texttt{ASCC\_13} & -- & -- & -1.43 & -0.77 & 78.1006 & 44.3596 & -0.37 & -1.86 & 15.810 & 1.164 & 1.013 & 6007 & 1010.5 & 1023.7 \\
    202573987944702208 & \texttt{ASCC\_13} & -- & -- & -1.02 & -0.37 & 78.1311 & 44.3805 & -0.67 & -2.04 & 17.694 & 1.731 & 0.731 & 4230 & 1312.0 & 1241.8 \\
    202575671571115648 & \texttt{ASCC\_13} & -- & -- & -8.51 & -5.32 & 77.5761 & 44.1153 & -0.62 & -1.81 & 16.187 & 1.595 & 0.944 & 4677 & 1146.4 & 1134.3 \\
    202575985107124096 & \texttt{ASCC\_13} & -- & -- & -8.29 & -4.74 & 77.5917 & 44.1464 & -0.50 & -1.80 & 14.184 & 0.864 & 1.375 & 6786 & 1086.2 & 1088.5 \\
    202576736724799360 & \texttt{ASCC\_13} & -- & -- & -6.32 & -3.96 & 77.7376 & 44.1887 & -0.51 & -1.57 & 18.926 & 2.223 & 0.609 & 3532 & 1519.3 & 1052.1 \\
    202578527728103936 & \texttt{ASCC\_13} & -- & -- & -9.27 & -4.77 & 77.5190 & 44.1443 & -0.51 & -1.51 & 14.567 & 0.993 & 1.276 & 6775 & 1114.9 & 1104.6 \\
    202579657300937344 & \texttt{ASCC\_13} & -- & -- & -9.24 & -3.87 & 77.5211 & 44.1925 & -0.44 & -1.84 & 16.885 & 1.723 & 0.835 & 4498 & 1205.3 & 1145.6 \\
    202582307298898304 & \texttt{ASCC\_13} & -- & -- & -6.21 & -1.41 & 77.7446 & 44.3245 & -0.82 & -1.87 & 11.237 & 0.276 & 3.014 & 7563 & 1053.0 & 1057.3 \\
    202583406810503936 & \texttt{ASCC\_13} & -- & -- & -3.79 & -2.35 & 77.9249 & 44.2748 & -0.41 & -1.78 & 17.591 & 2.196 & 0.740 & 4413 & 1210.3 & 1173.7 \\
    202583578605150592 & \texttt{ASCC\_13} & -- & -- & -3.30 & -1.72 & 77.9613 & 44.3086 & -0.65 & -1.92 & 20.075 & 1.482 & 0.492 & -- & 6946.1 & 5294.2 \\
     \ldots & \ldots & \ldots & \ldots & \ldots & \ldots & \ldots & \ldots & \ldots & \ldots & \ldots & \ldots & \ldots & \ldots & \ldots & \ldots \\
    3312637842237882880 & \texttt{Hyades} & 1004 & 1 & 0.66 & -0.20 & 67.9661 & 15.8515 & 99.04 & -33.68 & 5.916 & 0.468 & 2.600 & 6920 & 47.5 & 47.5 \\
    3312644885984344704 & \texttt{Hyades} & 1004 & 1 & 0.88 & -0.23 & 68.2482 & 15.8189 & 100.39 & -25.53 & 8.469 & 0.912 & 1.400 & 5531 & 46.8 & 46.7 \\
    3312709379213017728 & \texttt{Hyades} & 1004 & 1 & -0.20 & -0.43 & 66.9000 & 15.5891 & 104.99 & -24.07 & 7.274 & 0.700 & 1.749 & 6069 & 46.9 & 46.9 \\
    3312751882209460864 & \texttt{Hyades} & 1004 & 1 & -0.50 & -0.48 & 66.5249 & 15.5242 & 131.02 & -24.15 & 7.336 & 0.722 & 1.749 & 5984 & 42.5 & 42.4 \\
    3312837025641272320 & \texttt{Hyades} & 1004 & 1 & 0.44 & 0.04 & 67.6955 & 16.1485 & 114.13 & -33.76 & 6.386 & 0.589 & 2.249 & 6476 & 44.5 & 44.5 \\
    3312842042162993536 & \texttt{Hyades} & 1004 & 1 & 0.06 & -0.10 & 67.2190 & 15.9815 & 110.61 & -27.25 & 11.246 & 1.905 & 0.899 & -- & 45.1 & 45.1 \\
    3312882483575093760 & \texttt{Hyades} & 1004 & 1 & 0.52 & 0.25 & 67.7964 & 16.3958 & 98.72 & -24.69 & 13.815 & 2.751 & 0.650 & -- & 49.1 & 49.1 \\
    3312899491645515776 & \texttt{Hyades} & 1004 & 1 & 0.18 & 0.12 & 67.3796 & 16.2447 & 104.18 & -21.07 & 9.927 & 1.384 & 1.050 & 4465 & 45.8 & 45.8 \\
    3312904233289409152 & \texttt{Hyades} & 1004 & 1 & 0.08 & 0.21 & 67.2511 & 16.3462 & 104.87 & -27.38 & 13.095 & 2.646 & 0.700 & -- & 46.1 & 46.1 \\
    3313207698500079360 & \texttt{Hyades} & 1004 & 1 & 1.98 & 0.80 & 69.6319 & 17.0406 & 105.98 & -31.21 & 13.642 & 2.810 & 0.650 & -- & 43.7 & 43.6 \\
    3313259169388356608 & \texttt{Hyades} & 1004 & 1 & 1.01 & 0.56 & 68.4086 & 16.7624 & 97.96 & -25.51 & 9.153 & 1.064 & 1.200 & 5032 & 48.4 & 48.4 \\
    3313285110990793472 & \texttt{Hyades} & 1004 & 1 & 1.07 & 0.65 & 68.4868 & 16.8689 & 101.61 & -28.33 & 13.767 & 2.763 & 0.650 & -- & 46.4 & 46.4 \\
    3338650225766660992 & \texttt{Hyades} & 1004 & 1 & 9.74 & -5.26 & 79.0663 & 9.4466 & 58.90 & -7.03 & 14.274 & 2.777 & 0.609 & -- & 53.8 & 53.8 \\
    3339182389394565760 & \texttt{Hyades} & 1004 & 1 & 11.60 & -4.07 & 81.4374 & 10.7341 & 40.49 & -6.97 & 12.838 & 2.172 & 0.700 & -- & 66.6 & 66.7 \\
    3339344704798578944 & \texttt{Hyades} & 1004 & 1 & 11.85 & -3.39 & 81.7963 & 11.5258 & 40.87 & -5.41 & 8.365 & 0.746 & 1.400 & 5914 & 62.8 & 62.8 \\
    3340855571214473600 & \texttt{Hyades} & 1004 & 1 & 12.27 & -3.08 & 82.3392 & 11.8585 & 38.60 & -11.55 & 15.711 & 3.072 & 0.450 & -- & 70.6 & 70.5 \\
    3387381646261643776 & \texttt{Hyades} & 1004 & 1 & 10.51 & -3.41 & 80.1062 & 11.6098 & 52.26 & -13.99 & 8.840 & 0.923 & 1.300 & 5456 & 55.6 & 55.6 \\
    3387573300585597184 & \texttt{Hyades} & 1004 & 1 & 9.91 & -3.30 & 79.3672 & 11.7907 & 50.33 & -12.24 & 11.860 & 1.884 & 0.800 & -- & 58.2 & 58.3 \\
    3390181861859258240 & \texttt{Hyades} & 1004 & 1 & 10.17 & -1.50 & 79.7962 & 13.9248 & 29.40 & -14.51 & 14.821 & 2.746 & 0.589 & -- & 86.5 & 86.5 \\
    3391712038447198080 & \texttt{Hyades} & 1004 & 1 & 6.98 & -1.83 & 75.7822 & 13.7306 & 66.20 & -18.02 & 8.786 & 0.927 & 1.300 & 5407 & 53.8 & 53.8 \\
    \enddata
    \tablecomments{This table is published in its entirety in machine-readable format. We just show 40 rows with key stellar parameters for cluster members. The first 20 rows show data from \texttt{ASCC\_13} cluster, followed by 20 rows from \texttt{Hyades}.}
\end{deluxetable*}
\clearpage

\section{Cross-matching with \cite{QIN2023}} \label{Qin2023}
We also cross-matched the Open Cluster of Solar Neighborhood (OCSN) catalog from \cite{QIN2023} with our planet catalog. This survey, which used Gaia DR3 data to identify 324 open clusters within 500 pc, yielded three confirmed planets (HD 31253 b in OCSN 220, Kepler-1643 b in OCSN 256, and USco CTIO 108 b in OCSN 98), one planetary candidate (PATHOS 38/TIC 123755508.01 in OCSN 199 \citep{QIN2023}), and four false positives (TIC 96246348.01 in OCSN 163, TIC 21755546.01 in OCSN 152, TIC 129991079.01 in OCSN 302, and TIC 405484221.01 in OCSN 219). Given its small size, this additional sample does not significantly affect our overall conclusions.

\bibliography{reference}{}
\bibliographystyle{aasjournal}


\end{document}